\documentclass[]{aastex631}
\usepackage{longtable}
\usepackage{multirow}

\newcommand{\eps}{erg s$^{-1}$}
\newcommand{\nh}{$N_{\rm H}$}

\newcommand{\M}{$M_{\odot}$}
\newcommand{\led}{$\log(\rm L_{\rm Edd})$}
\newcommand{\lse}{$\log(\rm L_{\rm SE}/L_{\rm Edd})$}

\newcommand{\lpc}{$\log(\rm L_{\rm PC}/L_{\rm Edd})$}

\newcommand{\gpc}{$\Gamma^{\rm PC}$}
\newcommand{\gse}{$\Gamma^{\rm SE}$}
\shorttitle{Soft excess study for Bare AGNs}
\shortauthors{Nandi et al.}
\graphicspath{{./}{figures/}}

\begin{document}

\title{Survey of Bare Active Galactic Nuclei in the local universe ($z<0.2$): I. On the origin of Soft-Excess}

\correspondingauthor{Prantik Nandi, Arka Chatterjee}
\email{prantiknandi007@gmail.com, arka.chatterjee@umanitoba.ca}

\author[0000-0003-3840-0571]{Prantik Nandi}
\affiliation{Physical Research Laboratory, Navrangpura, Ahmedabad, 380009, India}

\author[0000-0003-3932-6705]{Arka Chatterjee}
\affiliation{Department of Physics \& Astronomy, University of Manitoba, Winnipeg, R3T 2N2, Canada}

\author[0000-0001-7500-5752]{Arghajit Jana}
\affiliation{Institute of Astronomy, National Tsing Hua University, Hsinchu, 300044, Taiwan}

\author[0000-0002-0193-1136]{Sandip K. Chakrabarti}
\affiliation{Indian Centre for Space Physics, Netaji Nagar,  Kolkata, 700099, India}

\author[0000-0003-2865-4666]{Sachindra Naik}
\affiliation{Physical Research Laboratory, Navrangpura, Ahmedabad, 380009, India}

\author[0000-0001-6189-7665]{Samar Safi-Harb}
\affiliation{Department of Physics \& Astronomy, University of Manitoba, Winnipeg, R3T 2N2, Canada}

\author[0000-0002-5617-3117]{Hsiang-Kuang Chang}
\affiliation{Institute of Astronomy, National Tsing Hua University, Hsinchu, 300044, Taiwan}

\author[0000-0001-9739-367X]{Jeremy Heyl}
\affiliation{Department of Physics and Astronomy, University of British Columbia, Vancouver, V6T 1Z1, Canada}



\begin{abstract}

We analyse a sample of 21 `bare' Seyfert~1 Active Galactic Nuclei (AGNs), a sub-class of Seyfert~1s, with intrinsic absorption $\mathrm{N_{H}} \sim 10^{20}~ \mathrm{cm}^{-2}$, in the local universe (z $<$ 0.2) using {\it XMM-Newton} and {\it Swift}/XRT observations. The luminosities of the primary continuum, the X-ray emission in the 3 to 10 keV energy range and the soft-excess, the excess emission that appears above the low-energy extrapolation of the power-law fit of 3 to 10 keV X-ray spectra, are calculated. Our spectral analysis reveals that the long-term intrinsic luminosities of the soft-excess and the primary continuum are tightly correlated $(L_{PC}\propto L_{SE}^{1.1\pm0.04})$. We also found that the luminosities are correlated for each source. This result suggests that both the primary continuum and soft excess emissions exhibit a dependency on the accretion rate in a similar way.

\end{abstract}



\section{Introduction} \label{sec:intro}
Most massive galaxies nurture supermassive black holes (SMBHs), having mass M$_{BH}\sim10^6-10^{9.5}$ M$_\odot$, at their centre \citep{Kormendy1995,Kormendy2013}. The accretion of matter onto the SMBHs is one of the most efficient mechanisms to transform gravitational potential energy into electromagnetic radiation. The radiation spans the entire range of the electromagnetic spectrum, from radio to $\gamma$-rays. Using the X-ray band, it is possible to explore the innermost regions of the accretion disk \citep{Shakura1973} and the Compton cloud or corona \citep{Fabian2015, Fabian2017} around these SMBHs. A galactic nucleus becomes active when the radiation in any energy band surpasses significantly its stellar or thermal radiation and is classified as an Active Galactic Nucleus (AGN). The primary source of emission in the X-ray band from the AGNs is an optically thin and hot $(T\sim10^9~K)$ corona through the process of inverse Compton scattering \citep{Sunyaev1980} of the UV photons, originating from the accretion disk \citep{Shakura1973}. The scatterings produce a power-law spectrum with a sharp cut-off in the X-ray band \citep{Sunyaev1980, Haardt1991, Haardt1993}. A fraction of coronal continuum photons could get reprocessed in the colder circumnuclear matter, like a dusty torus, broad line region (BLR), and narrow line region (NLR), producing several spectral features, like absorption lines and iron lines having various shapes \citep{Jana2020}.  The X-ray spectra of AGNs are often associated with an excess emission below 2 keV, known as soft-excess \citep{Halpern1984, Arnaud1985, Fabian2009, Done2012, Garcia2014, Nandi2021} and several ﬂuorescent emission lines in the soft and hard X-rays. Among them, the most prominent and ubiquitous is the Fe~K$_\alpha$ line at $\sim6.4$ keV \citep{Ross2005, Fabian2009, Garcia2010}. The Fe~K$_\alpha$ line could be broadened and distorted by the relativistic effects due to the strong gravitational field around the black hole \citep{Fabian1989, Laor1991}. However, the broadening of the Fe~K$_\alpha$ line is not omnipresent. The narrow K$_\alpha$ line is believed to originate far from the black hole in the broad-line (BLR) or in the distant torus region \citep{George1991,Matt1991}. The Comptonized photons could also be reflected from the ionized accretion disc which produces a Compton hump above 15 keV \citep{Magdziarz1995, Murphy2009}. The Compton hump is generally found in the energy band of 15 to 50 keV and peaks between 20 to 30 keV. The low-energy part of this hump is generally shaped by the photoelectric absorption of iron in the reﬂector. In contrast,  the high-energy counterpart is formed by the process of down-scattering of high-energy photons from the Compton cloud reprocessed in the accretion disc or distant matter 
\citep{Pounds1990,Nandra1991}.  
	
The soft-excess, an excess emission below 2 keV, is an extraordinary feature in the X-ray spectra for most of the Seyfert~1 AGNs \citep{Pravdo1981, Halpern1984, Arnaud1985, Turner1989}. The origin of soft-excess is one of the major open questions in AGN research \citep{Turner2009}, even about four decades after its discovery. The excess emission appears above the low-energy extrapolation of the power-law fit of 3 to 10 keV X-ray spectra. Historically associated with the high-energy counterpart of the blackbody radiation coming from the accretion disk, it has been shown that modelling soft-excess with thermal continuum \citep{Singh1985, Pounds1986, Leighly1999, Czerny2003, Gierlinski2004, Piconcelli2005, Crummy2006} indicates a characteristic temperature which is much higher than the expectations from the standard disk \citep{Shakura1973}. Besides, the temperature remains remarkably constant across a range of AGNs despite the widespread black hole mass and AGN luminosity \citep{Bechtold1987, Vaughan2002, Piconcelli2005}. Moreover, the blackbody luminosity-temperature ($\sigma T^4$) relation is not followed by the soft-excess emission from the bright and variable AGNs \citep{Ponti2006}. It was also observed that the ratio between the soft excess at 0.5~keV and the extrapolation of the high-energy power-law emission has a small scatter \citep{Piconcelli2005, Miniutti2009}. This departs from the Galactic black holes in their bright soft state, where the radiation is mostly dominated by the disk black-body emission \citep{Done2007}. Moreover, considering the timing aspect of the soft-excess with respect to the primary continuum, if a thermal blackbody generates the soft-excess, then 0.5 to 2 keV from the standard accretion disk should lead to the 3-10 keV Comptonized primary continuum in time. However, this was not observed for all cases \citep{Nandi2021}. These discoveries provoked the alternate origins of the soft-excess. One of the popular ideas suggests that Comptonization might take place in the upper layer of the accretion disc \citep{Czerny2003, Sobolewska2007, Jiang2018, Garcia2019}. This model explains some of the characteristics of soft-excess, such as the shape of the soft excess, its high temperature, coronal-disc feedback etc. On the other hand, the observed high temperature of the soft-excess suggests a nature tied to the atomic processes. If the upper layer of the disc is ionized, the reflection component will contain many X-ray lines \citep{Ross2005} which will be broadened and distorted due to relativistic effects. Ionized absorption features are also imprinted into the X-ray band. Numerical simulations show \citep{Schurch2008} that the outflowing absorbers could produce soft-excess emission. However, velocities of the absorbers of the order of $\sim 0.9c$ are needed to reproduce the excess emission in the soft X-ray band. 

The other proposed model to explain the soft-excess is a warm Comptonizing corona model \citep{Czerny1987, Middleton2009, Done2012, Kubota2018, Petrucci2018}. In this scenario, the UV photons are Compton up-scattered in a warm $(kT_e\sim0.1-1~\text{keV})$ and optically thick $(\tau\sim10-40)$ corona which is somewhat sandwiching the inner region of the disk. Recently, \citet{Nandi2021} reported a strong correlation (a Pearson coefficient of 0.9) between the soft-excess and the primary continuum luminosities in the 0.5--10 keV energy band for Ark~120, a well-known `bare' AGN. Their finding suggests the origin of both luminosities could be linked to a similar radiation process. Further, \citet{Bechtold1987, Vaughan2002} suggested a plausible cause of soft-excess emission which depends on the inverse Compotization in the Compton cloud or hot corona. From the Monte-Carlo simulations, it was observed that the fewer scatterings in the corona could provide the steeper power-law spectrum for the soft X-ray regime while the higher number of scatterings could produce the primary continuum \citep{Nandi2021}. 
	
In this paper, we apply a holistic approach to probe the origin of the soft-excess in `bare' AGNs, a sub-class of Seyfert~1s, with intrinsic neutral and ionized absorption $\mathrm{N_{H}} \sim 10^{20}~ \mathrm{cm}^{-2}$, in the local universe (at a redshift z $<$ 0.2). We utilize a large sample of archival data obtained with {\it XMM-Newton} and {\it Swift}/XRT. Our study focused on investigating the origin of the soft excess from an observational standpoint. Instead of employing physical or phenomenological models to fit the observed spectra, we exclusively utilized a power law to parameterize the observed spectrum. The purpose of our work was focused to explore the physical mechanism or drivers of the soft excess emission, rather than to validate or establish any specific model. The paper is organized as follows. Section~\ref{observations & data reductions} describes the observations; the details on the source selection and the data reduction procedures are presented in Section~\ref{sec:sample selection} and Section~\ref{sec:data reduction}, respectively. Then, in Section~\ref{sec:RandD}, we discuss the global results of our analysis, where we also discuss the inter-dependency of various parameters, such as spectral indices (Section~\ref{sec:gpc}), luminosities (Section~\ref{sec:luminosity}), and a plausible explanation of the origin of the soft-excess in our sample in Section~\ref{sec:se}. Finally, the conclusions are summarized in Section~\ref{sec:Conclusions}.

	
\section{Observations and Data Reductions}
\label{observations & data reductions}
	
\subsection{Sample Selection}
\label{sec:sample selection}
\begin{figure*}
\centering
  \includegraphics[trim={0 0 0 4cm}, width=1.0\textwidth, angle =0]{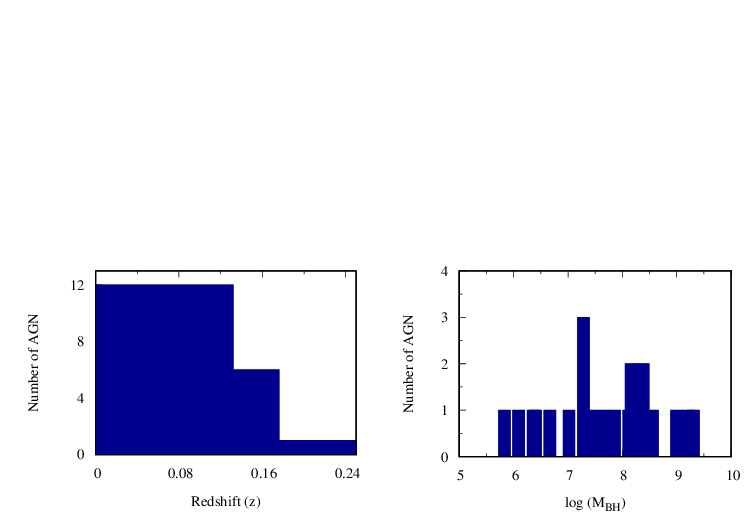}
\caption{Distribution of sources with respect to redshift (z), mass of central black holes (M$_\mathrm{BH}$) are shown in the left, middle, and right histograms, respectively.}
\label{fig:source_dist}
\end{figure*}

To understand the soft-excess emission, we focused on the `bare' type of AGNs, a sub-class of Seyfert~1s, with intrinsic neutral and ionized absorption $\mathrm{N_{H}} \sim 10^{20}~ \mathrm{cm}^{-2}$. They exhibit excess emission below $\sim$2.0 keV energy. In this study, we followed the same procedures described by \citep{Walton2013}. Initially, we fitted the 3.0 to 10.0 keV X-ray spectra with the {\tt powerlaw} model along with the Galactic and extragalactic absorption using {\tt TBabs} and {\tt zTbabs} models, respectively. Then the values for Galactic hydrogen column densities are calculated using NASA's HEASARC online tool\footnote{\url{https://heasarc.gsfc.nasa.gov/cgi-bin/Tools/w3nh/w3nh.pl}}. For some cases, we found the presence of Fe K-$\alpha$ line within the energy range of 6--7 keV. We modelled the line with a {\tt Gaussian} function which tackles the iron emission present in the spectra and provides reliable determination of the continuum. Following that, we selected our sources which have relatively smoother and cleaner excess emissions below 2.0 keV. The resulting sample contains 21 sources which have no less than a total of five observations with {\it XMM-Newton} and {\it Swift}/XRT. The observational details of the sources are provided in Table~\ref{table:source_details}. The distribution of sources with respect to various intrinsic parameters, such as redshift ($z$) and the mass of the central black hole ($\mathrm{M_{BH}}$) are shown in Figure~\ref{fig:source_dist}. The sample considers the `bare' type of AGNs in the local universe, where most of the sources have redshifts $\mathrm{z} \leq 0.1$. The sample, however, has only three sources with $0.1 < \mathrm{z} < 0.2$. The Eddington luminosity ($\mathrm{L_{Edd}}$) \citep{Rybicki1979} is calculated using the relation
	\begin{equation}
			\mathrm{L_{\rm Edd}}=\frac{4\pi \mathrm{GM}_{\mathrm {BH}} \mathrm{m_p c}}{\sigma_\mathrm{T}} \\
			\approx 1.26 \times 10^{38} \big(\frac{\mathrm{M}_{\mathrm {BH}}}{\mathrm{M}_\odot}\big) ~\text{erg/s},
	\end{equation}
where, $\mathrm{G}$ is the universal gravitational constant, $\mathrm{m_p}$ is the mass of proton, $\mathrm{c}$ is the speed of light in vacuum, $\sigma_\mathrm{T}$ is the Thompson scattering cross-section, and $\mathrm{M}_{\odot}$ is the Solar mass.  
	
\begin{table*}
\caption{Information on the sample used in this work is presented. The positions and the redshifts ($z$) are obtained from the NASA Extra-galactic Database. The Galactic hydrogen column densities ($N_{\rm H}$) are collected from \citep{Kalberla2005} and used as constant values for {\tt TBabs} during the spectral fitting. The references for the black hole mass are addressed with numerical annotations.}
\begin{center}
\label{table:source_details}
			\begin{tabular}{c c c c c c c}
				\hline\hline
				Source & RA & Dec.& $z$& $N_{\rm H,Gal}$ & $\log(M_{\rm BH}/M_{\odot})$ & $\log(L_{\rm Edd})$ \\
				& (h:m:s) & (d:m:s) &    & $(10^{20}$ cm$^{-2}) $ & & (erg $s^{-1}$)  \\
				
				\hline
				1H 0323+342       &03:24:41.1 &+34:10:46 & 0.0610 & 12.7 & $7.30~(1)$ & 45.40\\
				1H 0419--577       &04:26:00.8 &-57:12:00 & 0.1040 & 1.26 & $8.58~(2)$ & 46.68\\
				1H 0707--495       &07:08:41.5 &-49:33:06 & 0.0406 & 4.31 & $6.37~(3)$ & 44.40\\
				3C 382            &18:35:03.4 &+32:41:47 & 0.0579 & 6.98 & $8.98~(4)$ & 47.08\\
				3C 390.3          &18:42:09.0 &+79:46:17 & 0.0561 & 3.47 & $9.30~(5)$ & 47.40\\
				Ark 120           &05:16:11.4 &-00:08:59 & 0.0327 & 9.78 & $8.18~(6)$ & 46.28\\
				Ark 564           &22:42:39.3 &+29:43:31 & 0.0247 & 4.34 & $6.40~(7)$ & 44.50\\
				Fairall 9         &01:23:45.8 &-58:48:20 & 0.0470 & 3.16 & $8.40~(6)$ & 46.51\\
				IRAS 13224--3809   &13:12:19.4 &-38:24:53 & 0.0658 & 5.34 & $6.30~(8)$ & 44.40\\
				Mrk 1018          &02:06:16.0 &-00:17:29 & 0.0424 & 2.43 & $7.85~(9)$ & 45.95\\
				Mrk 110           &09:25:12.9 &+52:17:11 & 0.0353 & 1.30 & $8.14~(10)$ & 46.36\\
				Mrk 335           &00:06:19.5 &+20:12:10 & 0.0258 & 3.56 & $7.43~(11)$ & 45.53\\
				Mrk 359           &01:27:32.5 &+19:10:44 & 0.0174 & 4.26 & $6.65~(12)$ & 44.46\\
				Mrk 509           &20:44:09.7 &-10:43:25 & 0.0344 & 4.25 & $8.15~(6)$ & 46.45\\
				Mrk 841           &15:04:01.2 &+10:26:16 & 0.0364 & 2.22 & $7.30~(13)$ & 45.40\\
				PDS 456           &17:28:19.8 &-14:15:56 & 0.1854 & 19.6 & $9.20~(14)$ & 47.30\\
				PKS 0558--504      &05:59:47.4 &-50:26:52 & 0.1372 & 3.36 & $8.4~(15)$ & 46.50\\
				SWIFT J0501.9--3239&05:19:35.8 &-32:39:38 & 0.0124 & 1.75 & $7.65~(16)$ & 45.79\\
				NGC 7469          &23:03:15.6 &+08:52:26 & 0.0163 & 4.45 & $7.00~(6)$ & 45.10\\
				Ton S180          &00:57:19.9 &-22:22:59 & 0.0620 & 1.36 & $7.30~(17)$ & 45.40\\
				UGC 6728          &11:45:16.0 &+79:40:53 & 0.0065 & 4.42 & $5.85~(18)$ & 43.95\\
				\hline
			\end{tabular}
		\end{center}
		(1) \cite{Landt2017}; (2) \cite{Nandra2005}; (3) \cite{Zhau2005}; (4) \cite{Fausnaugh2017}; (5) \cite{Sergeev2011}; (6) \cite{Peterson2004}; (7) \cite{Zhang2006}; (8) \cite{Alston2019}; (9) \cite{Ezhikode2017}; (10) \cite{Liu2017}; (11) \cite{Grier2012}; (12) \cite{Middei2020}; (13) \cite{Ross1992}; (14) \cite{Nardini2015}; (15) \cite{Gliozzi2010}; (16) \cite{Agis2014}; (17) \cite{Turner2002}; (18) \cite{Bentz2016}. 
	\end{table*}
	
\subsection{Data Reduction}
\label{sec:data reduction}
We employed publicly available archival data from the {\it XMM-Newton} and {\it Swift} observatories using HEASARC\footnote{\url{https://heasarc.gsfc.nasa.gov/}}. We reprocessed all the data using the {\tt HEAsoft} version {\tt 6.29c} \citep{Arnaud1996}, which includes {\tt XSPEC v12.12.0}.

We included the observations from {\it XMM-Newton} \citep{Jansen2001} in the X-ray range of 0.2--10 keV obtained with the EPIC-pn detector \citep{Struder2001}. We reprocessed the raw data to level~1 data for EPIC-pn by Scientific Analysis System ({\tt SAS v16.1.0}\footnote{\url{https://www.cosmos.esa.int/web/xmm-newton/sas-threads}}) \citep{Gabriel2004} with calibration files released on February 2, 2018. We followed the standard prescription outlined in the {\it XMM-Newton} ABC online guide\footnote{\url{https://heasarc.gsfc.nasa.gov/docs/xmm/abc/}}. Calibrated, cleaned event files were created from the raw data files of EPIC-pn detector using the SAS command {\tt epchain}. We used {\tt FLAG==0} to avoid the flagged events. We also excluded the bad pixels and the edge of the CCD. Apart from that, we also used {\tt PATTERN $\leq$ 4} for single and double pixels. We excluded the photon flares by using appropriate {\tt GTI} files to acquire the maximum signal-to-noise ratio. We chose an annular region with outer and inner radii of 30\arcsec~ and  5\arcsec, respectively, centred at the source to extract the source events. Circular regions of 60\arcsec~ radii were considered for the backgrounds on the same CCD chip far from the source to avoid contamination. Source spectra were extracted from the cleaned event files using the SAS task {\tt xmmselect}. For the pile-up correction, we used the SAS task {\tt EPATPLOT}. To remove the pile-up effect from the data, we adjusted the inner and outer radii of the annular extraction region. We generated instrumental response files using the SAS tasks {\tt rmfgen} and {\tt arfgen}. The details of the {\it XMM-Newton}/EPIC-pn observations of each source are listed in Table~\ref{table:obs_log}.

Apart from the {\it XMM-Newton} observations, we also used data from the X-ray Telescope (XRT, \citep{Burrows2005}) on board the Neil Gehrels Swift Observatory or {\it Swift}. The sample data were obtained between January 2005 and December 2021. {\it Swift}/XRT observed each source regularly as well as in non-regular intervals in both photon counting (PC) and window timing (WT) modes. Depending on the exposure time, we combined observation IDs to get a reasonable spectrum in the 0.5--10 keV energy band.
The details of the observation log are provided in Table~\ref{table:obs_log}. We used the online tool ``XRT product builder''\footnote{\url{https://www.swift.ac.uk/user_objects/}} \citep{Evans2009} to extract the spectra of each source. This product builder performs all necessary processing and calibration and produces the final spectra for the PC and WT modes. 
	
In the present study, we considered 171 {\it XMM-Newton} and 134 binned {\it Swift}/XRT spectra, a total of 305 observations, for 21 `bare' AGNs. 
	
\section{Results and Discussion}
\label{sec:RandD}
We performed our spectral analysis in 0.5 to 10.0 keV energy range, obtained from the {\it XMM-Newton} and {\it Swift} observations of the selected sources (see Table~\ref{table:source_details}) using  {\tt XSPEC v12.12.0} \citep{Arnaud1996}. We initially used the {\tt powerlaw} model to fit the data in the energy range of 3.0 to 10.0 keV to constrain the primary continuum. Thereafter, the excess counts below 2.0 keV were fitted using another {\tt powerlaw} component \citep{WF1993, Nandi2021}. Along with these components, we used {\tt Tbabs} and {\tt zTBabs} \citep{Wilms2000} for Galactic and extra-galactic absorptions. The component {\tt TBabs} was applied for the Galactic absorption, where the hydrogen column density ($N_{\rm H,Gal}$) was kept frozen. The other absorption component, {\tt zTBabs}, was utilized for the extra-galactic hydrogen column density ($N_{\rm H}$). We kept the {\tt zTBabs} component free to vary for various epochs of observations. The basic model in {\tt XSPEC} reads as: {\tt TBabs*zTBabs(powerlaw+powerlaw)}.  
	
\begin{figure}
		\centering
		\includegraphics[trim={0 0 0 4cm}, width=1.0\textwidth]{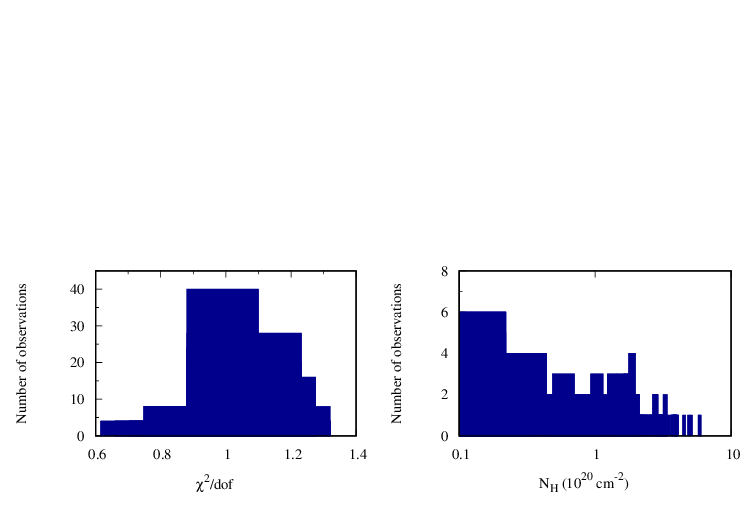}
				\caption{{\tt Left:} The number distribution of reduced $\chi^2$ ($\chi_{red}^{2}=\chi^2$/dof) obtained from the X-ray spectral fitting with the {\tt powerlaw+powerlaw} model on 21 `bare' AGNs. {\tt Right:} The number distribution of hydrogen column density $N_{\rm H}$ obtained from the X-ray spectral fitting of data of 21 `bare' AGNs with the composite model.}
		\label{fig:hist_chi_nh}
	\end{figure}

Along with this model, where needed, we used a {\tt Gaussian} component for the Fe fluorescent emission line near 6.4 keV. Thus, the model in {\tt XSPEC} reads as: {\tt TBabs*zTBabs(powerlaw+powerlaw+zGauss)}. For IRAS~13224-3809 and Mrk~335, we encountered an absorption feature near 1.8 keV \citep{Jiang2018} and 0.8 keV, respectively. For that, we used {\tt gabs} component to the above model, which in {\tt XSPEC} reads as: {\tt TBabs*zTBabs*gabs(powerlaw+powerlaw+zGauss)}. For each spectral fitting, we considered the final model if the reduced $\chi^2\sim 1$. All the 305 spectra were well fitted with the model and the distribution of $\chi^2$ normalised with respect to the degrees of freedom (DOF) is shown in the left panel of Figure~\ref{fig:hist_chi_nh}.

We used the following cosmological parameters throughout this work: $H_0$ = 70 km s$^{-1}$ Mpc $^{-1}$, $\Lambda_0$ = 0.73, $\Omega_{\rm M}$ = 0.27 \citep{Bennett2003}.

From model fitting, we estimated the extra-galactic hydrogen column densities from {\tt zTBabs} component for each source with 90\% confidence level. We calculated the intrinsic luminosities using {\tt clumin} command for soft excess ($L_{SE}$) and primary continuum ($L_{PC}$) for the energy range of 0.5--10.0 keV and then normalised them with respect to \led~ for each source. To compute the errors for each parameter, we used the `{\tt error}' command, which estimates error with 90\% confidence, available in {\tt XSPEC}. The uncertainties on the power-law indices were determined using {\tt STEPPAR} commend in {\tt XSPEC}. The contours were estimated for 1$\sigma$, 2$\sigma$, and 3$\sigma$ confidence ranges. For our current analysis, the errors are quoted with 90\% confidence level or mentioned otherwise.

	\begin{table*}
		\caption{Correlations between primary continuum and soft excess for different sources.}
		\begin{center}
			\label{table:correlation_all}
			\begin{tabular}{c c c c c c c}
				\hline\hline
                \multirow{2}{*}{Parameters} &
				\multicolumn{2}{c}{Pearson Correlation} &
			\multicolumn{2}{c}{Spearmann Correlation} &
			\multicolumn{2}{c}{Kendall Correlation} \\
				 & $\rho$ & p &  R  & p & $\tau$ & p  \\
				
				\hline
				$\Gamma^{PC} vs. \log(L_{PC}/L_{Edd})$ &0.38 &$<0.001$ &0.40 &$<0.001$ &0.30 &$<0.001$ \\
				$\Gamma^{SE} vs. \log(L_{SE}/L_{Edd})$ &0.38 &$<0.001$ &0.40 &$<0.001$ &0.28 &$<0.001$ \\
				$\Gamma^{PC} vs. \Gamma^{SE}$ &0.27 &$<0.001$ &0.35 &$<0.001$ &0.25 &$<0.001$ \\
				\hline \hline
			\end{tabular}
		\end{center}
	\end{table*}

\subsection{$N_{\rm H}$ distribution of selected sample}
\label{sec:nh}
\begin{figure*}
\centering
  \includegraphics[trim={0 0 0 4cm},width=1.0\textwidth, angle =0]{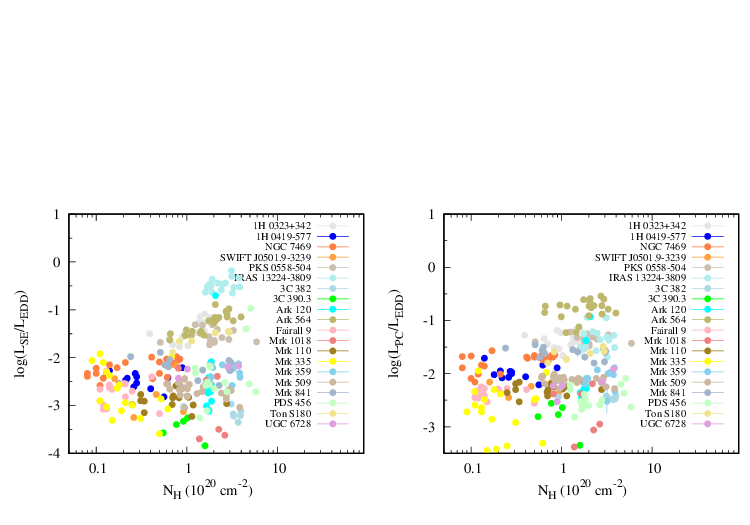}
\caption{Different colours are used for different sources. {\it Left} panel shows the variation of the intrinsic luminosity of the soft-excess with $N_{\rm H}$ and {\it right} panel shows the variation of the intrinsic luminosity  of primary continuum with $N_{\rm H}$. These plots indicate that there is no correlation between the intrinsic luminosities and $N_{\rm H}$ in long-term observations for different sources.}
\label{fig:nh_vs_lum}
\end{figure*}

The `bare' AGNs are defined by their little or negligible hydrogen column density along our line of sight \citep{Walton2013}. We computed the intrinsic $N_{\rm H}$ for the selected 21 sources using the {\tt zTBabs} model. The sources exhibited variations of $N_{\rm H}$ ranging within 2 orders of magnitude and are presented in the right column of Figure~\ref{fig:hist_chi_nh}. We found that the hydrogen column density remained less than $10^{21} \mathrm{cm}^{-2}$ for all sources. The maximum $N_{\rm H}$ was observed for PKS~0558-504 having a value of $5.91\pm0.51\times10^{20}$ cm$^{-2}$. The systematic survey revealed that the nearby population has a mean of $1.41\pm0.31\times10^{20}$ cm$^{-2}$, the median of $1.07\pm0.09\times10^{20}$ cm$^{-2}$ having a minimum at $0.05\times10^{20}$ cm$^{-2}$ and a standard deviation ($\sigma$) of $1.15\pm0.1\times10^{20}$. In the left panel of Figure~\ref{fig:nh_vs_lum}, we plotted the variation of \nh~ with respect to \lse. 
We failed to notice any global correlation between these two parameters. Similar to the \nh~ vs \lse, we plotted \nh~ vs \lpc~ in the right panel of Figure~\ref{fig:nh_vs_lum}. We did not observe any  significant correlation between these two parameters as well. This is a likely scenario for `bare' AGNs as, by definition, the line of sight hydrogen both ionized and neutral column density around them, is little to none.  However, there remains a possibility that the circumnuclear material could have higher column density away from our line of sight \citep{Reeves2016}. It should be noted that the current sample consists of sources that are in the sub-Eddington accretion regime.

\subsection{\gpc~and~\gse}
\label{sec:gpc}
\begin{figure*}
\centering
\includegraphics[width=0.9\textwidth]{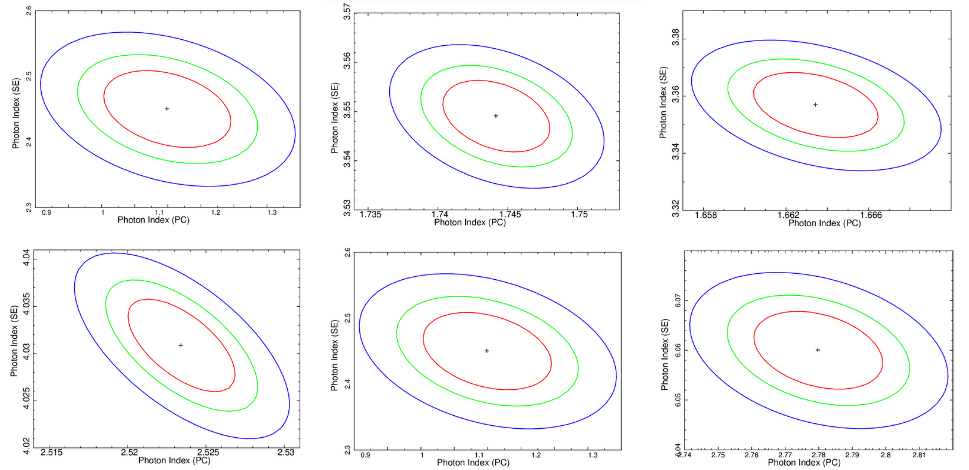}
\hspace{0mm}
\caption{The confidence contours between the photon index $(\Gamma)$ of primary continuum and soft-excess are shown for 1H~0419-577 (MJD-52635; top left), 3C~382 (MJD-54584; top middle), 3C~390.3 (MJD-53286; top right), Ark~564 (MJD-55705; bottom left), Mrk~1018 (MJD-55174; bottom middle), and IRAS~13224-3809 (MJD-52019); bottom right). The red, green and blue contours represent the 1~$\sigma$, $2~\sigma$ and 3~$\sigma$ levels, respectively.}
\label{fig:contour}
\end{figure*}

We examined the photon indices of the {\tt powerlaw} components used in the spectral fitting. We found that the photon index of the 3--10 keV primary {\tt powerlaw} continuum varied from $1.03\pm0.04$ to $2.77\pm0.03$ with a mean of $1.77\pm0.03$ and a median of $1.73\pm0.04$. The $\sigma$ for \gpc is $0.34\pm0.02$. The lower value of \gpc is expected to be low as, by definition, the contribution of soft excess appears above the primary continuum below 2 keV. From our survey, most of the `bare' AGN population exhibits a hard {\tt powerlaw} tail suggesting a hotter Comptonizing region \citep{Sunyaev1980}. We employed another {\tt powerlaw} component to fit the soft-excess below 2 keV. The photon indices of \gse varied in a wide range, starting from $1.86\pm0.07$ and reaching up to $7.34\pm0.10$. The mean, median, and $\sigma$ of \gse are found to be $3.97\pm0.11$, $3.87\pm0.09$, and $1.13\pm0.08$, respectively. We observed higher values of \gse as the index was only used to fit a relatively narrow energy range between $0.5-2.0$ keV. It should be noted that the mean and median for \gse are higher than that of \gpc. This is expected as the excess emission below 2 keV has a higher photon index and is often referred to as originating from a distinct colder Comptonizing region than the Corona \citep{Mehdipour2011, Done2012}. In Figure~\ref{fig:contour}, we present the confidence contours for the two power-law indices of six sources from different observations. The bottom-middle panel shows the contours for the extreme value $\Gamma^{PC}\sim1.05$ of the last two plots, representing one of the nearly extreme values of $\Gamma^{PC}\sim1.05$ and $\Gamma^{SE}\sim6.08$ respectively.   

\begin{figure*}
\centering
  \includegraphics[trim={0 1cm 0 4cm},width=1.0\textwidth, angle =0]{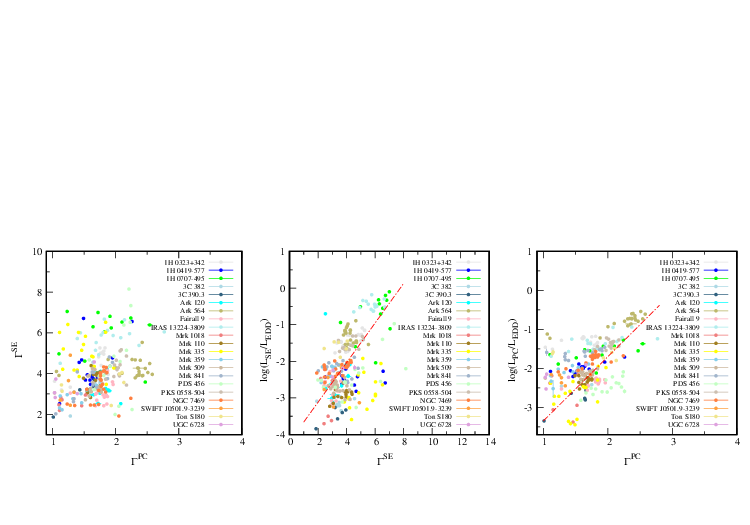}
  \caption{The variation of the spectral indices is plotted on the left panel. The variation between the normalised luminosity of soft-excess $(L_{SE}/L_{Edd})$ and corresponding spectral slope $(\Gamma^{SE})$ is plotted in the middle panel. The right panel shows the variation of the normalised luminosity of primary continuum $(L_{PC}/L_{Edd})$ and spectral slope $(\Gamma^{PC})$. The red dotted line represents the linear correlation of corresponding parameters.}
\label{fig:correlation_all}
\end{figure*}

In the left panel of Figure~\ref{fig:correlation_all}, we plotted the \gpc\ against \gse\ for all the sources. The absence of correlation between these parameters is evident as the Pearson correlation coefficient (PCC) is found to be 0.27 with {\it p-value $<0.001$}, tabulated in Table~\ref{table:correlation_all}. In some cases, we encountered $\Gamma^{PC}<1.3$. We carefully identified those observations and found that the majority of them were from Swift/XRT observations with exposure time less than 15 ks (as seen in the observation log). Among these observations, $36$ ($\sim10\%$ of the total data) met this criterion, with $34$ ($\sim95\%$ of the $10\%$ of total data) from Swift/XRT and 2 ($\sim5\%$ of the $10\%$ of total data) from XMM-Newton. Consequently, the high-energy data points (above $\sim6$ keV) for these observations were not well-constrained, leading to high uncertainty in Gamma for the primary continuum.

\begin{figure*}
\centering
  \includegraphics[trim={0 1cm 0 5cm},width=1.0\textwidth, angle =0]{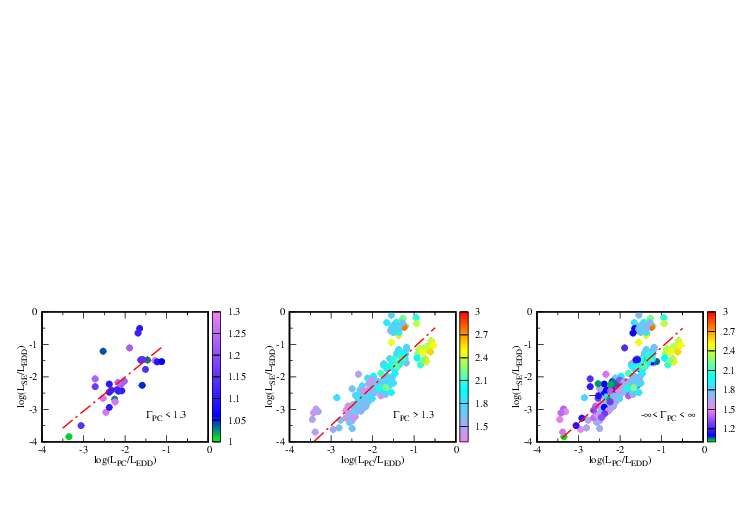}
  \caption{ The Correlation between the observed intrinsic normalised luminosities of primary continuum (\lpc) and soft-excess (\lse) for 21 bare AGNs with $\Gamma_{PC}$ for different scenarios: (i) when $\Gamma_{PC}<1.3$, the variation of the \lpc~ vs \lse~ is presented in the left panel; (ii)  when $\Gamma_{PC}>1.3$, the variation of the \lpc~ vs \lse~ is presented in the middle panel and (iii) for all values of $\Gamma_{PC}$, the variation of the \lpc~ vs \lse~ is presented in the right panel. The colour bar represents the variation of $\Gamma_{PC}$.}

\label{fig:corr_gamma}
\end{figure*}

In addition, we investigated whether the correlation between the luminosities of the primary continuum and soft excess still exists under the criteria $\Gamma^{PC} < 1.3$. We divided the luminosities into three categories: (i) for $\Gamma^{PC} < 1.3$, (ii) for $\Gamma^{PC} > 1.3$, and (iii) for all $\Gamma^{PC}$ values and presented the correlations in the left, middle, and right panels of Figure~\ref{fig:corr_gamma}, respectively. The correlation was found to persist under these criteria. Therefore, we conclude that the occurrence of $\Gamma^{PC} < 1.3$ is a result of poor data quality due to low exposure time, and this does not affect our final results.

\subsection{\lpc~and \lse}
\label{sec:luminosity}
\begin{figure*}
\centering
  \includegraphics[trim={0 0 0.5cm 4cm}, width=1.0\textwidth, angle =0]{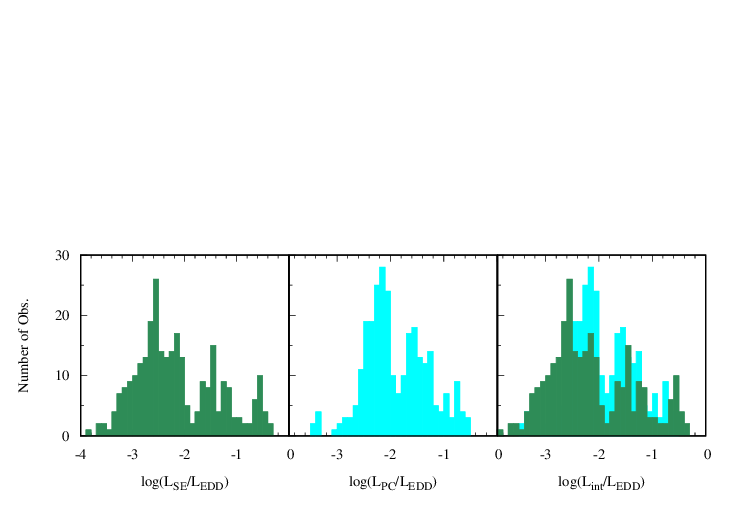}
\caption{The distribution of the number of observations with respect to normalised intrinsic luminosities. {\it Left} panel shows the number distribution of intrinsic luminosity of normalised soft-excess, {\it middle} panel represents the distribution of intrinsic luminosity of normalised primary continuum and {\it right} panel shows the comparison of these intrinsic luminosities.}
		\label{fig:hist_lum}
\end{figure*}

AGNs are considered luminous objects in the observable universe. We surveyed `bare' AGNs in the local universe ($z<0.2$), where the luminosity in the 0.5--10 keV energy band is in the range of $10^{42}-10^{45}$ \eps.  We calculated the intrinsic luminosities of the primary continuum and soft excess of each spectrum of all sources using the {\tt clumin} command on the {\tt powerlaw} component for the energy range of 0.5 to 10.0 keV. The result is plotted in the top left panel of Figure~\ref{fig:all}. Here, we observe that the mass-dependent intrinsic luminosities of the primary continuum and soft excess are highly correlated (PCC=0.83 with $p-value < 0.001$). Then, we opted for a mass-independent form, \led, which varied from 43.42 to 47.40 for the sample of our sources. From the spectral analysis of 21 `bare' AGNs, we found that the \lpc~ and \lse~ varied from -3.45 to -0.53 and -3.46 to -0.05, respectively. We chose \lpc~ and \lse~ over $\log(L_{PC})~$ and $\log(L_{SE})$ because the mass-dependent luminosities may appear due to two reasons: (i) high accretion rate with low black hole mass, and (ii) low accretion rate with high black hole mass. To break this degeneracy, one should normalize the luminosity with respect to mass. Hence, we normalised the luminosity by dividing it with the Eddington luminosity $(= L/L_{Edd})$. The normalised luminosity is basically the proxy of the mass accretion rate, which is more fundamental than the luminosity \citep{Jana2023}. We also cross-checked the dependence of normalised luminosity on the mass of our sample of AGNs and did not find any significant correlation among them. For the primary continuum, the mean value of \lpc~ is found to be $-1.86\pm0.05$ with median and standard deviation of $-2.01\pm0.06$ and $-0.53\pm0.05$, respectively. In the case of the soft excess, the luminosity (\lse) varies from -3.46 to -0.05 with a mean value of $-2.10\pm0.08$. The median and the standard deviation of \lse~ for this energy range are $2.01\pm0.06$ and $0.65\pm0.05$, respectively. It is clear that all the sources were in the sub-Eddington regime of accretion. Figure~\ref{fig:hist_lum} represents the number distribution of the normalised intrinsic luminosities for all the observations and their comparison. From that, it could be shown that all observations have sub-Eddington luminosity.

\subsection{Soft-excess}
\label{sec:se}
\begin{figure}
		\centering
		\includegraphics[width=1.0\textwidth]{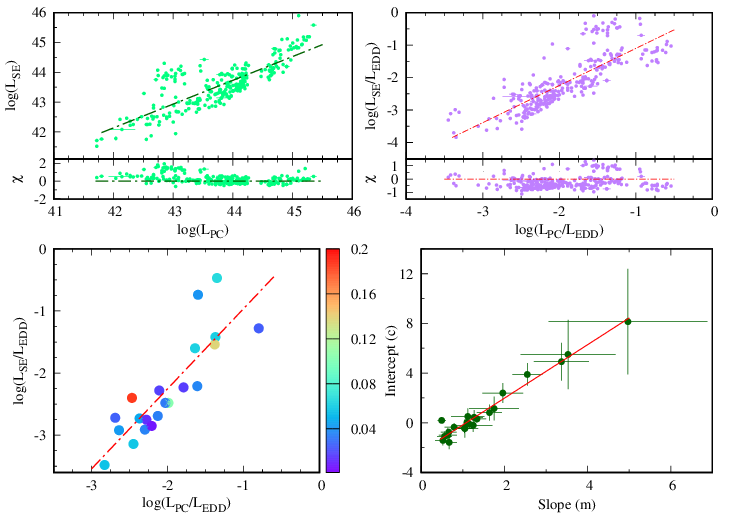}
		\caption{ {\it Top left:} Correlation between the observed intrinsic luminosities of the soft excess and the primary continuum in 0.5--10 keV range, estimated from 21 bare AGNs. A linear fit is shown on the data set by a black dash-dotted line and the corresponding variation of $\chi$ is shown in the bottom panel. {\it Top right:} Correlation between the observed intrinsic normalised luminosities of the primary continuum and soft excess in the energy range from 0.5 to 10 keV estimated from 21 bare AGNs. A linear fit is shown on the data set by a red dash-dotted line and the corresponding variation of $\chi$ is shown in the bottom panel. {\it Bottom left:} The average luminosity distribution of the primary continuum and soft excess across all sources with respect to redshift $(z)$. {\it Bottom right:} Correlation between slope and intercept of individual sources are plotted.}
		\label{fig:all}
	\end{figure}

Soft-excess emission is a common feature of most of the Seyfert~1 AGNs. The presence of strong soft-excess was found for all of the sample sources that we surveyed. As we opted for bare AGNs, the soft-excess is free or nearly free from absorption. Figure~\ref{fig:nh_vs_lum} shows that there are no correlation between the intrinsic luminosities of the primary continuum (\lpc) and soft-excess (\lse) with the extra-galactic hydrogen column density $(N_{\rm H})$. The luminosities are calculated using {\tt clumin} task in the {\tt powerlaw+powerlaw} model for each observation of each source. We found a correlation between these two luminosities for each source. The correlation coefficient is calculated using three different algorithms,  Pearson correlation method\footnote{\url{https://www.socscistatistics.com/tests/pearson/default2.aspx}}, Spearman Correlation method\footnote{\url{https://www.socscistatistics.com/tests/spearman/default2.aspx}} and Kendall Correlation method\footnote{\url{http://www.wessa.net/rwasp_kendall.wasp}} and the corresponding results are shown in Table~\ref{table:correlation}. The Pearson correlation coefficient values varied between 0.56 and 0.95, indicating that the luminosities of the primary continuum and soft-excess (\lpc~and~\lse) are tightly correlated for most of the sources, whereas two sources (1H~0323+342 and NGC~7469 have the PCC of 0.56 and 0.52, respectivtely) exhibit relatively weaker correlations (see Table~\ref{table:correlation}). The overall correlation between these two luminosities is represented in Figure~\ref{fig:all}. We calculated the correlation coefficients using three different methods which are presented in Table~\ref{table:correlation}. From the Pearson correlation calculation, we found that the overall correlation coefficient $(\rho)$ is 0.85 and for the Spearman Correlation method and Kendall Correlation method, the correlation coefficients are $R=0.88$ and $\tau=0.71$, respectively, with $p$-values of $<0.001$ for all cases. For our global sample, the connection between the soft-excess and primary continuum is established. Due to their strong correlation, we can argue that their origin could be the same. We normalised each calculated luminosity using the Eddington luminosity. Once we convert the luminosity into the Eddington unit, the parameter becomes the accretion rate ($\lambda  = L/L_{EDD})$, which is independent of the mass of the central object. Use of $\lambda$ in case of accretion physics, both in AGNs and stellar mass black holes, is profound \citep{Fabian2009, Done2012, Netzer2019, Mahmoud2023, Middei2023}. Hence, we can conclude that the normalised luminosities are independent of mass. We constructed a luminosity-luminosity plot incorporating distance (redshift z) to assess any potential dependency on other parameters. Our findings indicate that the average luminosity distribution of the primary continuum and soft-excess, across all sources, does not exhibit a dependence on $z$ (see Figure~\ref{fig:all}). Furthermore, we conducted correlation calculations based on long-term X-ray observations of individual sources, which are presented in the Appendix. Consequently, we deduce that the correlation between the luminosities of the primary continuum and soft excess holds true at both the individual level and the global scale for our source sample (refer to Figure~\ref{fig:all}).

In essence, we observed that the ratio of primary continuum and soft-excess luminosities remains unaffected by mass, distance, and other parameters associated with the sources. This suggests that the soft excess is not generated by a physical process that relies on these particular parameters.

There are many proposed theories on the origin of ubiquitous soft-excess \citep{Arnaud1985,Singh1985,Fabian2002,Gierlinski2004}. The soft-excess could be generated by the process of reflection \citep{Sobolewska2007, Fabian2009} or due to Comptonization by a warm optically thick region surrounding the accretion disc \citep{Mehdipour2011} or could be generated from the high disc accretion rate \citep{Done2012}. In the case of Fairall~9, \citep{Lohfink2012} showed that the origin of the soft-excess is linked with the source that produces the broad iron line. They implied that another source of Comptonization could be responsible for generating the soft-excess. The origin of soft-excess has been investigated by \citep{Fukumura2016} on the basis of radiative transfer and hydrodynamics around an AGN. They proposed that the soft-excess could be generated by the process of shock heating near the ISCO  (Innermost stable circular orbit). However, using a statistical survey of 120 Seyfert~1 AGNs, \citep{Boissay2016} suggested that the origin of soft-excess could be related to the thermodynamical properties of the Compton cloud and associated medium. Considering timing studies, \cite{Fabian2009} proposed that the soft-excess should be delayed compared to the continuum as the model gauge reflection to be the origin of soft-excess. However, for Ark~120, \cite{Nandi2021} reported that the soft-excess could produce zero, positive, and negative delays depending on the spectral state of the AGN. Thus, contrary to other models, we argue that the corona or Compton cloud itself could produce the soft-excess.

In the case of supermassive black holes, emission from the standard disk \citep{Shakura1973} peaks in the UV. The hot corona reprocesses the soft emission from the disk to produce the power-law continuum \citep{Sunyaev1980}. The density and temperature profiles within the Compton cloud or corona are believed to be non-homogeneous \cite{CT95}. According to earlier simulations \citep{PSS1983, Ghosh2011, Chatterjee2017, Nandi2021}, the number of scatterings within the corona plays is crucial in determining the spectral index. The photons, which have suffered less, could produce a steeper spectral slope than the continuum. For Ark~120, \citet{Nandi2021} investigated the possible scattering-dependent spectrum in the X-ray regime. In that work, they showed the variation of spectral components with respect to the number of scatterings and found that the spectral slope decreases with the increase in the number of scatterings. The primary continuum (above 3.0 keV) is mostly dominated by the photons where the number of scatterings is $\geq10$. For the soft-excess regime (below 2.0 keV), the dominating contribution of photons comes from those which have suffered $\leq10$ scatterings. The correlation between the luminosities could also be obtained by varying the accretion rates. A similar feature for all selected `bare' AGNs could be observed in the present work.  

To examine the correlation in individual sources, we employed a linear fit $(y=mx+c)$ between the luminosities in the $\mathrm{log}$ scale and the fitted parameters are shown in Table~\ref{table:correlation}. We found that the slope ($m$) varied from $0.49\pm0.10$ to $3.63\pm1.14$ with a mean value of $1.43\pm0.49$. The standard deviation and median of this slope are $0.86\pm0.19$ and $1.16\pm0.48$, respectively. The intercept ($c$) ranges between $-1.58\pm0.54$ to $5.50\pm2.79$ with the mean at $0.55\pm0.22$ and median at $-0.11\pm0.05$. The standard deviation of the intercept ($c$) for our data-set is $1.96\pm0.42$. For the overall analysis of all data-sets, we found the luminosity of the primary continuum (\lpc) is highly correlated with the luminosity of soft-excess (\lse). The parameters for the linear fit of all data points are shown in Figure~\ref{fig:all} and the corresponding values of fitted parameters are presented in Table~\ref{table:correlation}. We found that the overall slope ($\mathrm{m}$) is $1.10\pm0.04$ and intercept ($\mathrm{c}$) is $0.04\pm0.08$ for all data-sets and corresponding $\chi_{red}^2\sim1.20$. Thus, we suggest that the relationship between these two parameters is  $L_{PC} \propto L_{SE}^{1.1\pm0.04}$ for `bare' AGNs. The global fit indicates that the continuum provides a slightly higher luminosity than the soft-excess over $-3.5 < \rm {log(L/L_{Edd})} < -0.5$ regime. For individual sources, the variations of the slope and intercept are expected as their inclination and accretion state are variable with respect to the observer frame. 

\begin{figure}
		\centering
		\includegraphics[ width=0.9\textwidth]{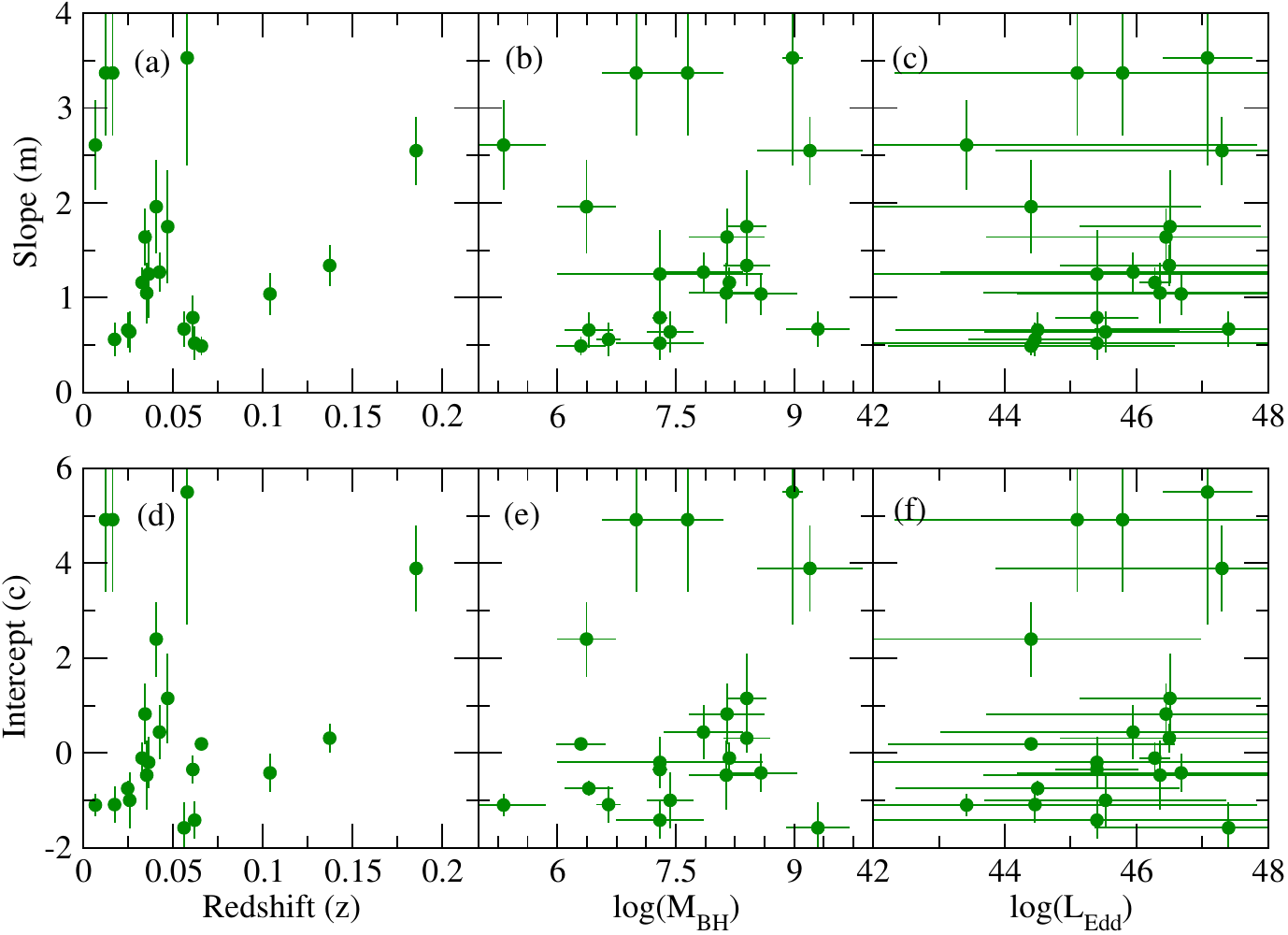}
		\caption{Correlation between slope and intercept of individual sources are plotted with intrinsic parameters, such as (a) \& (d) redshift, (b) \& (d) $\log(M_{BH})$, (c) \& (d) $\log(L_{Edd})$, respectively.}
		\label{fig:corr_par}
\end{figure}

From the linear fit of individual sources, we found that the slope ($m$) and intercept ($c$) are correlated. The correlation is presented in Figure~\ref{fig:all}. From this figure, we find that the soft-excess could vanish for a weaker primary continuum. We examined the dependency of `$m$' and `$c$' on the intrinsic parameters, such as redshift, mass, and Eddington luminosity and found that `$m$' and `$c$' are not dependent on these parameters (Figure~\ref{fig:corr_par}).

	\begin{table*}
		\caption{Correlations between primary continuum and soft excess for different sources.}
		\begin{center}
			\label{table:correlation}
			\begin{tabular}{c c c c c c c c c}
				\hline\hline
				\multirow{2}{*}{Source} &
			\multicolumn{2}{c}{Pearson Correlation} &
			\multicolumn{2}{c}{Spearmann Correlation} &
			\multicolumn{2}{c}{Kendall Correlation} &
            \multicolumn{2}{c}{Linear fit} \\
				 & $\rho$ & p &  R  & p & $\tau$ & p & m & c  \\
				
				\hline
				1H 0323+342       &0.56 &0.017 &0.52 &0.027 &0.39 &0.030 &$0.79\pm0.23$ &$-0.35\pm0.30$ \\
				1H 0419--577       &0.82 &0.001 &0.80 &0.001 &0.64 &0.003 &$1.04\pm0.22$ &$-0.42\pm0.41$ \\
				1H 0707--495       &0.78 &0.005 &0.60 &0.043 &0.43 &0.062 &$1.96\pm0.49$ &$2.40\pm0.79$ \\
				3C 382            &0.78 &0.022 &0.59 &0.040 &0.69 &0.024 &$3.53\pm1.14$ &$5.50\pm2.79$ \\
				3C 390.3          &0.90 &0.037 &0.59 &0.043 &0.60 &0.220 &$0.67\pm0.19$ &$-1.58\pm0.54$ \\
				Ark 120           &0.90 &$<0.001$&0.80 &0.001 &0.68 &0.001 &$1.16\pm0.16$ &$-0.11\pm0.34$ \\
				Ark 564           &0.60 &0.003 &0.56 &0.007 &0.44 &0.005 &$0.66\pm0.19$ &$-0.75\pm0.16$ \\
				Fairall 9         &0.63 &0.011 &0.63 &0.012 &0.49 &0.013 &$1.75\pm0.60$ &$1.15\pm0.95$ \\
				IRAS 13224--3809   &0.79 &$<0.001$&0.80 &$<0.001$&0.62 &0.001 &$0.49\pm0.10$ &$0.19\pm0.13$ \\
				Mrk 1018          &0.93 &0.002 &0.68 &0.089 &0.49 &0.171 &$1.27\pm0.21$ &$0.44\pm0.58$ \\
				Mrk 110           &0.67 &0.006 &0.55 &0.033 &0.41 &0.040 &$1.05\pm0.32$ &$-0.47\pm0.73$ \\
				Mrk 335           &0.59 &0.010 &0.62 &0.006 &0.46 &0.010 &$0.64\pm0.22$ &$-1.00\pm0.60$ \\
				Mrk 359           &0.73 &0.016 &0.45 &0.197 &0.34 &0.206 &$0.56\pm0.18$ &$-1.09\pm0.39$ \\
				Mrk 509           &0.75 &$<0.001$&0.71 &$<0.001$&0.55 &$<0.001$&$1.64\pm0.30$ &$0.82\pm0.64$ \\
				Mrk 841           &0.62 &0.018 &0.59 &0.026 &0.46 &0.024 &$1.25\pm0.46$ &$-0.20\pm0.54$ \\
				PDS 456           &0.88 &$<0.001$&0.82 &$<0.001$&0.68 &$<0.001$ &$2.55\pm0.36$ &$3.89\pm0.91$\\
				PKS 0558--504      &0.87 &$<0.001$&0.84 &$<0.001$&0.68 &0.001 &$1.34\pm0.22$ &$0.31\pm0.31$\\
				SWIFT J0501.9--3239&0.95 &0.013 &0.90 &0.037 &0.80 &0.086 &$3.37\pm0.66$ &$4.92\pm1.52$ \\
				NGC 7469          &0.52 &0.010 &0.46 &0.026 &0.35 &0.023 &$0.52\pm0.19$ &$-1.42\pm0.39$ \\
				Ton S180          &0.85 &0.032 &0.83 &0.041 &0.73 &0.060 &$1.12\pm0.40$ &$0.51\pm0.65$ \\
				UGC 6728          &0.83 &0.081 &0.80 &0.057 &0.75 &0.086 &$4.97\pm1.91$ &$8.14\pm4.24$\\
				\hline
				Overall           &0.85 &$<0.001$&0.88 &$<0.001$&0.71 &$<0.001$ &$1.10\pm0.04$ &$0.04\pm0.01$ \\
				\hline \hline
			\end{tabular}
		\end{center}
	\end{table*}
		
 \section{Conclusions}
	\label{sec:Conclusions}
 We studied 21 Seyfert~1 AGNs in the energy range of 0.5 to 10.0 keV by using observations spanned over a long duration of time. These sources are previously reported as `bare' AGNs and we find similar characteristics of each source. The following are the significant findings from our presented work. 

\begin{enumerate}

  \item In most cases, the intrinsic luminosities vary a lot. However, these variations are not related to the hydrogen column density $(N_{\rm H})$ along the line of sight. From the spectral analysis of all the observations, we found the mean value of extra-galactic hydrogen column density $(N_{\rm H})$ at $1.41\pm0.31\times10^{20}$ $\mathrm{cm}^{-2}$ with a standard deviation of $1.15\pm0.01\times10^{20}$ $\mathrm{cm}^{-2}$.
  
  \item The power-law slope of the soft-excess $(\Gamma^{\rm SE})$ and the power-law slope of the primary continuum $(\Gamma^{\rm PC})$ are not constant and vary for all sources. From our survey, we found that the mean for the 3--10 keV primary continuum power-law index is $1.77\pm0.03$ with a standard deviation of $0.34\pm0.02$. For the soft-excess, the mean of the power-law index is $3.97\pm0.11$ with a standard deviation of $1.13\pm0.08$. Therefore, it indicates that the spectral slope for soft-excess $(\Gamma^{\rm SE})$ is higher than the spectral slope of the primary continuum $(\Gamma^{\rm PC})$.
  
  \item From the analysis of long-term data of 21 `bare' AGNs, we found that the intrinsic luminosity of soft-excess ($L_{SE}$) and the intrinsic luminosity of primary continuum ($L_{PC}$) are widely variable and the variation is different for different sources. The mean value of ~\lpc~ is found to be $-1.86\pm0.05$ with a standard deviation of $0.65\pm0.05$. Whereas the ~\lse~ has a mean value of $-2.10 \pm 0.08$ with a standard deviation of $0.65\pm0.05$. We found that these two luminosities are correlated for individual sources. For the overall picture, the normalised intrinsic luminosities (normalised by Eddington luminosity) are also tightly correlated. While extending the previous work of \citep{Nandi2021}, we found that the thermodynamical properties of the Compton cloud could attribute to the origin of soft-excess in the `bare' AGNs. On a global scale, we found that the accretion rate drives the luminosity of soft excess as well as the primary continuum.
  
  \item The slope $\mathrm{m}$ and intercept$\mathrm{c}$ for the individual sources exhibit a correlation between these two parameters which indicates that the primary continuum will be present whether or not the soft-excess is observed.

  \end{enumerate}

 In future, high-resolution spectroscopic missions, such as \textit{XRISM} \citep{Tashiro18} and  \textit{ATHENA} \citep{Nandra2013} would provide detailed line emission in the soft-excess regime. While working in a similar energy range, the large field of view of \textit{AXIS} \citep{Mushotzky2018} would be able to extend the sample size as well as provide crucial information related to the origin of the soft excess. On the other hand, the large effective area and high throughput of \textit{Colibr\`i} \citep{Heyl2019, Caiazzo2019} could provide serendipitous detection of low luminosity ($log(L/L_{Edd}) < -3.5$) `bare' AGNs. Additionally, the hard X-ray properties of the soft-excess \citep{Boissay2016} could be explored by \textit{HEX-P} \citep{Madsen2018}.

\section*{acknowledgments}
 We would like to express our sincere gratitude to the reviewers for their valuable insights and comprehensive evaluation of our work, greatly contributing to its improved quality and clarity. PN and SN acknowledge support from the Physical Research Laboratory, Ahmedabad, India, funded by the Department of Space, Government of India, for this work. AC, SSH and JH are supported by the Canadian Space Agency (CSA) and the Natural Sciences and Engineering Research Council of Canada (NSERC) through the Discovery Grants and the Canada Research Chairs programs. AJ and HK acknowledge the support of the grant from the Ministry of Science and Technology of Taiwan with the grand numbers MOST 110-2811-M-007-500 and  MOST 111-2811-M-007-002. HK acknowledge the support of the grant from the Ministry of Science and Technology of Taiwan with the grand number MOST 110-2112-M-007-020. This research is based on observations obtained with XMM-Newton, an ESA science mission with instruments and contributions directly funded by ESA Member States and NASA. This work made use of XRT data supplied by the UK Swift Science Data Centre at the University of Leicester, UK.


%

\vspace{5mm}
\facilities{{\it XMM-Newton}, {\it Swift}/XRT}


\software{{\tt HEASoft v6.29c} \citep{Arnaud1996}, `XRT product builder'\footnote{\url{https://www.swift.ac.uk/user_objects/}} \citep{Evans2009}, Scientific Analysis System ({\tt SAS v16.1.0}\footnote{\url{https://www.cosmos.esa.int/web/xmm-newton/sas-threads}}) \citep{Gabriel2004}.}



\appendix
\section{Observation log}
The details of the {\it XMM-Newton}/EPIC-pn and {\it Swift}/XRT observations of each source are listed in Table~\ref{table:obs_log}. We have considered 171 {\it XMM-Newton} observations and 134 binned {\it Swift}/XRT observations for 21 `bare' AGNs. So, the total number of observations is 305 for this work. The details data reduction procedure is given in Section~\ref{sec:data reduction}.

 \label{table:obs_log}
		\begin{longtable*}{c c c c c c c}
        \caption{The observational log for each source.}\\
			\hline
			Source&{\it XMM-Newton}& date        &Exposure&{\it Swift}/XRT& date         &Exposure\\
			&    observations&(yyyy:mm:dd) & (ks)   & observations  &(yyyy:mm:dd)  & (ks)   \\
			\hline
			\endfirsthead
			\multicolumn{7}{c}%
			{\tablename\ \thetable\ : Observation log.} \\
			\hline
			Source&{\it XMM-Newton}& date        &Exposure&{\it Swift}/XRT& date         &Exposure\\
			&    observations&(yyyy:mm:dd) & (ks)   & observations  &(yyyy:mm:dd)  & (ks)   \\
						\hline
			\endhead
			\hline \multicolumn{7}{r}{\textit{Continued on next page}} \\
			\endfoot
			\hline
			\endlastfoot
1H 0323+342 &0764670101 &2015:08:23 &80.9& 00036533001-00036533009 & 2007:07:20-2007:12:28 & 32.5 \\
			&0823780201 &2018:08:14 &54.2& 00036533010-00036533012 & 2008:01:04-2008:11:16 & 10.6 \\
			&0823780301 &2018:08:18 &49.3& 00036533013-00036533018 & 2009:07:24-2009:08:08 & 18.5 \\
			&0823780401 &2018:08:20 &49.1& 00090415001-00090415031 & 2010:10:28-2010:11:30 & 91.5 \\
			&0823780501 &2018:08:24 &49.6& 00036533019-00036533027 & 2011:07:06-2011:12:28 & 16.1  \\
			&0823780601 &2018:09:05 &51.9& 00036533028-00036533030 & 2012:01:02-2012:03:03 & 04.9  \\
			&0823780701 &2018:09:09 &50.6& 00036533031-00036533051 & 2013:01:13-2013:10:02 & 62.2  \\
			&           &           &    & 00036533052-00036533053 & 2014:12:10-2014:12:12 & 05.8  \\
			&           &           &    & 00036533054-00036533080 & 2015:08:02-2015:12:24 & 50.1  \\
			&           &           &    & 00036533081-00036533104 & 2018:07:05-2018:12:13 & 24.2  \\
			&           &           &    & 00036533105-00036533109 & 2019:10:03-2019:12:21 & 08.4  \\
			&&&&&&\\
1H 0419--577&0148000201 &2002:09:25 &15.1& 00037559001-00037559002 & 2008:10:22-2008:11:12 & 16.1 \\
			&0148000301 &2002:12:27 &18.0& 00091621001             & 2013:06:17            & 01.1 \\
			&0148000401 &2003:03:30 &14.0& 00081695001-00081695004 & 2015:06:03-2015:06:09 & 04.2 \\
			&0148000501 &2003:06:25 &13.2& 00093031001-00093031002 & 2017:05:28-2017:05:31 & 02.5 \\
			&0148000601 &2003:09:16 &13.9& 00088681001             & 2018:11:13            & 02.0 \\
			&0148000701 &2003:11:15 &12.2&  & & \\
			&0604720401 &2010:05:28 &60.9&  & & \\
			&0604720301 &2010:05:30 &106.7&  & & \\
			&0820360101 &2018:05:16 &52.0&  & & \\
			&0820360201 &2018:11:13 &53.3&  & & \\
			&&&&&&\\
1H 0707--495&0148010301 &2002:10:13 &80.0& 00090393001-00090393070 & 2010:04:03-2010:12:31 & 78.9 \\
			&0506200301 &2007:05:14 &41.0& 00090393071-00090393104 & 2011:01:04-2011:03:29 & 35.8 \\
			&0506200201 &2007:05:16 &10.9& 00091623001-00091623002 & 2013:05:19-2013:06:19 & 03.2 \\
			&0506200501 &2007:06:20 &47.0& 00080720001-00080720004 & 2014:05:09-2014-06-28 & 04.9 \\
			&0506200401 &2007:07:06 &42.9& 00080720005-00090393105 & 2018:01:01-2018:04:30 & 52.6 \\
			&0511580101 &2008:01:29 &123.8& &  & \\
			&0511580201 &2008:01:31 &123.7&  & & \\
			&0511580301 &2008:02:02 &122.5&  & & \\
			&0511580401 &2008:02:04 &121.9&  & & \\
			&0653510301 &2010:09:13 &116.6&  & & \\
			&0653510401 &2010:09:15 &128.2&  & & \\
			&0653510501 &2010:09:17 &127.6&  & & \\
			&0653510601 &2010:09:19 &129.0&  & & \\
			&0554710801 &2011:01:12 &89.3&  & & \\
			&0853000101 &2019:10:11 &60.7&  & & \\
			&&&&&&\\                  
3C 382      &0506120101 &2008:04:28 &29.4& 00080216001 & 2013:12:18 & 01.9 \\
			&0790600101 &2016:08:29 &31.0& 00081830001-00081830002 & 2016:09:11-2016:10:17 & 03.6 \\
			&0790600201 &2016:09:11 &23.0& 00014251001-00096110004 & 2021:04:09-2021:06:29 & 03.1 \\
			&0790600301 &2016:09:22 &28.0&  &  &  \\
			&0790600401 &2016:10:05 &23.0&  &  &  \\
			&0790600501 &2016:10:17 &24.0&  &  &  \\
			&&&&&&\\
3C 390.3    &0203720201 &2004:10:08 &70.4& 00037596001-00037596004 & 2008:05:30-2008:06:13 & 19.5 \\
			&0203720301 &2004:10:17 &52.8& 00080221001             & 2013:05:25            & 02.1 \\
			&           &           &    & 00080221002-00096111001 & 2021:03:18-2021:05:18 & 03.2 \\
			&&&&&& \\
Ark 120     &0147190101 &2003:08:24 &112.1&00037593001-00037593003 & 2008:07:24-2008:08:03 & 10.9  \\
			&0693781501 &2013:02:18 &130.5&00091909002-00091909022 & 2014:09:04-2014:10:19 & 22.8  \\
			&0721600201 &2014:03:18 &132.7&00091909023-00091909044 & 2014:10:22-2014:12:05 & 20.2  \\
			&0721600301 &2014:03:20 &131.8&00091909045-00091909068 & 2014:12:09-2015:01:26 & 23.5  \\
			&0721600401 &2014:03:22 &133.3&00091909069-00091909090 & 2015:01:26-2015:03:15 & 21.7  \\
			&0721600501 &2014:03:24 &133.3&00010379001-00010379048 & 2017:12:07-2018:01:24 & 44.1 \\
			&&&&&& \\
Ark 564     &0006810101 &2000:06:17 &34.5& 00035062001-00035062003 & 2005:04:19-2005:12:09 & 18.4 \\
			&0006810301 &2001:06:09 &16.2& 00033282001-00033282002 & 2014:05:17            & 02.7 \\
			&0206400101 &2005:01:05 &101.8&00081687001             & 2015:05:22            & 02.5 \\
			&0670130201 &2011:05:24 &19.5& 00092237001-00092237015 & 2016:05:05-2016:12:27 & 13.8 \\
			&0670130301 &2011:05:30 &55.9& 00093158001-00093158017 & 2017:05:04-2017:12:19 & 14.8 \\
			&0670130401 &2011:06:05 &63.6& 00094000001-00094000018 & 2018:05:08-2018:12:22 & 14.7 \\
			&0670130501 &2011:06:11 &67.3& 00095000001-00095000016 & 2019:05:08-2019:12:22 & 14.0 \\
			&0670130601 &2011:06:17 &60.9& 00095653001-00095653017 & 2020:05:08-2020:12:22 & 13.3 \\
			&0670130701 &2011:06:25 &64.4& 00096050001-00096113003 & 2021:04:22-2021:04:30 & 05.7 \\
			&0670130801 &2011:06:29 &58.2&  & &  \\
			&0670130901 &2011:07:01 &55.9&  & &  \\
			&0830540101 &2018:12:01 &114.9&  & &  \\
			&0830540201 &2018:12:03 &114.4&  & &  \\
			&&&&&& \\
Fairall 9   &0101040201 &2000:07:05 &33.0 & 00037595001             & 2008:09:10            & 07.1 \\
			&0605800401 &2009:12:09 &130.1& 00037595002-00037595045 & 2013:04:16-2013:12:30 & 45.1 \\
			&0721110101 &2013:12:19 &73.0 & 00037595046-00091908098 & 2014:01:01-2014:10:21 & 49.5 \\
			&0721110201 &2014:01:02 &51.2 & 00037595052-00037595071 & 2015:02:08-2015:04:23 & 23.0 \\
			&0741330101 &2014:05:09 &141.4& 00094060001-00094060100 & 2018:05:13-2018:08:26 & 55.1 \\
			& & &                         & 00094060101-00094060225 & 2018:08:27-2018:12:31 & 67.5 \\
			& & &                         & 00094060226-00011108100 & 2019:01:01-2019:05:20 & 69.9 \\
			& & &                         & 00011108101-00011108200 & 2019:05:21-2019:09:07 & 82.1 \\
			& & &                         & 00011108201-00095400055 & 2019:09:08-2019:12:31 & 94.8 \\
			& & &                         & 00095400056-00037595140 & 2020:01:01-2020:12:28 & 88.8 \\
			&&&&&& \\
IRAS 13224--3809&0673580101 &2011:07:19 &133.1& 00090394001-00090394049 & 2010:04:03-2010:12:31 & 54.2 \\
			&0673580201 &2011:07:21 &132.4& 00090394050-00090394072 & 2011:01:04-2011:03:29 & 23.4 \\
			&0673580301 &2011:07:25 &129.4& 00091635002-00091635004 & 2014:02:18-2014:03:17 & 06.1 \\
			&0673580401 &2011:07:29 &134.7& 00034597001-00034597030 & 2016:07:08-2016:08:18 & 50.0 \\
			&0780560101 &2016:07:08 &141.3&  &  & \\
			&0780561301 &2016:07:10 &141.0&  &  & \\
			&0780561401 &2016:07:12 &138.1&  &  & \\
			&0780561501 &2016:07:20 &140.8&  &  & \\
			&0780561601 &2016:07:22 &140.8&  &  & \\
			&0780561701 &2016:07:24 &140.8&  &  & \\
			&0792180101 &2016:07:26 &141.0&  &  & \\
			&0792180201 &2016:07:30 &140.5&  &  & \\
			&0792180301 &2016:08:01 &140.5&  &  & \\
			&0792180401 &2016:08:03 &140.8&  &  & \\
			&0792180501 &2016:08:07 &138.0&  &  & \\
			&0792180601 &2016:08:09 &136.0&  &  & \\
			&&&&&& \\
Mrk 1018    &0201090201 &2005:01:15 &11.9& 00035166001 & 2005:08:05 & 05.2 \\
			&0554920301 &2008:08:07 &17.6& 00035776001 & 2008:06:11 & 04.7 \\
			&0821240201 &2018:07:23 &74.8& 00080898001-00080898002 & 2016:02:11-2016:02:11 & 06.8 \\
			&0821240301 &2019:01:04 &67.7& 00088207001-00035776058 & 2018:01:06-2018:11:24 & 53.8 \\
			&0864350101 &2021:02:04 &65.0&  &  &  \\
			&&&&&& \\
Mrk 110     &0840220701 &2019:11:03 &43.6& 00037561001-00037561002 & 2010:01:06-2010:01:12 & 13.8 \\
			&0840220801 &2019:11:05 &43.0& 00091850001-00091850003 & 2014:04:02-2014:04:09 & 01.1 \\
			&0840220901 &2019:11:07 &40.6& 00037561004             & 2015:03:23            & 01.2 \\
			&0852590101 &2019:11:17 &44.5& 00092396001-00037561009 & 2016:04:24-2016:05:21 & 50.4 \\
			&0852590201 &2020:04:06 &48.5& 00081846001-00093255100 & 2017:01:23-2017:11:29 & 59.2 \\
			&           &           &    & 00093255101-00093255193 & 2017:11:29-2017-12-31 & 95.6 \\
			&           &           &    & 00093255194-00010538012 & 2018:01:01-2018:01:25 & 68.3 \\
			&           &           &    & 00011136001-00011136061 & 2019:02:20-2019:06:22 & 55.4 \\
			&           &           &    & 00011136062-00088896002 & 2019:09:02-2019:11:17 & 62.6 \\
			&           &           &    & 00095040001-00011136158 & 2020:01:12-2020:06:07 & 24.8 \\
			&           &           &    & 00011136159-00096116002 & 2021:02:21-2021:09:09 & 12.2 \\
			&&&&&& \\
Mrk 335     &0101040101 &2000:12:25 &36.9 & 00035755001-00035755025 & 2007:05:17-2007:12:26 & 89.0 \\
			&0306870101 &2006:01:03 &133.3& 00035755026-00090006034 & 2008:01:02-2008:12:28 & 77.0 \\
			&0510010701 &2007:07:10 &22.6 & 00090006035-00035755040 & 2009:01:05-2010:12:31 & 79.6 \\
			&0600540601 &2009:06:11 &132.3& 00035755042-00035755080 & 2011:01:24-2011:12:30 & 40.5 \\
			&0600540501 &2009:06:13 &82.6 & 00035755081-00035755145 & 2012:01:03-2012:12:28 & 92.6 \\
			&0741280201 &2015:12:30 &140.4& 00035755146-00035755217 & 2013:01:05-2013:12:31 & 58.2 \\
			&0780500301 &2018:07:11 &114.5& 00035755218-00033420031 & 2014:01:04-2014:12:30 & 40.2 \\
			&0831790601 &2019:01:08 &117.8& 00033420032-00092239013 & 2015:01:07-2016:12:31 & 71.7 \\
			&0854590401 &2019:12:27 &105.9& 00033420114-00033420202 & 2017:01:08-2018:12:28 & 86.1 \\
			&           &           &     & 00033420203-00095118242 & 2019:01:04-2019:12:31 & 95.6 \\
			&           &           &     & 00095118243-00013544077 & 2020:01:01-2020:12:30 & 52.7 \\
			& &  & & &  &  \\
Mrk 359     &0112600601 &2000:07:09 &27.2 & 00045920001-00045920007 & 2012:06:05-2012:08:16 & 13.3 \\
			&0655590501 &2010:07:29 &64.0 & 00088710001-00088710002 & 2019:01:26-2019:01:28 & 03.6 \\
			&0830550801 &2019:01:25 &60.0 &  &  & \\
			&0830550901 &2019:01:26 &55.8 &  &  & \\
			&0830551001 &2019:01:28 &53.0 &  &  & \\
			&0830551101 &2019:01:31 &62.7 &  &  & \\
			&0830551201 &2019:02:02 &61.2 &  &  & \\
			&&&&&& \\
Mrk 509     &0130720101 &2000:10:25 &31.6& 00035469001-00035469003 & 2006:03:18-2006:04:20 & 08.3 \\
			&0130720201 &2001:04:20 &44.4& 00035469004             & 2007:03:26            & 03.0 \\
			&0306090201 &2005:10:18 &84.9& 00035469005-00035469023 & 2009:09:04-2009:12:12 & 17.3 \\
			&0306090301 &2005:10:20 &47.1& 00081459001-00035469026 & 2015:04:28-2015:09:08 & 10.6 \\
			&0306090401 &2006:04:25 &70.0& 00092240001-00092240008 & 2016:05:03-2016:11:01 & 07.7 \\
			&0601390201 &2009:10:15 &60.9& 00093157001-00093157241 & 2017:03:17-2017:12:15 & 232.7 \\
			&0601390301 &2009:10:19 &63.8& 00094002001-00094002011 & 2018:04:11-2018:11:09 & 11.2 \\
			&0601390401 &2009:10:23 &60.9& 00095002001-00095002013 & 2019:04:10-2019:11:09 & 11.3 \\
			&0601390501 &2009:10:29 &60.9& 00095655001-00095655012 & 2020:04:10-2020:11:09 & 10.2 \\
			&0601390601 &2009:11:02 &62.8& 00096119001-00096444006 & 2021:06:05-2021:11:07 & 06.5 \\
			&0601390701 &2009:11:06 &63.1& & & \\
			&0601390801 &2009:11:10 &60.9& & & \\
			&0601390901 &2009:11:14 &60.9& & & \\
			&0601391001 &2009:11:18 &65.5& & & \\
			&0601391101 &2009:11:20 &62.8& & & \\
			&&&&&& \\
Mrk 841     &0112910201 &2001:01:13 &10.1& 00035468002 & 2007:01:01 & 10.4 \\
			&0070740101 &2001:01:13 &12.3& 00081590001 & 2015:07:14 & 01.4  \\
			&0070740301 &2001:01:14 &14.8& 00092241001-00092241008 & 2016:06:18-2016:12:27 & 07.6 \\
			&0205340201 &2005:01:16 &72.7& 00092241009-00093162008 & 2017:01:04-2017:12:29 & 13.6 \\
			&0205340401 &2005:07:17 &29.5& 00093094001-00094003010 & 2018:01:04-2018:12:27 & 15.8 \\
			&0763790501 &2015:07:14 &29.5& 00094003011-00095003010 & 2019:01:03-2019:12:27 & 15.9 \\
			&           &           &    & 00095003011-00095656010 & 2020:01:03-2020:12:27 & 15.1 \\
			&           &           &    & 00095656011-00096445009 & 2021:01:03-2021:12:26 & 18.1 \\
			&&&&&& \\
PDS 456     &0041160101 &2001:02:26 &46.5& 00090078001-00090078015 & 2009:04:17-2009:10:02 & 72.2 \\
			&0501580101 &2007:09:12 &92.4& 00090078016-00090078019 & 2010:01:29-2010:03:20 & 25.9 \\
			&0501580201 &2007:09:14 &89.7& 00093145001-00093145047 & 2017:03:23-2017:10:09 & 152.6 \\
			&0721010201 &2013:08:27 &111.2&00010383001-00010383033 & 2018:08:22-2018:10:06 & 87.9 \\
			&0721010301 &2013:09:06 &113.5&00037748003-00010383058 & 2019:04:15-2019:09:26 & 56.5 \\
			&0721010401 &2013:09:15 &120.5&00010383059-00010383103 & 2021:07:01-2021:10:07 & 69.7 \\
			&0721010501 &2013:09:20 &112.1& & &  \\
			&0721010601 &2014:02:26 &140.8& & &  \\
			&0780690201 &2017:03:23 &82.3& & &  \\
			&0780690301 &2017:03:25 &89.3& & &  \\
			&0830390101 &2018:09:20 &86.0& & &  \\
			&0830390201 &2019:09:02 &83.0& & &  \\
			&0830390401 &2019:09:24 &94.3& & &  \\
			&&&&&& \\
PKS 0558--504&0116700301 &2000:02:07 &22.5& 00090020001-00090020025 & 2008:09:07-2008:12:25 & 52.8 \\
			&0117710701 &2000:02:12 &51.8& 00090020026-00090020050 & 2009:01:01-2009:06:15 & 47.5 \\
			&0119100201 &2000:03:01 &45.4& 00090020051-00090020081 & 2009:06:22-2009:12:28 & 53.5 \\
			&0125110101 &2000:05:24 &57.3& 00090020082-00090020093 & 2010:01:04-2010:03:30 & 15.2 \\
			&0129360201 &2000:10:10 &26.4& 00080990001             & 2016:11:19            & 06.7 \\
			&0137550201 &2001:06:26 &14.8&  &  &  \\
			&0137550601 &2001:10:19 &14.8&  &  &  \\
			&0555170201 &2008:09:07 &126.9&  &  &  \\
			&0555170301 &2008:09:09 &129.0&  &  &  \\
			&0555170401 &2008:09:11 &129.2&  &  &  \\
			&0555170501 &2008:09:13 &128.6&  &  &  \\
			&0555170601 &2008:09:15 &126.7&  &  &  \\
            &&&&&& \\
SWIFT J0501.9--3239&0312190701 &2006:01:28 &119.1&00035234001-00035234002 & 2005:10:29-2005:11:26 & 08.9 \\ 
			&0610180101 &2010:01:29 &76.9& 00040311001-00040311002 & 2011:01:07-2011:01:23 & 02.3 \\
			&0790810101 &2016:09:24 &120.8&00040311003-00040311004 & 2020:04:03-2020:04:17 & 01.8 \\
			&&&&&& \\
NGC 7469    &0112170101 &2000:12:26 &19.0& 00035470001-00035470005 & 2006:04:27-2006:08:02 & 08.9 \\
			&0112170301 &2000:12:26 &24.6& 00035470006-00035470007 & 2007:04:26-2007:05:17 & 02.2 \\
			&0207090101 &2004:11:30 &85.0& 00035470008-00035470060 & 2013:04:28-2013:06:03 & 32.4 \\
			&0207090201 &2004:12:03 &79.1& 00035470061-00035470121 & 2013:06:03-2013:07:05 & 37.9 \\
			&0760350201 &2015:06:12 &90.8& 00035470122-00035470171 & 2013:07:06-2013:08:01 & 29.3 \\
			&0760350301 &2015:11:24 &87.0& 00035470172-00035470210 & 2013:08:02-2013:08:20 & 25.0 \\
			&0760350401 &2015:12:15 &85.9& 00081531001-00092214049 & 2015:06:12-2015:12:31 & 67.4 \\
            &0760350501 &2015:12:23 &90.9& 00092214050-00092244014 & 2016:01:01-2016:12:19 & 36.5 \\
			&0760350601 &2015:12:24 &95.5& 00093165001-00093165013 & 2017:05:06-2017:12:16 & 12.1 \\
			&0760350701 &2015:12:26 &98.0& 00094006001-00094006016 & 2018:05:02-2018:12:16 & 12.3 \\
			&0760350801 &2015:12:28 &101.6&00095006001-00095006015 & 2019:05:02-2019:12:16 & 12.7 \\
			&           &         &      & 00095670001-00095670040 & 2020:04:27-2020:12:26 & 42.2 \\
			&&&&&& \\
Ton S180    &0110890401 &2000:12:14 &31.0& 00093003001-00093003002 & 2017:05:28-2017:06:03 & 01.3 \\
			&0110890701 &2002:06:30 &18.4& 00096062001-00096062002 & 2021:05:04-2021:05:06 & 03.3 \\
			&0764170101 &2015:07:03 &141.3&  &  &  \\
			&0790990101 &2016:06:13 &32.0&  &  &  \\
			&&&&&&\\
UGC 6728    &0312191601 &2006:02:23 &11.9& 00081098001 & 2016:07:10 & 06.9 \\
			&           &           &    & 00088256001 & 2017:10:13 & 07.0 \\
			&           &           &    & 00013662001 & 2020:09:05 & 01.8 \\
			&           &           &    & 00013662002-00096132008 & 2021:04:13-2021:10:25 & 02.2 \\

		\end{longtable*}

\section{Individual Source Details}
	\label{sec:isd}
	\textbf{1H~0323+342} is one of the closest ($z$=0.0629) exotic NLS1 AGN that exhibits superluminal motion of its relativistic outflow. The origin of the X-ray emission from this source could be explained as due to interactions between the disc and the Compton cloud \citep{Paliya2019}. From the method of single-epoch spectrum for several broad emission lines, \citep{Landt2017} estimated its mass as $\sim2\times10^7~M_\odot$. 
	
\begin{figure}
		\centering
    \includegraphics[width=1.0\textwidth, angle =0]{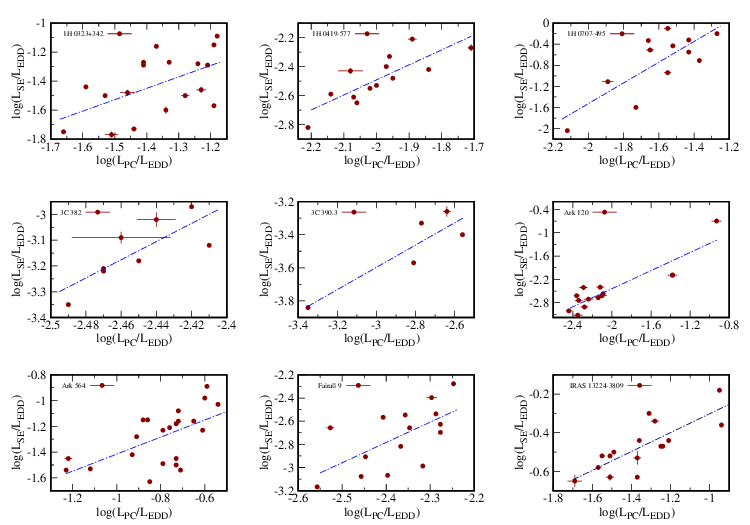}

		\caption{Correlation between the intrinsic luminosities of primary continuum (PC) and soft excess (SE) in 0.5-10.0 keV range, obtained from the {\tt powerlaw+powerlaw} model fitting for different sources. Each data point is normalised by the Eddington luminosity of the source. The blue line shows the linear correlation fit between the intrinsic luminosities of primary continuum (PC) and soft excess (SE).}
		\label{fig:figure_1}
	\end{figure}

	1H~0323+342 has been observed with {\it XMM-Newton} and {\it Swift}/XRT, and the X-ray spectra showed the nature of a bare-AGN, such as soft-excess, low $N_{\rm H}$ along the line of sight etc. The details of the observation log are given in Table~\ref{table:obs_log}. The procedures followed in fitting the X-ray data are described in Section~\ref{sec:RandD}. Our composite model for fitting the data in 0.5 to 10.0 keV range is:
	
	\begin{center}
		{\tt TBabs*zTBabs*(powerlaw+powerlaw)}
	\end{center}
	The results obtained from our spectral fitting are quoted in Table~\ref{tab:all_obs}. From the spectral fitting, we found that the variation of spectral slope of the primary continuum ($\Gamma^{\rm PC}$) is in the range of 1.0 to 1.9, whereas for the soft excess, the slope ($\Gamma^{\rm SE}$) varies in the range of 3.3 to 5.3. Corresponding luminosity for primary continuum ($\log(L^{\rm PC})$) varies from 43.8 to 44.3 and for soft excess ($\log(L^{\rm SE})$) varies from 43.6 to 44.5. The details of the parameter variations are presented in Table~\ref{tab:all_obs}. In Figure~\ref{fig:figure_1}, we presented the correlation between the normalised intrinsic luminosities of soft-excess $(L_{\rm SE}/L_{\rm Edd})$ and primary continuum $(L_{\rm PC}/L_{\rm Edd})$, where $L_{\rm Edd}$ is the Eddington luminosity (45.40; see Table~\ref{table:source_details}) for this source and the corresponding correlation coefficients are presented in Table~\ref{table:correlation}.

	\textbf{1H~0419--577} is a well-known nearby (z=0.104) Seyfert~1 AGN that has been observed with almost all X-ray observatories. This AGN shows complex X-ray broad-band spectrum \citep{Page2002, Walton2010}: strong soft-excess below 2 keV, reflection hump above 10 keV and an Fe~K$_\alpha$ line near 6 keV \citep{Pal2013, Walton2010}. The estimated mass of the SMBH, harboured in its nucleus, is $3.8\times10^8$ M$_\odot$\citep{Nandra2005}.
	
	{\it XMM-Newton} and {\it Swift}/XRT observed 1H~0419--577 in multiple epochs in nearly regular intervals from 2002 to 2018. The details of the observation log are given in Table~\ref{table:obs_log}. The details of the procedure followed for spectral fitting of the data are described in Section~\ref{sec:RandD}. We added a {\tt Gaussian} component as {\tt zGauss} with the {\tt (powerlaw+powerlaw)} model while fitting. The composite model for this source is:
	\begin{center}
		{\tt TBabs*zTBabs*(powerlaw+powerlaw+zGauss)}
	\end{center}
	The {\tt zGauss} component was used to take care of the Fe~K$_\alpha$ line near 6.4 keV with z=0.104. We found the Fe K$_\alpha$ line at $6.4\pm0.11$ keV with width $(\sigma)$=$160\pm110$ eV. The spectral fitting results are shown in Table~\ref{tab:all_obs}. From the spectral analysis, we found that the spectral slope for the primary continuum ($\Gamma^{\rm PC}$) varies from 1.0 to 2.3 and the spectral slope for the soft-excess component ($\Gamma^{\rm SE}$) varies from 2.6 to 6.7. We also calculated the luminosities corresponding to each spectral component. The luminosity of the primary continuum ($\log(L^{\rm PC})$) varies from 44.5 to 45.0 and for the soft excess, it ($\log(L^{\rm SE})$) varies from 44.1 to 44.9. The details of the parameter variations are presented in Table~\ref{tab:all_obs} and the variations of normalised intrinsic luminosities of soft-excess $(L_{\rm SE}/L_{\rm Edd})$ and primary continuum $(L_{\rm PC}/L_{\rm Edd})$ are shown in Figure~\ref{fig:figure_1}. As the intrinsic luminosities of soft-excess $(L_{\rm SE}/L_{\rm Edd})$ and the primary continuum $(L_{\rm PC}/L_{\rm Edd})$ are correlated to each other, we calculated the correlation coefficients  using different methods which are presented in Table~\ref{table:correlation}.
		
	\textbf{1H~0707--495} is a low redshift (z=0.04) narrow-line Seyfert~1 (NLS1) galaxy, which has extreme variability \citep{Turner1999, Boller2002} and spectral shape with a strong soft-excess and relativistic broad iron emission line \citep{Fabian2009}. This was the first AGN where an X-ray reverberation lag was detected \citep{Fabian2009} and this lag was shown to have a strong Fe line feature in the lag-energy spectrum \citep{Kara2013}. We adopted the black hole mass as $2\times10^6$ M$_\odot$ from \citep{Zhau2005}.
	
	1H~0707--495 was observed with {\it XMM-Newton} and {\it Swift}/XRT multiple times over a period of 20 years (Table~\ref{table:obs_log}). We analysed these observations using a composite model which reads in {\tt XSPEC} as,
	\begin{center}
		{\tt TBabs*zTBabs*(powerlaw+powerlaw+zGauss)}.
	\end{center}
	The details of the spectral fitting procedure are discussed in Section~\ref{sec:RandD}. The {\tt zGauss} component was used for the broad Fe~K$_\alpha$ line detected at $6.59\pm0.63$ keV with a line width ($\sigma$) of $112\pm90$ eV for z=0.04. In Table~\ref{tab:all_obs}, we represent the spectral fitting results. We found the variation of the spectral slope for primary continuum ($\Gamma^{\rm PC}$) from 1.0 to 3.5 and the spectral slope for soft-excess ($\Gamma^{\rm SE}$) from 3.5 to 7.0. Corresponding luminosities vary from 42.1 to 43.2 and 42.3 to 44.3 for the primary continuum ($\log(L^{\rm PC})$) and soft-excess ($\log(L^{\rm SE})$), respectively. The correlation between these two intrinsic luminosities (normalised by Eddington luminosity $L_{\rm Edd}$) are also plotted in Figure~\ref{fig:figure_1} with linear interpolation. Corresponding correlation coefficients, using different methods, are also shown in Table~\ref{table:correlation}.
		
	\textbf{3C~382} is a nearby (z=0.058) broad-line radio-loud AGN having a supermassive black hole of mass $(1.0\pm0.3)\times10^9$ M$_\odot$ \citep{Fausnaugh2017}. In the X-ray regime, this source has a strong soft-excess and \citep{Wozniak1998, Grandi2001} showed that the soft-excess can not be explained by extended thermal emission.
	
	3C~382 have been observed with {\it XMM-Newton} and {\it Swift}/XRT from 2008 to 2021. For the spectral analysis, we used the composite model as,
	\begin{center}
		{\tt TBabs*zTBabs*(powerlaw+powerlaw+zGauss)}.
	\end{center}
	The component {\tt zGauss} was used to fit the Fe K$_\alpha$ line at $6.42\pm0.14$ keV with a width of $98\pm59$ eV for z=0.058. The details of the spectral fitting procedure are discussed in Section~\ref{sec:RandD} and the results obtained from fitting the data with the composite model are presented in Table~\ref{tab:all_obs}. From Table~\ref{tab:all_obs}, it can be seen that the variations of spectral indices for the primary continuum ($\Gamma^{\rm PC}$) and the soft-excess ($\Gamma^{\rm SE}$) are from 1.2 to 1.8 and 1.5 to 3.6, respectively. The primary continuum luminosity ($\log(L^{\rm PC})$) is found to vary from 43.5 to 44.7, whereas the soft-excess ($\log(L^{\rm SE})$) varies from 43.7 to 44.4. The correlation between the intrinsic normalised luminosities of the primary continuum ($\log(L_{\rm PC}/L_{\rm Edd})$) and the soft-excess ($\log(L_{\rm SE}/L_{\rm Edd})$) is shown in Figure~\ref{fig:figure_1} with linear interpolation. The correlation coefficients, calculated from the different methods, are represented in Table~\ref{table:correlation}.
	
	\textbf{3C~390.3} is a radio-loud Seyfert~1 nearby (z=0.056) AGN \citep{Afanasiev2015}. However, from timing analysis, this source is classified as a radio-quiet Seyfert and it was found that there is no noticeable contribution from the jet to the X-ray emission \citep{Sambruna2009}. In the X-ray regime, detection of soft-excess below 2 keV and narrow Fe lines are reported from {\it XMM-Newton} observations of the source \citep{Sambruna2009}. Later, from the modelling of Fe lines and ionized reflection, it is shown that the source is ambiguous in broad-line radio galaxies \citep{Tombesi2013}. The mass of the central black hole is estimated as $2.0\times10^9$ M$_\odot$ \citep{Sergeev2011}. 
	
	For this work, we considered {\it XMM-Newton} and {\it Swift}/XRT observations of this source. {\it XMM-Newton} observed the source twice in 2010 (Table~\ref{table:obs_log}) and we binned the {\it Swift}/XRT observation into three bins (Table~\ref{table:obs_log}). We used
	\begin{center}
		{\tt TBabs*zTBabs*(powerlaw+powerlaw+zGauss)}
	\end{center}
	as the composite model to analyse the X-ray spectrum of this source. Considering the redshift of the source $z$=0.056, the component {\tt zGauss} was used to take care of the Fe-line at $6.6\pm0.18$ keV with a line width of $490\pm360$ eV. The detailed procedure of spectral analysis is described in Section~\ref{sec:RandD} and the {\tt powerlaw+powerlaw} model-fitted results are presented in Table~\ref{tab:all_obs}. The power-law index for the primary continuum ($\Gamma^{\rm PC}$) varies from 1.0 to 1.6 and the corresponding luminosity ($\log(L^{\rm PC})$) varies from 44.0 to 44.8. Similarly, the power-law index for the soft-excess ($\Gamma^{\rm SE}$) varies from 1.9 to 3.9 and the corresponding luminosity  ($\log(L^{\rm SE})$) varies from 43.6 to 44.8. The correlation between these two luminosities is shown in Figure=\ref{fig:figure_1}. Here, we normalised the primary continuum and soft excess luminosities using the Eddington luminosity of this source. We also calculated the correlation coefficients using various methods and presented in Table~\ref{table:correlation}.
	
	\textbf{Ark~120} is a nearby (z=0.0327) Seyfert~1 AGN, which has a `bare' nucleus \citep{Crenshaw1999, Vaughan2004}. The results obtained from a detailed study on this source \citep{Nandi2021} are used in the present work. We considered all X-ray observations of this source from {\it XMM-Newton}, {\it Swift}/XRT, and {\it Suzaku}. The {\tt powerlaw+powerlaw} model fitted results are presented in Table~3 of \citep{Nandi2021}. We found the power-law index for primary continuum ($\Gamma^{\rm PC})$) varies from 1.6 to 2.1 and the variation of corresponding luminosity ($\log(L^{\rm PC})$) is from 43.8 to 45.4. The power-law index for the soft-excess ($\Gamma^{\rm SE})$) is found to vary from 2.5 to 4.2 and the corresponding luminosity ($\log(L^{\rm SE})$) varies from 43.2 to 45.6. The variation of these two luminosities are plotted in Figure~\ref{fig:figure_1} and the correlation between them is shown by a blue dotted line in the same plot. We calculated the correlations of these two luminosities using different algorithms and the corresponding values of correlation coefficients are given in the Table~\ref{table:correlation}.

	\textbf{Ark~564} is an X-ray bright narrow-line Seyfert~1 AGN located at z=0.0247 with 3 to 10 keV luminosity of $\sim2.4\times10^{43}$ erg.s$^{-1}$ \citep{Turner2001}. This source is a rapid variable source in the X-ray domain with a strong soft excess and a steep spectrum in 0.5 to 10.0 keV band \citep{Turner1999,Turner2001}. From the analysis of {\it XMM-Newton} observations of this source, \citep{Sarma2015} showed that this high-Eddington NLS1 exhibits a prominent correlation between the spectral slope and flux. We considered the mass of the central black hole of this object as $2.5\times10^6$ M$_\odot$.
	
	{\it XMM-Newton} and {\it Swift}/XRT observed Ark~564 multiple times from 2000 to 2018. The details of the observation log are given in Table~\ref{table:obs_log}. The details of the spectral fitting procedure of the X-ray data are described in Section~\ref{sec:RandD}. For the spectral fitting, we used the model which reads in {\tt XSPEC} as:
	\begin{center}
		{\tt TBabs*zTBabs*(powerlaw+powerlaw)}
	\end{center}
	
 The power-law indices for primary continuum and soft-excess are presented in Table~\ref{tab:all_obs}. From the spectral fitting, we found that the power-law index for the primary continuum ($\Gamma^{\rm PC})$) varies from 1.0 to 2.6 and the corresponding luminosity ($\log(L^{\rm PC})$) varies from 43.2 to 43.9. The index for the soft-excess ($\Gamma^{\rm SE})$) is found to vary from 3.8 to 4.7 and the corresponding luminosity ($\log(L^{\rm SE})$) varies from 42.9 to 43.9. The variation of these two luminosities are plotted in Figure~\ref{fig:figure_1} and the correlation is shown by a blue dotted line in the figure. We also calculated correlations between these two luminosities using different algorithms and the corresponding values of correlation coefficients are given in Table~\ref{table:correlation}.   
		
	\textbf{Fairall~9} \citep{Fairall1977} is a nearby (z=0.047) Seyfert~1 AGN with a central black hole of mass $2.55\times10^8$ M$_\odot$ \citep{Peterson2004}. Several X-ray studies suggested that this source has low extinction and is free from warm absorbers around the central region \citep{Emmanoulopoulos2011}. The persistent nature of different spectral components of this source suggests a clear view of the inner flow around its central engine \citep{Lohfink2016}.
	
	Fairall~9 has been observed by almost all X-ray missions and we considered data from the {\it XMM-Newton} and {\it Swift}/XRT observations of the source for our analysis. From the spectral analysis, we found a strong soft-excess below 2 keV, low Hydrogen column density along the line of sight and Fe~K$_\alpha$ line near 6.4 keV. The details of the observations are given in Table~\ref{table:obs_log} and the spectral analysis procedures followed in our analysis are discussed in Section~\ref{sec:RandD}. For the spectral analysis, we used the composite model as,
	\begin{center}
		{\tt TBabs*zTBabs*(powerlaw+powerlaw+zGauss)}
	\end{center}
	where, the component {\tt zGauss} was used for the Fe line at $6.42\pm0.2$ keV with width of $84\pm30$ eV. The power-law indices obtained from our fitting the data with the {\tt powerlaw+powerlaw} model are given in Table~\ref{tab:all_obs}. We found that the power-law index for the primary continuum $(\Gamma^{\rm PC})$ is nearly constant at $\sim1.6$, whereas for soft excess, the index $(\Gamma^{\rm SE})$ varies from 1.5 to 3.6. We calculated the luminosities of these components and their correlation is plotted in Figure~\ref{fig:figure_1}. The luminosity of the primary continuum ($\log(L^{\rm PC})$) varies from 43.5 to 44.7, whereas the soft excess luminosity ($\log(L^{\rm SE})$) varies between 43.7 and 44.9. The correlation coefficients of these luminosities, calculated from different methods, are presented in Table~\ref{table:correlation}.
		
	\textbf{IRAS~13224--3809} is a radio-quiet narrow-line Seyfert~1 AGN, located at $z=0.0658$, with a central black hole of mass $1.26\times10^6$ M$_\odot$ \citep{Alston2019}. This is an X-ray bright AGN, considered one of the most variable Seyfert in X-ray band \citep{Boller2003,Dewangan2002,Jiang2018,Pinto2018}. From the X-ray data analysis, it was found that the X-ray source is located at $\sim3~r_g$ from the central black hole \citep{Emmanoulopoulos2014}.
	
	IRAS~13224--3809 was observed with the {\it XMM-Newton} and {\it Swift} observatories multiple times from 2010 to 2016. \citep{Alston2019} used the {\it XMM-Newton} data and found a piece of strong evidence for the non-stationary variability in the X-ray band. We used both {\it XMM-Newton} and {\it Swift}/XRT data (Table~\ref{table:obs_log}) in this work. For the X-ray spectral analysis, we used the composite model as,
	\begin{center}
		{\tt TBabs*zTBabs*(powerlaw+powerlaw)gabs},
	\end{center}
	where, the component {\tt gabs} was used for the absorption line at $1.81\pm0.4$ keV with a width of $154\pm69$ eV, as reported by \citep{Jiang2018}. The detailed spectral analysis is discussed in Section~\ref{sec:RandD} and the {\tt powerlaw+powerlaw} model fitting to the X-ray spectra are presented in Table~\ref{tab:all_obs}. We found that the variation of power-law indices for primary continuum ($\Gamma^{\rm PC}$) and soft-excess ($\Gamma^{\rm SE}$) are from 1.1 to 2.8 and 4.5 to 6.7, respectively and corresponding luminosities vary from 42.7 to 43.5 for primary continuum ($\log(L^{\rm PC})$) and from 43.2 to 44.1 for soft-excess ($\log(L^{\rm SE})$). The correlation between $\log(L^{\rm PC})$ and $\log(L^{\rm SE})$ is plotted in Figure~\ref{fig:figure_1} and corresponding correlation coefficients from different algorithms are presented in Table~\ref{table:correlation}.
		
	\textbf{Mrk~1018} is a Seyfert AGN which is mostly popular for its changing-look behaviour. It changed its behaviour from Seyfert 1.9 to Seyfert 1 between 1979 and 1984 \citep{Cohen1986} and then returned to its previous state in 2015 \citep{McElroy2016}. The AGN is located at a redshift of $z$=0.0424 and the central black hole mass is estimated as $6.91\times10^7$ \citep{Ezhikode2017}.
	
\begin{figure}
		\centering
  \includegraphics[width=1.0\textwidth, angle =0]{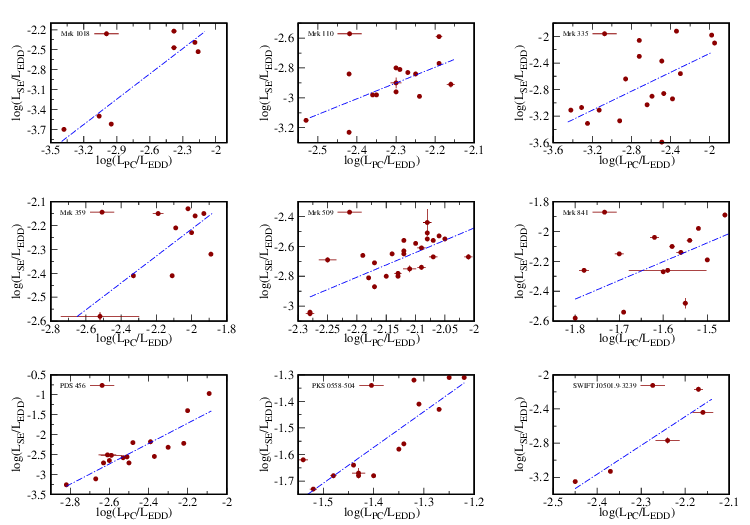}
		\caption{Correlation of intrinsic luminosities of the primary continuum (PC) and soft excess (SE) of 0.5-10.0 keV obtained from {\tt powerlaw+powerlaw} fitting for different sources. Each data point are normalised by the Eddington luminosity of the source. The blue line shows the linear correlation fit between the intrinsic luminosities of the primary continuum (PC) and soft excess (SE).}
		\label{fig:figure_2}
	\end{figure}

	Mrk~1018 was observed with {\it XMM-Newton} and {\it Swift}/XRT at multiple epochs. The details of the observation log are given in Table~\ref{table:obs_log}. We used the composite model 
	\begin{center}
		{\tt TBabs*zTBabs*(powerlaw+powerlaw)}
	\end{center}
	to fit the X-ray spectra of this source. The details of the spectral fitting procedure are described in Section~\ref{sec:RandD} and the corresponding results are shown in Table~\ref{tab:all_obs}. We found that the variation of power law indices for the primary continuum $(\Gamma^{\rm PC})$ and the soft-excess $(\Gamma^{\rm SE})$ from 1.0 to 1.5 and 2.2 to 3.0 respectively and the corresponding luminosities variation are 42.6 to 43.8 for primary continuum $(\log(L^{\rm PC}))$ and 42.2 to 43.7 for soft-excess $(\log(L^{\rm SE}))$). In Figure~\ref{fig:figure_2}, we have presented the correlation between the normalised intrinsic luminosities of soft-excess and primary continuum and the corresponding correlation coefficients are presented in Table~\ref{table:correlation}.
	
	\textbf{Mrk~110} is an X-ray bright radio-quiet narrow-line Seyfert~1 AGN located at $z$=0.0353 with a central black hole of mass $1.38\times10^8$ M$_\odot$ \citep{Liu2017}. From the studies of the variabilities of Hydrogen and Helium lines, \citep{Kollatschny2004} established the connection between the broad-line region (BLR) and accretion disc. From the X-ray studies, it was found that the soft-excess part (below 2 keV) is nearly absorption-free and moderately broad O~VII and Fe~K$_\alpha$ emission lines are present in the X-ray spectrum \citep{Porquet2021}. 
	
	Mrk~110 have been observed by {\it XMM-Newton} in 2019 and 2020 and by {\it Swift}/XRT in the period of 2010 to 2021 and nearly continuous from 2014 to 2021 (Table~\ref{table:obs_log}). We analysed the X-ray spectra from these observations using the composite model:
	\begin{center}
		{\tt TBabs*zTBabs*(powerlaw+powerlaw+zGauss)}
	\end{center}
	where, the component {\tt zGauss} was used to take care of the Fe~K$_\alpha$ line which was found at $6.42\pm0.05$ keV with a width of $79\pm50$ eV. The {\tt powerlaw+powerlaw} fitted results are presented in Table~\ref{tab:all_obs}. The details descriptions of spectral fitting are discussed in Section~\ref{sec:RandD}. The power-law indices for the primary continuum $(\Gamma^{\rm PC})$ and the soft-excess $(\Gamma^{\rm SE})$ vary from 1.5 to 1.7 and 2.8 to 4.4, respectively. We calculated corresponding luminosities for the primary continuum $(\log(L^{\rm PC}))$ and soft-excess $(\log(L^{\rm SE}))$, which vary from 43.8 to 44.2 and 43.1 to 43.6, respectively. The correlation between these two luminosities are plotted in Figure~\ref{fig:figure_2} and the correlation coefficients are quoted in Table~\ref{table:correlation}. 
	
	\textbf{Mrk~335} is a nearby (z=0.026) narrow-line Seyfert~1 AGN with a black hole mass of $2.7\times10^7$ M$_\odot$ \citep{Grier2012}. This source is popular for its extraordinary variation between high and low flux states in X-ray bands. Initially, it was known to be an X-ray bright source \citep{Halpern1982}. However, the X-ray intensity suddenly dropped from the brightest stage to the very low flux level in 2007 \citep{Grupe2007} and after that, the source remained mostly in a low flux state. In this low state, however, the source shows some X-ray flaring activities \citep{Gallo2018}.
	
	{\it XMM-Newton} and {\it Swift}/XRT observed Mrk~335 multiple times in the period from 2006 to 2020. The details of the observation log are given in Table~\ref{table:obs_log}. From X-ray studies, it was found that this source has a bare nucleus with strong soft-excess emission \citep{Longinotti2008}, complex absorption features \citep{Longinotti2013} and Fe~K$_\alpha$ emission line \citep{Keek2016}. We used the composite model as,
	\begin{center}
		{\tt TBabs*zTBabs*(powerlaw+powerlaw+zGauss)gabs}
	\end{center}
	to fit the overall X-ray spectra of Mrk~335. The components {\tt zGauss} and {\tt gabs} were used for the Fe~K$_\alpha$ line at $6.46\pm0.3$ keV with a width of $112\pm52$ eV and an absorption line which was found at $0.76\pm0.2$ keV with a width of $32\pm15$ eV, respectively. The detailed spectral fitting procedure is described in Section~\ref{sec:RandD} and the results obtained from the spectral fitting of the data with the {\tt powerlaw+powerlaw} model are quoted in Table~\ref{tab:all_obs}. From the spectral fitting, we found that the power-law indices for the primary continuum $(\Gamma^{\rm PC})$ and soft-excess $(\Gamma^{\rm SE})$ vary from 0.5 to 2.0 and 3.5 to 6.5, respectively and corresponding luminosities for the primary continuum $(\log(L^{\rm PC}))$ and soft-excess $(\log(L^{\rm SE}))$ vary from 42.1 to 43.6 and 41.9 to 43.6, respectively. The variations of normalised intrinsic luminosities ($L_{\rm PC}/L_{\rm Edd}$ vs $L_{\rm SE}/L_{\rm Edd}$) are plotted in Figure~\ref{fig:figure_2}. Corresponding correlation coefficients, calculated from  different algorithms, are shown in Table~\ref{table:correlation}.
		
	\textbf{Mrk~359} is a nearby (0.0174) narrow line Seyfert~1 AGN where the width of the broad emission line was reported to be smaller than 2000 km/s \citep{Elvis1992}. From the X-ray data analysis, \citep{OBrien2001} first reported the presence of prominent soft-excess below 2 keV without any significant neutral/warm intrinsic absorption along the line of sight. Besides this, a neutral Fe line of equivalent width $\sim200$ eV was also reported via X-ray data analysis. From the X-ray variability study, \citep{Middei2020} estimated the mass of the central object of this source as $3.6\times10^6$ M$_\odot$.
	
	Mrk~359 have been observed with {\it XMM-Newton} and {\it Swift}/XRT multiple times between 2000 to 2019 (Table~\ref{table:obs_log}). For the spectral analysis of X-ray  data from the {\it XMM-Newton} and {\it Swift}/XRT observations, we used
	\begin{center}
		{\tt TBabs*zTBabs*(powerlaw+powerlaw)}
	\end{center}
	as a composite model. We followed the same procedure, as described in Section~\ref{sec:RandD}, to fit the data. The result obtained from the spectral fitting is presented in Table~\ref{tab:all_obs}. Our spectral fitting indicates that the source was nearly in the same state throughout the observational period. We found that the power-law index for the primary continuum ($\Gamma^{\rm PC})$) varies from 1.0 to 1.7 and the corresponding luminosity ($\log(L^{\rm PC})$) varies from 42.1 to 42.8, whereas the power-law index for the soft-excess ($\Gamma^{\rm SE})$) varies from 2.2 to 3.4 and the corresponding luminosity ($\log(L^{\rm SE})$) varies from 42.1 to 42.5. The variation of these two luminosities are plotted in Figure~\ref{fig:figure_2}. We also calculated the correlations of these two luminosities from different algorithms and the corresponding values of correlation coefficients are given in Table~\ref{table:correlation}.
	
 	\textbf{Mrk~509} is a well-studied nearby (z=0.0344) Seyfert 1 AGN which is powered by a central black hole of mass $1.4\times10^8$ M$_\odot$ \citep{Peterson2004}. The soft-excess emission below 2 keV was first identified by \citet{Singh1985}. An iron line was detected in the X-ray spectrum \citep{Morini1987} which led to a detailed discussion of reflection features by \citet{Pounds1994}.
	
	In this work, we used the archival data of Mrk~509 from the {\it XMM-Newton} and {\it Swift}/XRT observations (Table~\ref{table:obs_log}). The data extraction procedures are described in Section~\ref{sec:data reduction} and the procedure for spectral fitting is discussed in Section~\ref{sec:RandD}. To fit the overall spectrum, we used the composite model as,
	\begin{center}
		{\tt TBabs*zTBabs*(powerlaw+powerlaw+zGauss)}
	\end{center}
	The component {\tt zGauss} was used for the Fe~K$_\alpha$ line at $6.37\pm0.24$ keV with a width of $350\pm87$ eV. The {\tt powerlaw+powerlaw} fitted results are presented in Table~\ref{tab:all_obs}. From the spectral fitting, we found that the variation of power-law indices for primary continuum $(\Gamma^{\rm PC})$ and soft-excess $(\Gamma^{\rm SE})$ vary from 1.0 to 1.8 and 1.7 to 4.3, respectively. Corresponding luminosities for the primary continuum $(\log(L^{\rm PC}))$ and the soft-excess $(\log(L^{\rm SE}))$ vary from 44.0 to 44.3 and 43.2 to 43.8, respectively. The variations of normalised intrinsic luminosities ($L_{\rm PC}/L_{\rm Edd}$ vs $L_{\rm SE}/L_{\rm Edd}$) are plotted in Figure~\ref{fig:figure_2} and the corresponding correlation coefficients, calculated from the different algorithms, are shown in Table~\ref{table:correlation}.
	
	\textbf{Mrk~841} is a nearby (z=0.00364) Seyfert~1 AGN which is known for its large spectral variability \citep{George1993}. This is the first object where a soft-excess \citep{Arnaud1985} was observed along Mrk~509. A variable Fe line was reported by \citep{George1993}. From the UV/X-ray observations, \citep{Ross1992} found that this source is a face-on source with a central black hole of mass $2\times10^7$ M$_\odot$. Recently, \cite{Mehdipour2023} reported a substantial decrement in the soft-excess band compared to its harder counterpart over the last 15 years.
	
	Mrk~841 has been observed 5 times during 3 different periods (January 2001, January 2005 and July 2005) with {\it XMM-Newton} and multiple times with {\it Swift}/XRT from 2007 to 2021. the details of the observations are given in Table~\ref{table:obs_log}. We used the composite model as,
	\begin{center}
		{\tt TBabs*zTBabs*(powerlaw+powerlaw+zGauss)}
	\end{center}
	to fit the X-ray spectra of this source. The {\tt zGauss} component was used for the Fe-line at $6.22\pm0.35$ keV with a width of $98\pm57$ eV. The detailed spectral analysis is discussed in Section~\ref{sec:RandD} and the {\tt powerlaw+powerlaw} fitting results are presented in Table~\ref{tab:all_obs}. We found that the variation of power-law indices for the primary continuum ($\Gamma^{\rm PC}$) and the soft-excess ($\Gamma^{\rm SE}$) are in the ranges of 1.0 to 2.0 and 2.0 to 4.8, respectively and corresponding luminosities vary in the range of 43.6 to 43.9 for primary continuum ($\log(L^{\rm PC})$) and 42.2 to 43.6 for soft-excess ($\log(L^{\rm SE})$). The correlations between $\log(L^{\rm PC})$ and $\log(L^{\rm SE})$ are plotted in Figure~\ref{fig:figure_2} and the corresponding correlation coefficients from different algorithms are presented in Table~\ref{table:correlation}.

	\textbf{PDS~456} is a radio-quiet AGN at a redshift of z=0.184 with a black hole mass of $1.58\times10^9$ M$_\odot$ \citep{Nardini2015}. This source is known for its ultra-fast wind with an outflow velocity of $0.25-0.3c$ \citep{Boissay2019,Nardini2015}. In the X-ray regime, the first flare was detected \citep{Reeves2000} and later, rapid variability has also been observed in long observations with different X-ray observatories. 
	
	In this work, we used {\it XMM-Newton} and {\it Swift}/XRT observations of PDS~456. The observation details are given in Table~\ref{table:obs_log}. For the X-ray spectral analysis, we used the composite model as,
	\begin{center}
		{\tt TBabs*zTBabs*(powerlaw+powerlaw+zGauss)},
	\end{center}
	where, the component {\tt zGauss} was used for the Fe line at $6.70\pm0.32$ keV with a width of $168\pm75$ eV. The spectral analysis procedure followed in this work is discussed in Section~\ref{sec:RandD} and the results obtained from the {\tt powerlaw+powerlaw} fitting are represented in Table~\ref{tab:all_obs}. We found that the variations of power-law indices for the primary continuum ($\Gamma^{\rm PC}$) and the soft-excess ($\Gamma^{\rm SE}$) are from 1.6 to 2.3 and 2.0 to 7.3, respectively and the corresponding luminosity variations are from 44.6 to 45.2 for the primary continuum ($\log(L^{\rm PC})$) and 44.0 to 46.3 for the soft-excess ($\log(L^{\rm SE})$). The correlation between $\log(L^{\rm PC})$ and $\log(L^{\rm SE})$ is plotted in Figure~\ref{fig:figure_2} and the corresponding correlation coefficients obtained from different algorithms are represented in Table~\ref{table:correlation}.
		
	\textbf{PKS~0558--504} (z=0.137) is a radio-loud narrow-line Seyfert~1 AGN whose spectrum is similar to that of radio-quiet narrow-line Seyfert~1 galaxies \citep{Komossa2006}. Radio observations (VLA) of the AGN revealed the two-sided radio structure \citep{Doi2012}. Multiwavelength observations constrained the central black hole mass to $2.5\times10^8$ M$_\odot$ \citep{Gliozzi2010}. From the spectral energy distribution (SED) of PKS~0550-504, it was reported that the SED is mostly dominated by optical-UV photons and the jet emission does not dominate beyond the radio band \citep{Gliozzi2010}.
	
	PKS~0558--504 was observed with {\it XMM-Newton} and {\it Swift}/XRT multiple times from 2000 to 2016 (see Table~\ref{table:obs_log}). We analysed these observations using the composite model,
	\begin{center}
		{\tt TBabs*zTBabs*(powerlaw+powerlaw)}
	\end{center}
	The details of the spectral fitting procedure are given in Section~\ref{sec:RandD}. The spectral analysis results are represented in Table~\ref{tab:all_obs}. Here, we found that the variation of the power-law index for primary continuum $(\Gamma^{\rm PC})$ is in the range of 1.8 to 2.2 and for the soft excess $(\Gamma^{\rm SE})$, it varies in the range of 3.43 to 5.17. Corresponding luminosity variations are in the range of 45.0 to 46.0 and 44.0 to 45.8 for primary continuum ($\log(L^{\rm PC})$) and soft-excess ($\log(L^{\rm SE})$), respectively. The variation of normalised intrinsic luminosities are plotted ($L_{\rm PC}/L_{\rm Edd}$ vs $L_{\rm SE}/L_{\rm Edd}$) in  Figure~\ref{fig:figure_2} and the corresponding correlation coefficients, calculated from the different algorithms, are shown in Table~\ref{table:correlation}.
	
	\textbf{SWIFT~J0501.9--3239} (also known as ESO~362-G18 or MCG~05-13-17) is a nearby (z=0.0124) Seyfert AGN which is popular for its short-timescale spectral variability \citep{Agis2014}. It was reported that this source hosts an extremely spinning supermassive black hole with a mass of $4.5\times10^7$ M$_\odot$ \citep{Agis2014}. It was also reported that the X-ray emission region is located within $\sim50$ $r_g$ (where $r_g=GM_{\rm BH}/c^2$) \citep{Agis2014}.
	
	For this work, we used data from the {\it XMM-Newton} and {\it Swift}/XRT observations of the source. The details of the observations are given in Table~\ref{table:obs_log} and the data reduction procedure is discussed in Section~\ref{sec:data reduction}. For the spectral fitting, we used the composite model as,
	\begin{center}
		{\tt TBabs*zTBabs*(powerlaw+powerlaw+zGauss)},
	\end{center}
	where, the component {\tt zGauss} was used for the Fe line at $6.37\pm0.35$ keV with a width of $69\pm25$ eV. The detailed spectral analysis is discussed in Section~\ref{sec:RandD} and the {\tt powerlaw+powerlaw} fitting results are represented in Table~\ref{tab:all_obs}. We found that the variation of power-law indices for the primary continuum ($\Gamma^{\rm PC}$) and soft-excess ($\Gamma^{\rm SE}$) are in the range of 1.0 to 1.3 and 2.4 to 4.9, respectively and the corresponding luminosity variation are 42.6 to 42.8 for primary continuum ($\log(L^{\rm PC})$) and 41.7 to 42.9 for soft-excess ($\log(L^{\rm SE})$). The correlations between the luminosities $\log(L^{\rm PC})$ and $\log(L^{\rm SE})$ are plotted in Figure~\ref{fig:figure_2} and the corresponding correlation coefficients from different algorithms are represented in Table~\ref{table:correlation}.
		
	\textbf{NGC~7469} is a well-studied nearby (z=0.0163) Seyfert~1 AGN which is most popular for its variability and excess brightness in X-ray domain \citep{Markowitz2004}. This source shows wavelength-dependent continuum delays of several light days, which indicates the presence of an accretion disk around the central black hole \citep{Collier1998} and  the reverberation mapping results in a black hole mass estimation of $1.0\times10^7$ M$_\odot$ \citep{Peterson2004}.
	
\begin{figure}
		\centering
  \includegraphics[trim={0 6cm 0 0}, width=1.0\textwidth, angle =0]{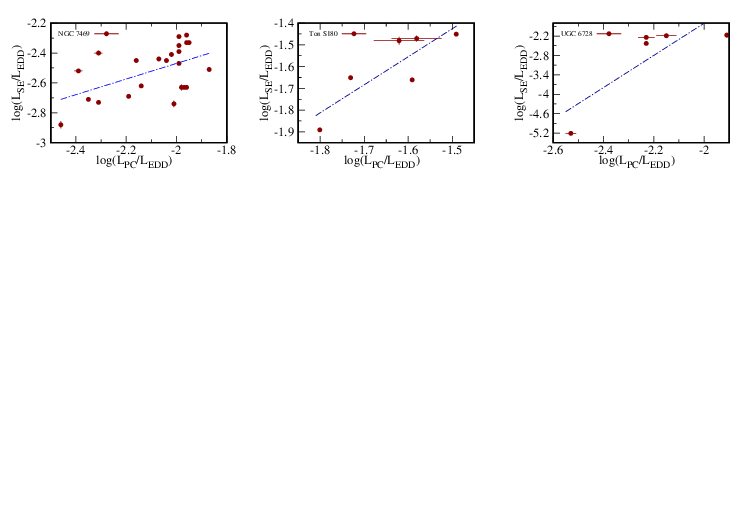}
		\caption{Correlation of intrinsic luminosities of the primary continuum (PC) and soft excess (SE) of 0.5-10.0 keV obtained from {\tt powerlaw+powerlaw} fitting for different sources. Each data point is normalised by the Eddington luminosity of the source. The blue line shows the linear correlation fit between the intrinsic luminosities of the primary continuum (PC) and the soft excess (SE).}
		\label{fig:figure_3}
	\end{figure}

	As NGC~7469 is a popular source in the X-ray domain, it is observed in almost all X-ray missions. For this work, we considered only the {\it XMM-Newton} and {\it Swift}/XRT observations. The details of the observations are given in Table~\ref{table:obs_log} and the data extraction procedure is described in Section~\ref{sec:data reduction}. We used the composite model as,
	\begin{center}
		{\tt TBabs*zTBabs*(powerlaw+powerlaw+zGauss)},
	\end{center}
	to fit the X-ray spectra of this source and the fitting procedure are described in Section~\ref{sec:data reduction}. The component {\it zGauss} was used to fit the Fe K$_\alpha$ line at $6.40\pm0.22$ keV with a width of $164\pm74$ eV. The {\tt powerlaw+powerlaw} fitted results are represented in Table~\ref{tab:all_obs}. From the spectral fitting, we found that the power-law indices for the primary continuum $(\Gamma^{\rm PC})$ and the soft-excess $(\Gamma^{\rm SE})$ vary from 1.2 to 2.0 and 1.1 to 4.3, respectively and the corresponding luminosities for the primary continuum $(\log(L^{\rm PC}))$ and soft-excess $(\log(L^{\rm SE}))$ vary from 43.0 to 44.0 and 42.5 to 43.1, respectively. The variation of the normalised intrinsic luminosities are plotted ($L_{\rm PC}/L_{\rm Edd}$ vs $L_{\rm SE}/L_{\rm Edd}$) in Figure~\ref{fig:figure_3} and the corresponding correlation coefficients, calculated from the different algorithms, are shown in Table~\ref{table:correlation}.
	
	\textbf{Ton~S180} or Tonantzinta~S180 is a local (z=0.062) narrow line Seyfert~1 AGN which is considered one of the prototypes `bare' AGN with no trace of absorption and a featureless and prominent soft-excess \citep{Vaughan2002}. From {\it Suzaku} observations, \citet{Takahashi2010} also suggested the presence of an intriguing hard excess in the 15--55 keV range for this source. The X-ray spectrum of this source contains a strong soft-excess \citep{Vaughan2002}, a broad Fe K emission line and hard X-ray emission up to 30 keV \citep{Nardini2012}. From the {\it XMM-Newton} observation, \citet{Parker2018} suggested that the X-ray spectrum favoured two Comptonization components with a reflection component from a disc around a black hole of low spin. \citet{Turner2002} estimated the mass of the central black hole as $2\times10^7$ M$_\odot$. 
	
	Ton~S180 have been observed with {\it XMM-Newton} and {\it Swift}/XRT in multiple times (Table~\ref{table:obs_log}). We used the composite model as,
	\begin{center}
		{\tt TBabs*zTBabs*(powerlaw+powerlaw+zGauss)}
	\end{center}
	to fit the X-ray spectra of this source. The {\tt zGauss} component was used for the Fe-line at $6.43\pm0.27$ keV with a width of $289\pm57$ eV. The detailed spectral analysis is discussed in Section~\ref{sec:RandD} and the {\tt powerlaw+powerlaw} fitting results are presented in Table~\ref{tab:all_obs}. We found that the variation of power-law indices for primary continuum ($\Gamma^{\rm PC}$) and soft-excess ($\Gamma^{\rm SE}$) are in the ranges of 1.0 to 2.1 and 1.68 to 4.1, respectively and the corresponding luminosity variation are in the ranges of 43.6 to 43.9 for primary continuum ($\log(L^{\rm PC})$) and 43.5 to 43.9 for soft-excess ($\log(L^{\rm SE})$). The correlation between the luminosities ($\log(L^{\rm PC})$ and $\log(L^{\rm SE})$) is plotted in Figure~\ref{fig:figure_3} and the corresponding correlation coefficients from different algorithms are presented in Table~\ref{table:correlation}.
	
	\textbf{UGC~6728} is a local (z=0.067) Seyfert~1.2 AGN which is yet to be explored. This source has an ultra-hard X-ray tail (14-195 keV) explored by \citep{Bernete2019}. We adopted the black hole mass of the central region as $7.1\times10^5$ M$_\odot$ \citep{Bentz2016}.
	
	\textit{XMM-Newton} observed UGC~6758 once in 2006 and \textit{Swift}/XRT observed the AGN multiple times from 2016 to 2021. The observation details are given in Table~\ref{table:obs_log}  and the data reduction procedure is discussed in Section~\ref{sec:RandD}. To fit the X-ray spectral data, we used
	\begin{center}
		{\tt TBabs*zTBabs*(powerlaw+powerlaw)}
	\end{center}
	as a composite model and the results are presented in Table~\ref{tab:all_obs}. The power-law index for the primary continuum ($\Gamma^{\rm PC}$) varies from 1.0 to 1.7 and the corresponding luminosity ($\log(L^{\rm PC})$) varies from 43.7 to 44.1. Similarly, the power-law index for soft-excess ($\Gamma^{\rm SE}$) varies from 2.8 to 4.6 and the corresponding luminosity ($\log(L^{\rm SE})$) varies from 42.6 to 44.0. The correlation between these two luminosities is shown in Figure~\ref{fig:figure_3}. Here we used normalised luminosity using the Eddington luminosity of the source. We also calculated the correlation coefficients using various methods and presented them in Table~\ref{table:correlation}.

 	\begin{center}
		\begin{longtable}{c c c c c c c c c}
			\caption{X-ray (0.5 to 10.0 keV) spectral fit parameters for all sources.} 
            \label{tab:all_obs} \\
			
			\hline \multicolumn{9}{c}{\textbf{1H 0323+342}} \\ \hline 
			\endfirsthead
			
			\multicolumn{9}{c}%
			{{ \tablename\ \thetable{} : X-ray (0.5 to 10.0 keV) spectral fit parameters for all sources.}} \\
			\hline
			Instrument       & MJD & $N_{\rm H}$                & $\Gamma^{\rm PC}$& $Norm^{PC}$ & $\log(L^{\rm PC})$  & $\Gamma^{\rm SE}$& $Norm^{SE}$ & $\log(L^{\rm SE})$ \\
			&     & $(10^{20}~cm^{-2})$  &              & $(10^{-3})$ &     (erg/s)     &              & $(10^{-3})$ & (erg/s)  \\
			\hline
			\endhead
			
			\hline \multicolumn{9}{r}{{\it Continued on next page}} \\
			\endfoot
			
			\hline \hline
			\endlastfoot
			
				Instrument       & MJD & $N_{\rm H}$                & $\Gamma^{\rm PC}$& $Norm^{PC}$ & $\log(L^{\rm PC})$  & $\Gamma^{\rm SE}$& $Norm^{SE}$ & $\log(L^{\rm SE})$ \\
&     & $(10^{20}~cm^{-2})$  &              & $(10^{-3})$ &     (erg/s)     &              & $(10^{-3})$ & (erg/s)  \\
\hline
{\it XMM-Newton} &57257& $1.35\pm0.50$        & $1.73\pm0.02$& $2.01\pm0.41$& $44.07\pm0.003$& $4.82\pm0.02$& $2.98\pm0.22$& $44.13\pm0.002$ \\
{\it XMM-Newton} &58344& $1.81\pm0.58$        & $1.83\pm0.02$& $2.04\pm0.42$& $44.03\pm0.002$& $5.26\pm0.03$& $3.14\pm0.25$& $44.24\pm0.003$ \\
{\it XMM-Newton} &58348& $0.96\pm0.50$        & $1.77\pm0.03$& $1.35\pm0.46$& $43.87\pm0.003$& $4.40\pm0.03$& $2.03\pm0.24$& $43.90\pm0.002$ \\
{\it XMM-Newton} &58350& $1.58\pm0.56$        & $1.81\pm0.02$& $3.12\pm0.47$& $44.22\pm0.002$& $5.10\pm0.04$& $3.96\pm0.19$& $44.31\pm0.002$ \\
{\it XMM-Newton} &58354& $1.34\pm0.53$        & $1.81\pm0.03$& $2.82\pm0.49$& $43.99\pm0.003$& $4.95\pm0.03$& $2.97\pm0.20$& $44.11\pm0.003$ \\
{\it XMM-Newton} &58366& $1.56\pm0.58$        & $1.79\pm0.04$& $1.79\pm0.49$& $43.97\pm0.002$& $5.04\pm0.04$& $2.69\pm0.22$& $44.13\pm0.002$ \\
{\it XMM-Newton} &58370& $1.57\pm0.50$        & $1.76\pm0.05$& $1.14\pm0.47$& $43.81\pm0.003$& $5.00\pm0.03$& $1.88\pm0.25$& $43.96\pm0.002$ \\
{\it Swift}/XRT  &54381& $1.03\pm0.55$        & $1.88\pm0.05$& $2.81\pm0.51$& $43.74\pm0.006$& $3.95\pm0.07$& $2.62\pm0.35$& $43.65\pm0.010$ \\
{\it Swift}/XRT  &54627& $0.83\pm0.55$        & $1.14\pm0.04$& $6.12\pm1.58$& $43.89\pm0.021$& $3.19\pm0.05$& $1.40\pm0.55$& $43.63\pm0.021$ \\
{\it Swift}/XRT  &55043& $1.34\pm0.56$        & $1.63\pm0.04$& $2.48\pm0.55$& $44.19\pm0.008$& $4.18\pm0.07$& $3.57\pm0.75$& $44.44\pm0.010$ \\
{\it Swift}/XRT  &55513& $0.89\pm0.55$        & $1.62\pm0.05$& $2.54\pm0.42$& $44.21\pm0.003$& $3.78\pm0.05$& $2.07\pm0.35$& $43.83\pm0.006$ \\
{\it Swift}/XRT  &55835& $0.64\pm0.58$        & $1.84\pm0.05$& $3.52\pm0.62$& $44.06\pm0.006$& $3.74\pm0.04$& $0.98\pm0.45$& $43.80\pm0.024$ \\
{\it Swift}/XRT  &55958& $1.26\pm0.54$        & $1.03\pm0.09$& $0.69\pm0.55$& $43.94\pm0.023$& $3.62\pm0.04$& $2.63\pm0.54$& $43.92\pm0.019$ \\
{\it Swift}/XRT  &56436& $1.44\pm0.59$        & $1.61\pm0.06$& $2.28\pm0.45$& $44.16\pm0.005$& $4.06\pm0.02$& $3.78\pm0.14$& $44.12\pm0.006$ \\
{\it Swift}/XRT  &57002& $1.02\pm0.50$        & $1.60\pm0.05$& $2.27\pm0.35$& $44.17\pm0.015$& $3.63\pm0.04$& $2.74\pm0.28$& $43.94\pm0.021$ \\
{\it Swift}/XRT  &57309& $1.40\pm0.05$        & $1.75\pm0.08$& $2.86\pm0.45$& $44.21\pm0.005$& $5.26\pm0.02$& $3.24\pm0.21$& $44.25\pm0.007$ \\
{\it Swift}/XRT  &58384& $0.79\pm0.51$        & $1.82\pm0.05$& $1.71\pm0.28$& $43.96\pm0.008$& $4.09\pm0.03$& $1.32\pm0.44$& $43.67\pm0.013$ \\
{\it Swift}/XRT  &58798& $0.39\pm0.58$        & $1.28\pm0.04$& $1.42\pm0.36$& $44.12\pm0.013$& $3.33\pm0.04$& $2.57\pm0.71$& $43.90\pm0.014$ \\
\hline \multicolumn{9}{c}{\textbf{1H 0419-577}} \\ \hline 
Instrument       & MJD & $N_{\rm H}$                & $\Gamma^{\rm PC}$& $Norm^{PC}$ & $\log(L^{\rm PC})$  & $\Gamma^{\rm SE}$& $Norm^{SE}$ & $\log(L^{\rm SE})$ \\
&     & $(10^{20}~cm^{-2})$  &              & $(10^{-3})$ &     (erg/s)     &              & $(10^{-3})$ & (erg/s)  \\
\hline
{\it XMM-Newton} &52542& $0.83\pm0.51$        & $1.49\pm0.02$& $1.58\pm0.56$& $44.54\pm0.003$& $6.71\pm0.03$& $0.48\pm0.15$& $44.09\pm0.003$ \\
{\it XMM-Newton} &52635& $0.55\pm0.59$        & $1.46\pm0.03$& $1.32\pm0.42$& $44.47\pm0.005$& $4.67\pm0.02$& $0.51\pm0.20$& $44.86\pm0.005$ \\
{\it XMM-Newton} &52728& $0.28\pm0.55$        & $1.62\pm0.02$& $2.52\pm0.63$& $44.68\pm0.006$& $3.86\pm0.03$& $1.21\pm0.25$& $44.15\pm0.005$ \\
{\it XMM-Newton} &52815& $0.26\pm0.55$        & $1.52\pm0.03$& $1.96\pm0.45$& $44.61\pm0.002$& $3.68\pm0.03$& $1.15\pm0.32$& $44.07\pm0.003$ \\
{\it XMM-Newton} &52898& $0.41\pm0.57$        & $1.42\pm0.04$& $1.79\pm0.48$& $44.62\pm0.004$& $4.55\pm0.05$& $0.79\pm0.21$& $44.03\pm0.004$ \\
{\it XMM-Newton} &53518& $0.27\pm0.59$        & $1.60\pm0.03$& $2.65\pm0.59$& $44.72\pm0.001$& $3.47\pm0.02$& $2.24\pm0.12$& $44.35\pm0.001$ \\
{\it XMM-Newton} &53520& $0.21\pm0.54$        & $1.64\pm0.01$& $2.90\pm0.42$& $44.73\pm0.001$& $3.77\pm0.02$& $1.51\pm0.33$& $44.20\pm0.001$ \\
{\it XMM-Newton} &58254& $0.18\pm0.55$        & $1.71\pm0.03$& $2.62\pm0.54$& $44.66\pm0.002$& $3.81\pm0.03$& $1.25\pm0.36$& $44.13\pm0.003$ \\
{\it XMM-Newton} &58435& $0.28\pm0.59$        & $1.59\pm0.02$& $2.66\pm0.68$& $44.71\pm0.001$& $3.68\pm0.02$& $1.84\pm0.54$& $44.28\pm0.002$ \\
{\it Swift}/XRT  &54771& $0.62\pm0.54$        & $1.95\pm0.04$& $4.95\pm0.87$& $44.84\pm0.005$& $4.77\pm0.05$& $0.81\pm0.56$& $44.26\pm0.013$ \\
{\it Swift}/XRT  &56460& $0.22\pm0.59$        & $1.11\pm0.10$& $1.06\pm0.53$& $44.60\pm0.038$& $2.53\pm0.08$& $3.33\pm0.48$& $44.25\pm0.020$ \\
{\it Swift}/XRT  &57179& $0.89\pm0.55$        & $1.55\pm0.09$& $3.03\pm0.79$& $44.79\pm0.013$& $3.94\pm0.04$& $2.68\pm0.42$& $44.47\pm0.016$ \\
{\it Swift}/XRT  &57902& $0.14\pm0.51$        & $2.26\pm0.05$& $7.84\pm1.15$& $44.97\pm0.010$& $6.56\pm0.05$& $0.67\pm0.18$& $44.41\pm0.023$ \\
\hline \multicolumn{9}{c}{\textbf{1H 0707-495}} \\ \hline 
Instrument       & MJD & $N_{\rm H}$                & $\Gamma^{\rm PC}$& $Norm^{PC}$ & $\log(L^{\rm PC})$  & $\Gamma^{\rm SE}$& $Norm^{SE}$ & $\log(L^{\rm SE})$ \\
&     & $(10^{20}~cm^{-2})$  &              & $(10^{-4})$ &     (erg/s)     &              & $(10^{-4})$ & (erg/s)  \\
\hline
{\it XMM-Newton} &52560& $1.12\pm0.54$        & $2.54\pm0.03$& $40.6\pm6.54$& $43.03\pm0.002$& $6.39\pm0.06$& $1.12\pm0.54$& $43.69\pm0.001$ \\
{\it XMM-Newton} &54234& $1.77\pm0.58$        & $1.82\pm0.05$& $1.56\pm0.11$& $42.67\pm0.006$& $4.61\pm0.05$& $3.94\pm0.24$& $42.80\pm0.002$ \\
{\it XMM-Newton} &54236& $0.91\pm0.53$        & $1.51\pm0.04$& $0.75\pm0.21$& $42.28\pm0.001$& $6.64\pm0.08$& $0.67\pm0.22$& $42.36\pm0.005$ \\
{\it XMM-Newton} &54494& $0.78\pm0.57$        & $2.10\pm0.03$& $5.43\pm0.58$& $42.97\pm0.007$& $6.53\pm0.04$& $1.91\pm0.53$& $43.85\pm0.002$ \\
{\it XMM-Newton} &54496& $0.68\pm0.51$        & $2.24\pm0.05$& $8.56\pm0.29$& $43.13\pm0.007$& $6.73\pm0.05$& $3.81\pm0.69$& $44.20\pm0.002$ \\
{\it XMM-Newton} &54498& $0.76\pm0.58$        & $1.93\pm0.04$& $4.72\pm0.37$& $42.97\pm0.010$& $6.24\pm0.07$& $3.71\pm0.85$& $44.08\pm0.002$ \\
{\it XMM-Newton} &54500& $0.89\pm0.55$        & $1.64\pm0.02$& $2.93\pm0.65$& $42.88\pm0.013$& $6.30\pm0.06$& $2.81\pm0.54$& $43.97\pm0.003$ \\
{\it XMM-Newton} &55091& $1.95\pm0.58$        & $1.91\pm0.05$& $2.77\pm0.37$& $42.74\pm0.004$& $6.81\pm0.05$& $2.70\pm0.62$& $44.07\pm0.001$ \\
{\it XMM-Newton} &55458& $0.72\pm0.54$        & $1.72\pm0.04$& $2.85\pm0.42$& $42.85\pm0.020$& $7.00\pm0.04$& $4.41\pm0.65$& $44.30\pm0.002$ \\
{\it Swift}/XRT  &55425& $1.05\pm0.56$        & $2.47\pm0.92$& $5.76\pm0.59$& $42.85\pm0.020$& $3.58\pm0.15$& $2.42\pm0.59$& $43.46\pm0.016$ \\
{\it Swift}/XRT  &55607& $1.52\pm0.54$        & $1.23\pm0.11$& $0.46\pm0.68$& $42.51\pm0.032$& $7.05\pm0.05$& $0.79\pm0.10$& $43.29\pm0.010$ \\
{\it Swift}/XRT  &58178& $1.53\pm0.52$        & $1.10\pm0.14$& $1.15\pm0.41$& $42.75\pm0.021$& $4.98\pm0.04$& $4.21\pm0.54$& $43.89\pm0.004$ \\
\hline \multicolumn{9}{c}{\textbf{3C 382}} \\ \hline 
Instrument       & MJD & $N_{\rm H}$                & $\Gamma^{\rm PC}$& $Norm^{PC}$ & $\log(L^{\rm PC})$  & $\Gamma^{\rm SE}$& $Norm^{SE}$ & $\log(L^{\rm SE})$ \\
&     & $(10^{20}~cm^{-2})$  &              & $(10^{-3})$ &     (erg/s)     &              & $(10^{-3})$ & (erg/s)  \\
\hline
{\it XMM-Newton} &54584& $3.10\pm0.55$        & $1.75\pm0.01$& $10.1\pm2.62$& $44.68\pm0.001$& $3.55\pm0.02$& $5.07\pm1.57$& $44.13\pm0.001$ \\
{\it XMM-Newton} &57629& $3.71\pm0.52$        & $1.67\pm0.01$& $7.68\pm1.51$& $44.61\pm0.001$& $3.62\pm0.01$& $2.11\pm0.18$& $43.75\pm0.002$ \\
{\it XMM-Newton} &57642& $3.33\pm0.54$        & $1.63\pm0.02$& $8.38\pm2.09$& $44.63\pm0.001$& $3.35\pm0.02$& $2.94\pm0.20$& $43.88\pm0.002$ \\
{\it XMM-Newton} &57653& $3.26\pm0.55$        & $1.64\pm0.02$& $8.35\pm1.52$& $44.65\pm0.001$& $3.32\pm0.01$& $3.22\pm0.52$& $43.92\pm0.002$ \\
{\it XMM-Newton} &57653& $3.93\pm0.55$        & $1.67\pm0.01$& $8.28\pm1.59$& $44.63\pm0.001$& $3.51\pm0.01$& $2.96\pm0.26$& $43.89\pm0.002$ \\
{\it XMM-Newton} &57678& $3.97\pm0.57$        & $1.66\pm0.01$& $9.40\pm1.72$& $44.69\pm0.001$& $3.49\pm0.01$& $3.63\pm0.47$& $43.98\pm0.002$ \\
{\it Swift}/XRT  &56644& $1.77\pm0.52$        & $1.28\pm0.15$& $2.18\pm0.78$& $44.64\pm0.028$& $1.91\pm0.06$& $3.90\pm0.45$& $44.01\pm0.022$ \\
{\it Swift}/XRT  &57660& $0.99\pm0.55$        & $1.64\pm0.08$& $8.62\pm0.85$& $44.66\pm0.011$& $4.56\pm0.05$& $3.57\pm0.49$& $44.08\pm0.028$ \\
{\it Swift}/XRT  &59370& $0.93\pm0.58$        & $1.68\pm0.07$& $1.20\pm0.24$& $43.52\pm0.072$& $1.54\pm0.07$& $4.57\pm0.58$& $44.43\pm0.014$ \\
\hline \multicolumn{9}{c}{\textbf{3C 390.3}} \\ \hline 
Instrument       & MJD & $N_{\rm H}$                & $\Gamma^{\rm PC}$& $Norm^{PC}$ & $\log(L^{\rm PC})$  & $\Gamma^{\rm SE}$& $Norm^{SE}$ & $\log(L^{\rm SE})$ \\
&     & $(10^{20}~cm^{-2})$  &              & $(10^{-3})$ &     (erg/s)     &              & $(10^{-3})$ & (erg/s)  \\
\hline
{\it XMM-Newton} &53286& $0.56\pm0.55$        & $1.65\pm0.01$& $8.64\pm2.66$& $44.59\pm0.001$& $3.37\pm0.01$& $2.45\pm1.59$& $43.83\pm0.002$ \\
{\it XMM-Newton} &53295& $0.92\pm0.56$        & $1.64\pm0.01$& $7.41\pm2.72$& $44.63\pm0.001$& $3.95\pm0.01$& $3.82\pm1.43$& $44.07\pm0.001$ \\
{\it Swift}/XRT  &54630& $0.77\pm0.58$        & $1.66\pm0.09$& $12.6\pm2.58$& $44.84\pm0.004$& $3.70\pm0.02$& $3.45\pm1.48$& $43.56\pm0.012$ \\
{\it Swift}/XRT  &56437& $1.59\pm0.52$        & $1.01\pm0.14$& $8.43\pm2.56$& $44.05\pm0.006$& $1.86\pm0.03$& $7.84\pm1.52$& $43.56\pm0.012$ \\
{\it Swift}/XRT  &59321& $1.02\pm0.55$        & $1.43\pm0.07$& $5.09\pm2.42$& $44.76\pm0.020$& $3.20\pm0.04$& $4.94\pm1.74$& $44.14\pm0.031$ \\
\hline \multicolumn{9}{c}{\textbf{Ark 564}} \\ \hline 
Instrument       & MJD & $N_{\rm H}$                & $\Gamma^{\rm PC}$& $Norm^{PC}$ & $\log(L^{\rm PC})$  & $\Gamma^{\rm SE}$& $Norm^{SE}$ & $\log(L^{\rm SE})$ \\
&     & $(10^{20}~cm^{-2})$  &              & $(10^{-3})$ &     (erg/s)     &              & $(10^{-3})$ & (erg/s)  \\
\hline
{\it XMM-Newton} &51712& $2.07\pm0.54$        & $2.39\pm0.02$& $16.2\pm3.54$& $43.91\pm0.001$& $4.65\pm0.02$& $7.77\pm2.56$& $43.61\pm0.001$ \\
{\it XMM-Newton} &52069& $2.80\pm0.57$        & $2.35\pm0.02$& $8.31\pm3.52$& $43.62\pm0.002$& $4.27\pm0.02$& $4.68\pm1.42$& $43.35\pm0.002$ \\
{\it XMM-Newton} &53375& $2.93\pm0.61$        & $2.37\pm0.03$& $15.8\pm3.68$& $43.90\pm0.001$& $4.10\pm0.03$& $7.20\pm2.85$& $43.52\pm0.002$ \\
{\it XMM-Newton} &53705& $2.72\pm0.59$        & $2.51\pm0.01$& $19.5\pm3.47$& $43.96\pm0.001$& $4.05\pm0.01$& $6.58\pm1.15$& $43.47\pm0.001$ \\
{\it XMM-Newton} &53711& $2.12\pm0.58$        & $2.43\pm0.01$& $13.2\pm3.48$& $43.74\pm0.001$& $3.91\pm0.01$& $4.43\pm1.47$& $43.29\pm0.001$ \\
{\it XMM-Newton} &53717& $2.27\pm0.51$        & $2.34\pm0.01$& $11.8\pm2.56$& $43.78\pm0.001$& $3.89\pm0.01$& $5.06\pm1.97$& $43.34\pm0.001$ \\
{\it XMM-Newton} &53723& $2.05\pm0.52$        & $2.42\pm0.01$& $14.5\pm2.51$& $43.85\pm0.001$& $3.90\pm0.01$& $5.07\pm1.97$& $43.37\pm0.001$ \\
{\it XMM-Newton} &53729& $2.78\pm0.54$        & $2.37\pm0.01$& $12.0\pm2.56$& $43.78\pm0.001$& $3.99\pm0.01$& $5.93\pm1.42$& $43.42\pm0.001$ \\
{\it XMM-Newton} &53737& $2.00\pm0.53$        & $2.31\pm0.01$& $7.48\pm1.94$& $43.59\pm0.001$& $3.79\pm0.01$& $3.88\pm1.44$& $43.22\pm0.001$ \\
{\it XMM-Newton} &53741& $2.01\pm0.54$        & $2.37\pm0.01$& $11.9\pm2.91$& $43.77\pm0.001$& $3.92\pm0.01$& $4.71\pm1.47$& $43.32\pm0.001$ \\
{\it XMM-Newton} &53743& $1.75\pm0.48$        & $2.58\pm0.02$& $16.8\pm2.45$& $43.89\pm0.001$& $3.96\pm0.02$& $3.35\pm1.12$& $43.27\pm0.003$ \\
{\it XMM-Newton} &53753& $3.98\pm0.59$        & $2.36\pm0.02$& $8.70\pm1.12$& $43.64\pm0.001$& $4.26\pm0.02$& $4.72\pm1.18$& $43.35\pm0.001$ \\
{\it XMM-Newton} &53755& $3.26\pm0.51$        & $2.47\pm0.02$& $10.8\pm1.28$& $43.71\pm0.002$& $4.25\pm0.02$& $3.93\pm1.17$& $43.27\pm0.001$ \\
{\it Swift}/XRT  &53596& $0.96\pm0.54$        & $2.42\pm0.03$& $12.3\pm1.54$& $43.79\pm0.006$& $5.53\pm0.03$& $10.1\pm2.18$& $43.96\pm0.005$ \\
{\it Swift}/XRT  &56784& $1.12\pm0.58$        & $1.10\pm0.13$& $11.1\pm2.42$& $43.27\pm0.005$& $3.91\pm0.04$& $16.6\pm3.29$& $42.96\pm0.014$ \\
{\it Swift}/XRT  &57164& $1.55\pm0.57$        & $1.06\pm0.08$& $10.8\pm2.43$& $43.38\pm0.004$& $4.10\pm0.04$& $16.0\pm3.84$& $42.97\pm0.012$ \\
{\it Swift}/XRT  &57631& $1.15\pm0.56$        & $2.31\pm0.05$& $11.6\pm2.51$& $43.77\pm0.008$& $4.67\pm0.03$& $13.2\pm3.52$& $43.05\pm0.007$ \\
{\it Swift}/XRT  &57991& $0.65\pm0.58$        & $2.10\pm0.05$& $7.75\pm2.05$& $43.65\pm0.008$& $4.04\pm0.03$& $13.3\pm3.55$& $42.87\pm0.005$ \\
{\it Swift}/XRT  &58360& $0.99\pm0.51$        & $2.00\pm0.08$& $5.76\pm2.41$& $43.57\pm0.010$& $4.28\pm0.04$& $12.6\pm3.47$& $43.08\pm0.006$ \\
{\it Swift}/XRT  &58725& $0.66\pm0.57$        & $2.54\pm0.09$& $13.0\pm2.71$& $43.77\pm0.006$& $4.49\pm0.04$& $3.91\pm1.20$& $43.00\pm0.012$ \\
{\it Swift}/XRT  &59091& $1.42\pm0.51$        & $2.26\pm0.08$& $9.71\pm2.56$& $43.71\pm0.009$& $4.69\pm0.04$& $15.0\pm4.52$& $43.01\pm0.007$ \\
{\it Swift}/XRT  &59330& $1.44\pm0.55$        & $1.88\pm0.07$& $3.41\pm1.50$& $43.28\pm0.017$& $4.69\pm0.05$& $10.4\pm4.68$& $43.05\pm0.009$ \\
\hline \multicolumn{9}{c}{\textbf{Fairall 9}} \\ \hline 
Instrument       & MJD & $N_{\rm H}$                & $\Gamma^{\rm PC}$& $Norm^{PC}$ & $\log(L^{\rm PC})$  & $\Gamma^{\rm SE}$& $Norm^{SE}$ & $\log(L^{\rm SE})$ \\
&     & $(10^{20}~cm^{-2})$  &              & $(10^{-3})$ &     (erg/s)     &              & $(10^{-3})$ & (erg/s)  \\
\hline
{\it XMM-Newton} &51730& $0.50\pm0.52$        & $1.68\pm0.02$& $2.55\pm0.72$& $43.95\pm0.001$& $2.98\pm0.01$& $1.26\pm0.58$& $43.34\pm0.002$ \\
{\it XMM-Newton} &55174& $0.12\pm0.55$        & $1.68\pm0.01$& $3.24\pm0.77$& $44.05\pm0.001$& $3.26\pm0.01$& $1.56\pm0.57$& $43.43\pm0.001$ \\
{\it XMM-Newton} &56645& $0.11\pm0.57$        & $1.70\pm0.02$& $3.33\pm0.71$& $44.06\pm0.002$& $3.04\pm0.02$& $2.26\pm0.52$& $43.60\pm0.002$ \\
{\it XMM-Newton} &56659& $0.13\pm0.57$        & $1.82\pm0.01$& $6.00\pm0.72$& $44.26\pm0.001$& $2.84\pm0.02$& $9.43\pm0.51$& $43.23\pm0.001$ \\
{\it XMM-Newton} &56786& $0.14\pm0.56$        & $1.80\pm0.03$& $5.44\pm0.70$& $44.23\pm0.002$& $2.95\pm0.03$& $4.32\pm0.50$& $43.88\pm0.002$ \\
{\it Swift}/XRT  &54719& $0.41\pm0.57$        & $1.70\pm0.05$& $4.70\pm0.75$& $44.21\pm0.012$& $2.69\pm0.02$& $6.90\pm0.51$& $44.11\pm0.001$ \\
{\it Swift}/XRT  &56527& $0.12\pm0.59$        & $1.87\pm0.03$& $8.39\pm0.72$& $44.19\pm0.002$& $4.14\pm0.02$& $8.27\pm0.58$& $43.52\pm0.010$ \\
{\it Swift}/XRT  &56804& $0.11\pm0.55$        & $1.81\pm0.03$& $4.64\pm0.71$& $44.16\pm0.004$& $2.42\pm0.02$& $3.45\pm0.54$& $43.85\pm0.006$ \\
{\it Swift}/XRT  &57098& $0.19\pm0.56$        & $1.28\pm0.08$& $1.73\pm0.74$& $43.98\pm0.009$& $3.04\pm0.03$& $4.07\pm0.55$& $43.85\pm0.006$ \\
{\it Swift}/XRT  &58303& $0.15\pm0.55$        & $1.68\pm0.09$& $3.66\pm0.77$& $44.10\pm0.004$& $2.42\pm0.04$& $4.26\pm0.56$& $43.94\pm0.004$ \\
{\it Swift}/XRT  &58420& $0.14\pm0.57$        & $1.85\pm0.08$& $5.44\pm0.71$& $44.11\pm0.002$& $2.92\pm0.04$& $0.99\pm0.24$& $43.44\pm0.011$ \\
{\it Swift}/XRT  &58553& $0.13\pm0.54$        & $1.86\pm0.04$& $4.86\pm0.72$& $44.15\pm0.003$& $2.50\pm0.05$& $4.66\pm0.49$& $43.96\pm0.004$ \\
{\it Swift}/XRT  &58678& $0.21\pm0.55$        & $1.87\pm0.03$& $5.81\pm0.70$& $44.23\pm0.003$& $2.87\pm0.06$& $3.59\pm0.52$& $43.81\pm0.005$ \\
{\it Swift}/XRT  &58791& $0.25\pm0.55$        & $1.95\pm0.03$& $7.65\pm0.78$& $44.22\pm0.002$& $3.59\pm0.05$& $0.85\pm0.51$& $43.97\pm0.011$ \\
{\it Swift}/XRT  &58030& $0.20\pm0.56$        & $1.96\pm0.04$& $8.10\pm0.71$& $44.14\pm0.003$& $3.46\pm0.02$& $2.24\pm0.50$& $43.69\pm0.006$ \\
\hline \multicolumn{9}{c}{\textbf{IRAS 13224-3809}}\\ \hline 
Instrument       & MJD & $N_{\rm H}$                & $\Gamma^{\rm PC}$& $Norm^{PC}$ & $\log(L^{\rm PC})$  & $\Gamma^{\rm SE}$& $Norm^{SE}$ & $\log(L^{\rm SE})$ \\
&     & $(10^{20}~cm^{-2})$  &              & $(10^{-3})$ &     (erg/s)     &              & $(10^{-3})$ & (erg/s)  \\
\hline
{\it XMM-Newton} &55761& $2.20\pm0.56$        & $2.22\pm0.03$& $0.35\pm0.15$& $43.19\pm0.004$& $5.96\pm0.04$& $0.98\pm0.18$& $43.96\pm0.001$ \\
{\it XMM-Newton} &55767& $1.77\pm0.58$        & $1.44\pm0.05$& $0.51\pm0.14$& $43.15\pm0.008$& $6.00\pm0.05$& $0.39\pm0.12$& $43.93\pm0.002$ \\
{\it XMM-Newton} &55771& $2.47\pm0.51$        & $2.77\pm0.03$& $1.03\pm0.35$& $43.16\pm0.003$& $6.05\pm0.02$& $2.38\pm0.50$& $43.93\pm0.001$ \\
{\it XMM-Newton} &57579& $3.15\pm0.54$        & $2.14\pm0.03$& $0.27\pm0.09$& $43.09\pm0.003$& $6.64\pm0.02$& $1.04\pm0.32$& $44.10\pm0.002$ \\
{\it XMM-Newton} &57581& $2.67\pm0.55$        & $1.69\pm0.04$& $0.17\pm0.10$& $43.04\pm0.004$& $5.52\pm0.03$& $1.31\pm0.35$& $43.96\pm0.001$ \\
{\it XMM-Newton} &57591& $3.77\pm0.52$        & $1.68\pm0.06$& $0.20\pm0.09$& $43.12\pm0.021$& $5.14\pm0.05$& $1.94\pm0.47$& $43.06\pm0.003$ \\
{\it XMM-Newton} &57593& $3.85\pm0.57$        & $1.82\pm0.03$& $0.14\pm0.07$& $42.89\pm0.006$& $6.67\pm0.03$& $0.62\pm0.14$& $43.88\pm0.002$ \\
{\it XMM-Newton} &57595& $2.41\pm0.50$        & $1.39\pm0.02$& $0.08\pm0.01$& $42.83\pm0.005$& $5.72\pm0.02$& $0.87\pm0.23$& $43.82\pm0.001$ \\
{\it XMM-Newton} &57599& $1.85\pm0.47$        & $1.87\pm0.04$& $0.15\pm0.07$& $42.91\pm0.008$& $6.12\pm0.03$& $0.86\pm0.28$& $43.90\pm0.002$ \\
{\it XMM-Newton} &57601& $1.69\pm0.41$        & $1.80\pm0.03$& $0.12\pm0.05$& $42.85\pm0.013$& $6.01\pm0.05$& $0.88\pm0.21$& $43.88\pm0.003$ \\
{\it XMM-Newton} &57603& $3.11\pm0.49$        & $2.80\pm0.02$& $0.59\pm0.09$& $43.45\pm0.004$& $5.76\pm0.04$& $2.17\pm0.56$& $43.22\pm0.001$ \\
{\it XMM-Newton} &57607& $1.89\pm0.52$        & $1.96\pm0.03$& $0.71\pm0.11$& $43.03\pm0.005$& $5.87\pm0.06$& $0.71\pm0.25$& $43.77\pm0.001$ \\
{\it XMM-Newton} &57609& $1.91\pm0.54$        & $2.37\pm0.01$& $0.72\pm0.18$& $43.46\pm0.003$& $5.38\pm0.02$& $1.72\pm0.14$& $44.04\pm0.001$ \\
{\it Swift}/XRT  &55425& $1.61\pm0.50$        & $1.77\pm0.03$& $0.12\pm0.05$& $42.89\pm0.020$& $5.89\pm0.03$& $0.53\pm0.10$& $43.77\pm0.010$ \\
{\it Swift}/XRT  &55607& $3.19\pm0.55$        & $1.12\pm0.09$& $0.06\pm0.01$& $42.71\pm0.037$& $4.50\pm0.04$& $0.51\pm0.14$& $43.75\pm0.031$ \\
{\it Swift}/XRT  &57597& $3.32\pm0.57$        & $1.73\pm0.08$& $0.16\pm0.08$& $43.03\pm0.020$& $4.92\pm0.03$& $1.02\pm0.35$& $43.87\pm0.035$ \\
\hline \multicolumn{9}{c}{\textbf{Mrk 1018}}\\ \hline 
Instrument       & MJD & $N_{\rm H}$                & $\Gamma^{\rm PC}$& $Norm^{PC}$ & $\log(L^{\rm PC})$  & $\Gamma^{\rm SE}$& $Norm^{SE}$ & $\log(L^{\rm SE})$ \\
&     & $(10^{20}~cm^{-2})$  &              & $(10^{-3})$ &     (erg/s)     &              & $(10^{-3})$ & (erg/s)  \\
\hline
{\it XMM-Newton} &53385& $0.81\pm0.54$        & $1.10\pm0.07$& $0.74\pm0.24$& $43.57\pm0.022$& $2.46\pm0.05$& $3.90\pm0.89$& $43.73\pm0.005$ \\
{\it XMM-Newton} &54685& $1.74\pm0.55$        & $1.54\pm0.02$& $2.22\pm0.89$& $43.79\pm0.002$& $3.00\pm0.04$& $2.24\pm0.87$& $43.42\pm0.020$ \\
{\it XMM-Newton} &58322& $2.63\pm0.58$        & $1.52\pm0.03$& $0.36\pm0.12$& $43.00\pm0.004$& $2.94\pm0.02$& $0.18\pm0.06$& $42.33\pm0.007$ \\
{\it XMM-Newton} &58487& $1.38\pm0.51$        & $1.46\pm0.02$& $0.13\pm0.04$& $42.57\pm0.005$& $2.42\pm0.02$& $0.13\pm0.06$& $42.25\pm0.005$ \\
{\it Swift}/XRT  &53587& $2.18\pm0.58$        & $1.13\pm0.09$& $1.33\pm0.14$& $43.76\pm0.016$& $2.82\pm0.03$& $2.98\pm0.05$& $43.56\pm0.015$ \\
{\it Swift}/XRT  &54776& $1.65\pm0.55$        & $1.11\pm0.17$& $0.76\pm0.15$& $43.57\pm0.025$& $2.19\pm0.04$& $1.91\pm0.04$& $43.48\pm0.015$ \\
{\it Swift}/XRT  &58285& $2.26\pm0.57$        & $1.15\pm0.09$& $0.17\pm0.04$& $42.89\pm0.013$& $2.71\pm0.05$& $0.23\pm0.03$& $42.45\pm0.016$ \\
\hline \multicolumn{9}{c}{\textbf{Mrk 110}}\\ \hline 
Instrument       & MJD & $N_{\rm H}$                & $\Gamma^{\rm PC}$& $Norm^{PC}$ & $\log(L^{\rm PC})$  & $\Gamma^{\rm SE}$& $Norm^{SE}$ & $\log(L^{\rm SE})$ \\
&     & $(10^{20}~cm^{-2})$  &              & $(10^{-3})$ &     (erg/s)     &              & $(10^{-3})$ & (erg/s)  \\
\hline
{\it XMM-Newton} &53320& $0.82\pm0.54$        & $1.70\pm0.01$& $7.03\pm1.29$& $44.11\pm0.001$& $3.91\pm0.01$& $3.35\pm0.54$& $43.52\pm0.001$ \\
{\it XMM-Newton} &58790& $0.70\pm0.53$        & $1.64\pm0.02$& $4.60\pm1.35$& $43.94\pm0.001$& $4.35\pm0.01$& $3.01\pm0.67$& $43.52\pm0.001$ \\
{\it XMM-Newton} &58792& $1.02\pm0.58$        & $1.71\pm0.01$& $6.53\pm1.72$& $44.07\pm0.001$& $4.02\pm0.01$& $3.57\pm0.84$& $43.55\pm0.003$ \\
{\it XMM-Newton} &58794& $1.05\pm0.59$        & $1.75\pm0.01$& $7.12\pm1.42$& $44.09\pm0.001$& $4.07\pm0.02$& $3.33\pm0.72$& $43.53\pm0.001$ \\
{\it XMM-Newton} &58804& $0.71\pm0.54$        & $1.70\pm0.01$& $6.36\pm1.81$& $44.06\pm0.001$& $4.01\pm0.02$& $3.61\pm0.49$& $43.56\pm0.001$ \\
{\it XMM-Newton} &58945& $0.69\pm0.58$        & $1.68\pm0.02$& $5.49\pm1.72$& $44.01\pm0.001$& $3.98\pm0.03$& $2.42\pm0.26$& $43.38\pm0.002$ \\
{\it Swift}/XRT  &55932& $0.76\pm0.57$        & $1.49\pm0.04$& $6.64\pm2.94$& $44.17\pm0.008$& $3.77\pm0.05$& $6.32\pm2.47$& $43.52\pm0.009$ \\
{\it Swift}/XRT  &56753& $0.31\pm0.56$        & $1.61\pm0.05$& $5.71\pm2.82$& $44.06\pm0.014$& $4.39\pm0.06$& $6.31\pm2.49$& $43.46\pm0.035$ \\
{\it Swift}/XRT  &57516& $0.34\pm0.54$        & $1.55\pm0.07$& $7.01\pm1.87$& $44.17\pm0.004$& $3.22\pm0.02$& $4.25\pm2.01$& $43.59\pm0.007$ \\
{\it Swift}/XRT  &57849& $0.31\pm0.50$        & $1.63\pm0.06$& $8.17\pm2.42$& $44.20\pm0.010$& $3.96\pm0.03$& $2.87\pm1.58$& $43.45\pm0.019$ \\
{\it Swift}/XRT  &58086& $0.36\pm0.55$        & $1.56\pm0.07$& $3.24\pm1.02$& $43.83\pm0.005$& $3.07\pm0.07$& $1.81\pm0.82$& $43.21\pm0.005$ \\
{\it Swift}/XRT  &58131& $0.25\pm0.50$        & $1.55\pm0.07$& $4.71\pm1.25$& $44.00\pm0.004$& $3.12\pm0.05$& $2.61\pm0.78$& $43.38\pm0.006$ \\
{\it Swift}/XRT  &58595& $1.18\pm0.58$        & $1.68\pm0.05$& $7.01\pm2.34$& $44.12\pm0.003$& $3.66\pm0.04$& $1.26\pm0.77$& $43.37\pm0.010$ \\
{\it Swift}/XRT  &58766& $2.76\pm0.51$        & $1.75\pm0.04$& $6.70\pm2.31$& $44.06\pm0.003$& $3.39\pm0.05$& $1.39\pm0.59$& $43.40\pm0.011$ \\
{\it Swift}/XRT  &59466& $1.14\pm0.52$        & $1.60\pm0.08$& $4.30\pm2.01$& $43.94\pm0.004$& $2.82\pm0.03$& $1.12\pm0.42$& $43.13\pm0.015$ \\
\hline \multicolumn{9}{c}{\textbf{Mrk 335}}\\ \hline
Instrument       & MJD & $N_{\rm H}$                & $\Gamma^{\rm PC}$& $Norm^{PC}$ & $\log(L^{\rm PC})$  & $\Gamma^{\rm SE}$& $Norm^{SE}$ & $\log(L^{\rm SE})$ \\
&     & $(10^{20}~cm^{-2})$  &              & $(10^{-3})$ &     (erg/s)     &              & $(10^{-3})$ & (erg/s)  \\
\hline
{\it XMM-Newton} &51903& $0.68\pm0.55$        & $1.97\pm0.02$& $4.94\pm1.01$& $43.55\pm0.003$& $4.27\pm0.03$& $6.74\pm1.05$& $43.55\pm0.002$ \\
{\it XMM-Newton} &53795& $0.12\pm0.52$        & $1.85\pm0.02$& $4.75\pm1.14$& $43.58\pm0.002$& $3.47\pm0.03$& $5.84\pm1.17$& $43.43\pm0.001$ \\
{\it XMM-Newton} &54291& $0.13\pm0.54$        & $1.17\pm0.02$& $0.37\pm0.09$& $42.81\pm0.005$& $5.21\pm0.04$& $0.53\pm0.08$& $43.23\pm0.003$ \\
{\it XMM-Newton} &54993& $0.50\pm0.58$        & $1.66\pm0.01$& $1.14\pm0.81$& $43.04\pm0.002$& $4.35\pm0.04$& $3.51\pm0.95$& $41.94\pm0.003$ \\
{\it XMM-Newton} &54995& $0.52\pm0.58$        & $1.62\pm0.03$& $1.16\pm0.75$& $43.06\pm0.002$& $4.85\pm0.05$& $0.26\pm0.09$& $42.67\pm0.002$ \\
{\it XMM-Newton} &57386& $0.31\pm0.50$        & $1.06\pm0.02$& $0.23\pm0.07$& $42.61\pm0.003$& $4.21\pm0.03$& $0.38\pm0.09$& $42.26\pm0.002$ \\
{\it XMM-Newton} &58429& $0.25\pm0.57$        & $1.16\pm0.02$& $0.06\pm0.02$& $42.17\pm0.005$& $5.81\pm0.04$& $0.38\pm0.07$& $42.54\pm0.003$ \\
{\it XMM-Newton} &58491& $0.62\pm0.55$        & $1.39\pm0.02$& $0.05\pm0.01$& $42.22\pm0.005$& $5.83\pm0.03$& $0.30\pm0.06$& $42.46\pm0.002$ \\
{\it XMM-Newton} &58844& $0.15\pm0.54$        & $1.52\pm0.04$& $0.03\pm0.01$& $42.08\pm0.006$& $5.43\pm0.03$& $0.21\pm0.05$& $42.22\pm0.002$ \\
{\it Swift}/XRT  &54349& $0.11\pm0.54$        & $1.49\pm0.03$& $1.31\pm0.19$& $42.19\pm0.005$& $5.14\pm0.04$& $5.71\pm0.42$& $43.61\pm0.003$ \\
{\it Swift}/XRT  &55148& $0.18\pm0.58$        & $1.46\pm0.04$& $3.33\pm0.25$& $43.23\pm0.003$& $6.03\pm0.04$& $0.91\pm0.05$& $42.97\pm0.004$ \\
{\it Swift}/XRT  &55198& $0.21\pm0.59$        & $1.11\pm0.09$& $0.75\pm0.26$& $43.15\pm0.004$& $4.28\pm0.06$& $0.73\pm0.06$& $42.59\pm0.005$ \\
{\it Swift}/XRT  &55937& $0.14\pm0.57$        & $1.64\pm0.04$& $1.12\pm0.38$& $43.04\pm0.004$& $6.04\pm0.05$& $0.31\pm0.07$& $42.16\pm0.005$ \\
{\it Swift}/XRT  &56477& $0.11\pm0.58$        & $1.41\pm0.03$& $0.69\pm0.29$& $42.94\pm0.006$& $6.02\pm0.04$& $0.41\pm0.08$& $42.63\pm0.009$ \\
{\it Swift}/XRT  &56841& $0.09\pm0.59$        & $1.24\pm0.08$& $0.42\pm0.31$& $42.81\pm0.010$& $6.41\pm0.04$& $0.81\pm0.06$& $43.47\pm0.008$ \\
{\it Swift}/XRT  &57391& $0.13\pm0.50$        & $1.34\pm0.06$& $0.32\pm0.19$& $42.89\pm0.007$& $4.17\pm0.04$& $0.61\pm0.07$& $42.50\pm0.006$ \\
{\it Swift}/XRT  &58120& $0.17\pm0.50$        & $1.84\pm0.07$& $0.58\pm0.18$& $42.67\pm0.007$& $6.52\pm0.04$& $0.26\pm0.07$& $42.89\pm0.008$ \\
{\it Swift}/XRT  &58667& $0.19\pm0.51$        & $1.42\pm0.08$& $0.11\pm0.08$& $42.11\pm0.014$& $3.85\pm0.14$& $0.59\pm0.10$& $42.42\pm0.006$ \\
\hline \multicolumn{9}{c}{\textbf{Mrk 359}}\\ \hline
Instrument       & MJD & $N_{\rm H}$                & $\Gamma^{\rm PC}$& $Norm^{PC}$ & $\log(L^{\rm PC})$  & $\Gamma^{\rm SE}$& $Norm^{SE}$ & $\log(L^{\rm SE})$ \\
&     & $(10^{20}~cm^{-2})$  &              & $(10^{-3})$ &     (erg/s)     &              & $(10^{-3})$ & (erg/s)  \\
\hline
{\it XMM-Newton} &51734& $3.44\pm0.57$        & $1.54\pm0.07$& $0.98\pm0.24$& $42.68\pm0.005$& $3.19\pm0.05$& $1.63\pm0.89$& $42.50\pm0.003$ \\
{\it XMM-Newton} &55402& $2.48\pm0.58$        & $1.45\pm0.03$& $1.03\pm0.27$& $42.73\pm0.005$& $3.01\pm0.03$& $1.64\pm0.88$& $42.51\pm0.004$ \\
{\it XMM-Newton} &55406& $3.84\pm0.51$        & $1.70\pm0.04$& $1.49\pm0.31$& $42.77\pm0.005$& $3.41\pm0.05$& $1.14\pm0.57$& $42.34\pm0.005$ \\
{\it XMM-Newton} &58508& $3.73\pm0.54$        & $1.23\pm0.05$& $0.64\pm0.37$& $42.64\pm0.006$& $2.96\pm0.04$& $1.72\pm0.41$& $42.53\pm0.003$ \\
{\it XMM-Newton} &58559& $1.94\pm0.55$        & $1.46\pm0.04$& $0.72\pm0.38$& $42.57\pm0.005$& $2.56\pm0.04$& $1.29\pm0.42$& $42.45\pm0.003$ \\
{\it XMM-Newton} &58561& $3.26\pm0.56$        & $1.60\pm0.03$& $1.02\pm0.39$& $42.66\pm0.004$& $3.13\pm0.04$& $1.37\pm0.43$& $42.43\pm0.003$ \\
{\it XMM-Newton} &58564& $3.88\pm0.57$        & $1.12\pm0.03$& $0.36\pm0.35$& $42.33\pm0.008$& $2.90\pm0.05$& $0.89\pm0.41$& $42.25\pm0.003$ \\
{\it XMM-Newton} &58566& $3.86\pm0.58$        & $1.32\pm0.02$& $0.58\pm0.36$& $42.55\pm0.006$& $2.76\pm0.02$& $0.86\pm0.42$& $42.25\pm0.004$ \\
{\it Swift}/XRT  &56119& $1.81\pm0.51$        & $1.11\pm0.08$& $0.33\pm0.37$& $42.47\pm0.030$& $2.31\pm0.05$& $2.60\pm0.84$& $42.51\pm0.008$ \\
{\it Swift}/XRT  &58511& $3.17\pm0.59$        & $1.15\pm0.15$& $0.08\pm0.54$& $42.14\pm0.222$& $2.22\pm0.55$& $2.33\pm1.84$& $42.08\pm0.017$ \\
\hline \multicolumn{9}{c}{\textbf{Mrk 509}}\\ \hline
Instrument       & MJD & $N_{\rm H}$                & $\Gamma^{\rm PC}$& $Norm^{PC}$ & $\log(L^{\rm PC})$  & $\Gamma^{\rm SE}$& $Norm^{SE}$ & $\log(L^{\rm SE})$ \\
&     & $(10^{20}~cm^{-2})$  &              & $(10^{-3})$ &     (erg/s)     &              & $(10^{-3})$ & (erg/s)  \\
\hline
{\it XMM-Newton} &51842& $3.85\pm0.52$        & $1.63\pm0.08$& $6.63\pm0.59$& $43.97\pm0.002$& $4.25\pm0.06$& $2.13\pm0.95$& $43.21\pm0.002$ \\
{\it XMM-Newton} &52019& $1.02\pm0.54$        & $1.64\pm0.01$& $8.59\pm0.55$& $44.07\pm0.001$& $3.55\pm0.02$& $4.25\pm0.84$& $43.44\pm0.001$ \\
{\it XMM-Newton} &53659& $2.74\pm0.57$        & $1.74\pm0.03$& $9.33\pm0.54$& $44.10\pm0.002$& $3.75\pm0.04$& $4.24\pm0.87$& $43.45\pm0.003$ \\
{\it XMM-Newton} &53661& $1.57\pm0.54$        & $1.74\pm0.02$& $9.13\pm0.60$& $44.15\pm0.001$& $3.82\pm0.03$& $6.89\pm0.79$& $43.67\pm0.001$ \\
{\it XMM-Newton} &53663& $1.48\pm0.54$        & $1.76\pm0.01$& $9.36\pm0.54$& $44.08\pm0.001$& $3.37\pm0.02$& $6.15\pm0.82$& $43.54\pm0.001$ \\
{\it XMM-Newton} &53850& $1.01\pm0.57$        & $1.73\pm0.01$& $10.5\pm0.55$& $44.12\pm0.001$& $3.33\pm0.01$& $4.42\pm0.84$& $43.45\pm0.001$ \\
{\it XMM-Newton} &55119& $1.02\pm0.55$        & $1.72\pm0.02$& $10.2\pm0.54$& $44.11\pm0.001$& $3.25\pm0.01$& $6.30\pm0.81$& $43.60\pm0.001$ \\
{\it XMM-Newton} &55123& $0.92\pm0.54$        & $1.75\pm0.01$& $10.8\pm0.55$& $44.13\pm0.001$& $3.17\pm0.01$& $6.51\pm0.82$& $43.62\pm0.001$ \\
{\it XMM-Newton} &55127& $0.84\pm0.54$        & $1.77\pm0.01$& $12.2\pm0.57$& $44.17\pm0.001$& $3.17\pm0.01$& $8.59\pm0.83$& $43.74\pm0.001$ \\
{\it XMM-Newton} &55133& $0.65\pm0.56$        & $1.73\pm0.01$& $9.07\pm0.56$& $44.06\pm0.001$& $3.31\pm0.01$& $7.66\pm0.84$& $43.59\pm0.001$ \\
{\it XMM-Newton} &55137& $0.89\pm0.55$        & $1.76\pm0.01$& $11.1\pm0.55$& $44.13\pm0.001$& $3.34\pm0.01$& $9.97\pm0.85$& $43.60\pm0.001$ \\
{\it XMM-Newton} &55141& $0.81\pm0.54$        & $1.76\pm0.01$& $12.7\pm0.59$& $44.19\pm0.001$& $3.22\pm0.01$& $8.17\pm0.88$& $43.72\pm0.001$ \\
{\it XMM-Newton} &55145& $0.60\pm0.57$        & $1.77\pm0.01$& $12.1\pm0.54$& $44.17\pm0.001$& $3.18\pm0.02$& $7.92\pm0.84$& $43.70\pm0.001$ \\
{\it XMM-Newton} &55149& $0.62\pm0.52$        & $1.77\pm0.02$& $13.2\pm0.61$& $44.20\pm0.001$& $3.08\pm0.01$& $7.88\pm0.82$& $43.70\pm0.001$ \\
{\it XMM-Newton} &55153& $1.07\pm0.53$        & $1.76\pm0.01$& $11.1\pm0.60$& $44.13\pm0.001$& $3.27\pm0.01$& $7.59\pm0.80$& $43.69\pm0.001$ \\
{\it XMM-Newton} &55155& $1.05\pm0.54$        & $1.72\pm0.01$& $12.1\pm0.57$& $44.18\pm0.001$& $3.20\pm0.01$& $7.70\pm0.83$& $43.69\pm0.001$ \\
{\it Swift}/XRT  &53828& $3.95\pm0.55$        & $1.58\pm0.03$& $6.26\pm0.87$& $43.97\pm0.007$& $3.90\pm0.04$& $2.29\pm0.94$& $43.20\pm0.016$ \\
{\it Swift}/XRT  &54185& $3.10\pm0.57$        & $1.03\pm0.08$& $3.34\pm0.88$& $44.00\pm0.015$& $3.05\pm0.04$& $5.68\pm1.05$& $43.56\pm0.016$ \\
{\it Swift}/XRT  &55127& $1.30\pm0.58$        & $1.58\pm0.09$& $9.23\pm0.89$& $44.14\pm0.011$& $2.89\pm0.05$& $4.62\pm0.95$& $43.50\pm0.022$ \\
{\it Swift}/XRT  &57206& $0.95\pm0.59$        & $1.56\pm0.09$& $9.86\pm0.94$& $44.17\pm0.007$& $3.87\pm0.05$& $9.47\pm0.98$& $43.81\pm0.090$ \\
{\it Swift}/XRT  &57454& $2.42\pm0.54$        & $1.60\pm0.07$& $10.0\pm1.05$& $44.16\pm0.007$& $3.27\pm0.07$& $5.10\pm0.99$& $43.51\pm0.015$ \\
{\it Swift}/XRT  &57965& $2.74\pm0.53$        & $1.61\pm0.08$& $8.65\pm1.01$& $44.08\pm0.002$& $3.10\pm0.05$& $3.77\pm1.01$& $43.38\pm0.003$ \\
{\it Swift}/XRT  &58325& $0.85\pm0.53$        & $1.67\pm0.09$& $10.0\pm1.20$& $44.12\pm0.006$& $4.20\pm0.05$& $4.00\pm0.95$& $43.47\pm0.013$ \\
{\it Swift}/XRT  &58689& $2.91\pm0.52$        & $1.52\pm0.08$& $7.43\pm1.04$& $44.18\pm0.007$& $2.71\pm0.03$& $5.51\pm0.99$& $43.58\pm0.010$ \\
{\it Swift}/XRT  &58949& $1.52\pm0.57$        & $1.61\pm0.09$& $7.79\pm0.95$& $44.16\pm0.007$& $3.12\pm0.03$& $5.25\pm0.89$& $43.64\pm0.012$ \\
{\it Swift}/XRT  &59447& $0.83\pm0.55$        & $1.84\pm0.09$& $15.3\pm0.98$& $44.24\pm0.007$& $4.02\pm0.03$& $5.46\pm1.05$& $43.58\pm0.017$ \\
\hline \multicolumn{9}{c}{\textbf{Mrk 841}}\\ \hline
Instrument       & MJD & $N_{\rm H}$                & $\Gamma^{\rm PC}$& $Norm^{PC}$ & $\log(L^{\rm PC})$  & $\Gamma^{\rm SE}$& $Norm^{SE}$ & $\log(L^{\rm SE})$ \\
&     & $(10^{20}~cm^{-2})$  &              & $(10^{-3})$ &     (erg/s)     &              & $(10^{-3})$ & (erg/s)  \\
\hline
{\it XMM-Newton} &51922& $0.24\pm0.58$        & $1.87\pm0.03$& $4.60\pm0.57$& $43.88\pm0.002$& $3.75\pm0.03$& $2.64\pm0.55$& $43.42\pm0.002$ \\
{\it XMM-Newton} &51923& $0.51\pm0.52$        & $1.95\pm0.03$& $5.67\pm0.58$& $43.94\pm0.002$& $4.55\pm0.03$& $2.59\pm0.56$& $43.51\pm0.002$ \\
{\it XMM-Newton} &51924& $3.01\pm0.54$        & $1.76\pm0.04$& $3.58\pm0.57$& $43.80\pm0.003$& $3.64\pm0.04$& $3.44\pm0.58$& $43.13\pm0.002$ \\
{\it XMM-Newton} &53386& $2.28\pm0.55$        & $1.39\pm0.02$& $1.48\pm0.56$& $43.60\pm0.001$& $4.08\pm0.02$& $0.61\pm0.57$& $42.82\pm0.002$ \\
{\it XMM-Newton} &53568& $1.80\pm0.52$        & $1.65\pm0.05$& $2.53\pm0.58$& $43.71\pm0.002$& $4.05\pm0.04$& $0.42\pm0.59$& $42.86\pm0.007$ \\
{\it XMM-Newton} &57217& $0.63\pm0.51$        & $1.90\pm0.06$& $5.03\pm0.67$& $43.90\pm0.005$& $4.81\pm0.06$& $1.20\pm0.63$& $43.21\pm0.007$ \\
{\it Swift}/XRT  &54101& $2.97\pm0.50$        & $1.34\pm0.09$& $1.43\pm0.52$& $43.62\pm0.011$& $3.09\pm0.03$& $1.44\pm0.51$& $43.14\pm0.013$ \\
{\it Swift}/XRT  &57223& $1.50\pm0.58$        & $1.05\pm0.09$& $0.50\pm0.54$& $43.81\pm0.088$& $2.02\pm0.07$& $4.75\pm0.59$& $43.14\pm0.015$ \\
{\it Swift}/XRT  &57640& $0.61\pm0.51$        & $1.63\pm0.05$& $3.37\pm0.49$& $43.84\pm0.010$& $4.07\pm0.03$& $1.70\pm0.43$& $43.26\pm0.016$ \\
{\it Swift}/XRT  &57937& $1.33\pm0.52$        & $1.91\pm0.08$& $4.48\pm0.40$& $43.85\pm0.006$& $4.37\pm0.03$& $0.30\pm0.41$& $42.92\pm0.034$ \\
{\it Swift}/XRT  &58299& $0.57\pm0.55$        & $1.33\pm0.08$& $1.67\pm0.44$& $43.70\pm0.010$& $2.62\pm0.04$& $3.38\pm0.48$& $42.25\pm0.007$ \\
{\it Swift}/XRT  &58665& $3.08\pm0.56$        & $1.64\pm0.05$& $3.23\pm0.47$& $43.82\pm0.006$& $3.44\pm0.03$& $2.09\pm0.51$& $43.30\pm0.010$ \\
{\it Swift}/XRT  &59210& $0.66\pm0.58$        & $1.34\pm0.05$& $2.11\pm0.43$& $43.78\pm0.010$& $2.76\pm0.03$& $3.56\pm0.55$& $43.36\pm0.008$ \\
{\it Swift}/XRT  &59402& $2.93\pm0.53$        & $1.69\pm0.08$& $3.76\pm0.49$& $43.86\pm0.006$& $3.31\pm0.02$& $2.27\pm0.56$& $43.34\pm0.010$ \\
\hline \multicolumn{9}{c}{\textbf{PDS 456}}\\ \hline
Instrument       & MJD & $N_{\rm H}$                & $\Gamma^{\rm PC}$& $Norm^{PC}$ & $\log(L^{\rm PC})$  & $\Gamma^{\rm SE}$& $Norm^{SE}$ & $\log(L^{\rm SE})$ \\
&     & $(10^{20}~cm^{-2})$  &              & $(10^{-3})$ &     (erg/s)     &              & $(10^{-3})$ & (erg/s)  \\
\hline
{\it XMM-Newton} &54355& $1.24\pm0.58$        & $2.07\pm0.03$& $2.70\pm0.57$& $45.08\pm0.002$& $4.59\pm0.03$& $2.12\pm0.59$& $45.08\pm0.002$ \\
{\it XMM-Newton} &54357& $1.78\pm0.57$        & $2.17\pm0.03$& $1.87\pm0.56$& $44.91\pm0.001$& $5.14\pm0.04$& $1.77\pm0.61$& $45.12\pm0.002$ \\
{\it XMM-Newton} &56531& $5.10\pm0.61$        & $2.24\pm0.01$& $3.90\pm0.23$& $45.21\pm0.001$& $7.34\pm0.01$& $7.25\pm0.32$& $46.33\pm0.001$ \\
{\it XMM-Newton} &56550& $1.63\pm0.52$        & $1.81\pm0.02$& $1.10\pm0.44$& $44.77\pm0.001$& $3.91\pm0.02$& $1.25\pm0.62$& $44.72\pm0.001$ \\
{\it XMM-Newton} &56555& $1.78\pm0.54$        & $1.74\pm0.03$& $1.10\pm0.43$& $44.79\pm0.001$& $3.87\pm0.02$& $1.31\pm0.57$& $44.74\pm0.002$ \\
{\it XMM-Newton} &56714& $2.00\pm0.57$        & $1.91\pm0.02$& $1.01\pm0.45$& $44.70\pm0.001$& $4.36\pm0.02$& $0.89\pm0.51$& $44.65\pm0.003$ \\
{\it XMM-Newton} &57835& $1.39\pm0.52$        & $1.58\pm0.03$& $0.50\pm0.59$& $44.48\pm0.002$& $4.67\pm0.03$& $0.20\pm0.57$& $44.04\pm0.008$ \\
{\it XMM-Newton} &57837& $1.67\pm0.53$        & $1.77\pm0.04$& $0.77\pm0.58$& $44.63\pm0.002$& $3.69\pm0.04$& $0.39\pm0.54$& $44.19\pm0.006$ \\
{\it XMM-Newton} &58381& $1.25\pm0.58$        & $1.97\pm0.02$& $3.65\pm0.51$& $45.00\pm0.001$& $4.85\pm0.03$& $1.48\pm0.45$& $44.98\pm0.002$ \\
{\it XMM-Newton} &58750& $1.86\pm0.59$        & $2.21\pm0.03$& $1.23\pm0.59$& $44.82\pm0.001$& $6.14\pm0.04$& $0.67\pm0.45$& $45.10\pm0.003$ \\
{\it Swift}/XRT  &55022& $4.90\pm0.81$        & $2.13\pm0.03$& $1.93\pm0.61$& $44.93\pm0.004$& $6.38\pm0.04$& $0.36\pm0.58$& $45.75\pm0.015$ \\
{\it Swift}/XRT  &55250& $5.81\pm0.84$        & $1.66\pm0.07$& $0.78\pm0.65$& $44.67\pm0.011$& $3.21\pm0.05$& $1.08\pm0.57$& $44.59\pm0.014$ \\
{\it Swift}/XRT  &57935& $1.69\pm0.92$        & $1.99\pm0.08$& $0.12\pm0.52$& $44.69\pm0.045$& $1.99\pm0.09$& $1.30\pm0.94$& $44.79\pm0.004$ \\
{\it Swift}/XRT  &58374& $2.78\pm0.94$        & $2.05\pm0.09$& $1.41\pm0.59$& $44.80\pm0.005$& $2.39\pm0.04$& $0.99\pm0.52$& $44.59\pm0.008$ \\
{\it Swift}/XRT  &58752& $4.52\pm0.82$        & $2.23\pm0.07$& $1.54\pm0.57$& $45.10\pm0.006$& $6.52\pm0.08$& $1.36\pm0.58$& $44.90\pm0.026$ \\
{\it Swift}/XRT  &59445& $2.63\pm0.85$        & $2.26\pm0.04$& $1.27\pm0.59$& $44.71\pm0.054$& $2.23\pm0.04$& $1.46\pm0.57$& $44.78\pm0.046$ \\
\hline \multicolumn{9}{c}{\textbf{PKS 0558-504}}\\ \hline
Instrument       & MJD & $N_{\rm H}$                & $\Gamma^{\rm PC}$& $Norm^{PC}$ & $\log(L^{\rm PC})$  & $\Gamma^{\rm SE}$& $Norm^{SE}$ & $\log(L^{\rm SE})$ \\
&     & $(10^{20}~cm^{-2})$  &              & $(10^{-3})$ &     (erg/s)     &              & $(10^{-3})$ & (erg/s)  \\
\hline
{\it XMM-Newton} &51586& $0.51\pm0.58$        & $2.06\pm0.03$& $5.41\pm0.58$& $45.10\pm0.004$& $3.88\pm0.02$& $3.30\pm0.59$& $44.82\pm0.004$ \\
{\it XMM-Newton} &51688& $2.92\pm1.57$        & $1.97\pm0.09$& $3.77\pm1.88$& $44.96\pm0.009$& $3.85\pm0.07$& $4.05\pm1.57$& $44.88\pm0.005$ \\
{\it XMM-Newton} &51827& $2.18\pm0.75$        & $2.09\pm0.04$& $5.37\pm0.67$& $45.19\pm0.003$& $5.17\pm0.04$& $3.57\pm0.51$& $45.09\pm0.002$ \\
{\it XMM-Newton} &52086& $1.79\pm0.54$        & $1.79\pm0.02$& $3.45\pm0.54$& $44.98\pm0.007$& $3.51\pm0.03$& $3.26\pm0.52$& $44.77\pm0.005$ \\
{\it XMM-Newton} &52201& $3.28\pm0.51$        & $2.17\pm0.03$& $12.4\pm0.58$& $45.23\pm0.004$& $3.43\pm0.03$& $6.25\pm0.55$& $44.07\pm0.005$ \\
{\it XMM-Newton} &54718& $2.02\pm0.56$        & $2.02\pm0.01$& $4.44\pm0.52$& $45.02\pm0.006$& $3.96\pm0.01$& $3.24\pm0.52$& $44.82\pm0.005$ \\
{\it XMM-Newton} &54720& $2.45\pm0.54$        & $2.07\pm0.02$& $6.06\pm0.53$& $45.15\pm0.001$& $4.43\pm0.01$& $3.39\pm0.53$& $44.92\pm0.005$ \\
{\it XMM-Newton} &54722& $1.81\pm0.53$        & $2.02\pm0.02$& $4.93\pm0.54$& $45.07\pm0.001$& $4.19\pm0.01$& $2.96\pm0.55$& $44.82\pm0.006$ \\
{\it XMM-Newton} &54726& $1.76\pm0.53$        & $2.04\pm0.01$& $6.23\pm0.57$& $45.16\pm0.001$& $4.28\pm0.02$& $3.77\pm0.59$& $44.94\pm0.005$ \\
{\it Swift}/XRT  &54740& $2.01\pm0.59$        & $1.84\pm0.03$& $4.27\pm0.51$& $45.06\pm0.003$& $3.72\pm0.03$& $3.76\pm0.58$& $44.86\pm0.003$ \\
{\it Swift}/XRT  &54921& $1.91\pm0.57$        & $2.01\pm0.02$& $7.42\pm0.52$& $45.25\pm0.003$& $4.65\pm0.02$& $5.69\pm0.54$& $45.19\pm0.003$ \\
{\it Swift}/XRT  &55105& $1.49\pm0.52$        & $2.07\pm0.02$& $6.59\pm0.58$& $45.18\pm0.003$& $5.11\pm0.02$& $4.53\pm0.54$& $45.18\pm0.004$ \\
{\it Swift}/XRT  &55242& $1.04\pm0.55$        & $1.98\pm0.03$& $7.88\pm0.57$& $45.28\pm0.004$& $4.42\pm0.03$& $6.37\pm0.59$& $45.19\pm0.005$ \\
{\it Swift}/XRT  &57711& $5.91\pm0.51$        & $2.17\pm0.05$& $5.57\pm0.59$& $45.07\pm0.013$& $4.86\pm0.05$& $2.31\pm0.61$& $45.83\pm0.019$ \\
\hline \multicolumn{9}{c}{\textbf{SWIFT J0501.9-3239}}\\ \hline
Instrument       & MJD & $N_{\rm H}$                & $\Gamma^{\rm PC}$& $Norm^{PC}$ & $\log(L^{\rm PC})$  & $\Gamma^{\rm SE}$& $Norm^{SE}$ & $\log(L^{\rm SE})$ \\
&     & $(10^{20}~cm^{-2})$  &              & $(10^{-3})$ &     (erg/s)     &              & $(10^{-3})$ & (erg/s)  \\
\hline
{\it XMM-Newton} &55225& $0.92\pm0.45$        & $1.28\pm0.02$& $1.18\pm0.49$& $42.63\pm0.001$& $4.90\pm0.02$& $0.32\pm0.49$& $41.87\pm0.002$ \\
{\it XMM-Newton} &57655& $0.26\pm0.42$        & $1.34\pm0.01$& $1.05\pm0.52$& $42.55\pm0.001$& $3.08\pm0.02$& $0.53\pm0.48$& $41.75\pm0.002$ \\
{\it Swift}/XRT  &53686& $0.77\pm0.43$        & $1.25\pm0.02$& $1.35\pm0.55$& $42.83\pm0.010$& $3.65\pm0.03$& $0.62\pm0.41$& $42.83\pm0.020$ \\
{\it Swift}/XRT  &54476& $0.16\pm0.41$        & $1.12\pm0.08$& $1.34\pm0.89$& $42.84\pm0.023$& $2.69\pm0.07$& $3.12\pm0.45$& $42.56\pm0.017$ \\
{\it Swift}/XRT  &57853& $0.46\pm0.46$        & $1.28\pm0.12$& $1.17\pm1.09$& $42.76\pm0.027$& $2.44\pm0.10$& $1.32\pm0.89$& $42.23\pm0.034$ \\
\hline \multicolumn{9}{c}{\textbf{NGC 7469}}\\ \hline
Instrument       & MJD & $N_{\rm H}$                & $\Gamma^{\rm PC}$& $Norm^{PC}$ & $\log(L^{\rm PC})$  & $\Gamma^{\rm SE}$& $Norm^{SE}$ & $\log(L^{\rm SE})$ \\
&     & $(10^{20}~cm^{-2})$  &              & $(10^{-3})$ &     (erg/s)     &              & $(10^{-3})$ & (erg/s)  \\
\hline
{\it XMM-Newton} &51940& $0.70\pm0.47$        & $1.80\pm0.05$& $7.07\pm0.58$& $43.38\pm0.002$& $4.12\pm0.05$& $4.86\pm0.52$& $42.99\pm0.002$ \\
{\it XMM-Newton} &51941& $0.80\pm0.46$        & $1.82\pm0.04$& $7.73\pm0.59$& $43.41\pm0.004$& $4.14\pm0.05$& $6.95\pm0.83$& $43.05\pm0.006$ \\
{\it XMM-Newton} &53339& $0.77\pm0.49$        & $1.85\pm0.01$& $8.52\pm0.57$& $43.44\pm0.003$& $4.30\pm0.02$& $6.23\pm0.84$& $43.12\pm0.005$ \\
{\it XMM-Newton} &53076& $0.84\pm0.45$        & $1.81\pm0.02$& $8.42\pm0.56$& $43.45\pm0.005$& $4.32\pm0.02$& $5.45\pm0.54$& $43.07\pm0.009$ \\
{\it XMM-Newton} &57185& $0.60\pm0.42$        & $1.86\pm0.02$& $8.66\pm0.53$& $43.44\pm0.004$& $4.04\pm0.02$& $5.93\pm0.53$& $43.07\pm0.005$ \\
{\it XMM-Newton} &57350& $0.10\pm0.43$        & $1.79\pm0.01$& $7.86\pm0.54$& $43.43\pm0.003$& $3.47\pm0.01$& $3.28\pm0.55$& $42.77\pm0.006$ \\
{\it XMM-Newton} &57371& $0.71\pm0.44$        & $1.76\pm0.01$& $6.04\pm0.55$& $43.33\pm0.002$& $4.04\pm0.01$& $4.60\pm0.51$& $42.96\pm0.007$ \\
{\it XMM-Newton} &57379& $0.62\pm0.49$        & $1.76\pm0.01$& $6.51\pm0.58$& $43.36\pm0.001$& $3.88\pm0.02$& $4.60\pm0.52$& $42.95\pm0.003$ \\
{\it XMM-Newton} &57380& $0.53\pm0.41$        & $1.77\pm0.02$& $7.24\pm0.57$& $43.41\pm0.005$& $3.67\pm0.03$& $5.48\pm0.57$& $43.01\pm0.008$ \\
{\it XMM-Newton} &57382& $0.61\pm0.47$        & $1.76\pm0.02$& $7.24\pm0.59$& $43.41\pm0.006$& $4.00\pm0.02$& $4.33\pm0.58$& $42.93\pm0.009$ \\
{\it XMM-Newton} &57384& $0.50\pm0.40$        & $1.77\pm0.02$& $7.80\pm0.50$& $43.44\pm0.002$& $3.37\pm0.02$& $3.27\pm0.58$& $42.77\pm0.007$ \\
{\it Swift}/XRT  &53899& $0.10\pm0.61$        & $1.23\pm0.04$& $1.56\pm0.68$& $43.01\pm0.016$& $2.50\pm0.04$& $5.69\pm0.74$& $42.88\pm0.001$ \\
{\it Swift}/XRT  &54704& $0.61\pm0.67$        & $1.54\pm0.02$& $2.54\pm0.67$& $43.05\pm0.005$& $3.22\pm0.03$& $2.70\pm0.73$& $42.69\pm0.007$ \\
{\it Swift}/XRT  &56528& $0.42\pm0.65$        & $1.84\pm0.01$& $7.44\pm0.61$& $43.39\pm0.004$& $3.09\pm0.02$& $2.48\pm0.71$& $42.66\pm0.022$ \\
{\it Swift}/XRT  &56462& $0.12\pm0.60$        & $1.60\pm0.01$& $2.94\pm0.62$& $43.09\pm0.005$& $2.45\pm0.02$& $2.14\pm0.72$& $42.67\pm0.009$ \\
{\it Swift}/XRT  &56491& $0.11\pm0.61$        & $1.73\pm0.02$& $2.42\pm0.63$& $43.94\pm0.006$& $2.48\pm0.02$& $0.63\pm0.74$& $42.52\pm0.024$ \\
{\it Swift}/XRT  &56514& $0.08\pm0.60$        & $1.76\pm0.02$& $4.59\pm0.68$& $43.21\pm0.005$& $3.72\pm0.02$& $2.76\pm0.70$& $42.71\pm0.010$ \\
{\it Swift}/XRT  &57286& $0.13\pm0.64$        & $2.05\pm0.01$& $6.58\pm0.67$& $43.26\pm0.003$& $1.91\pm0.03$& $0.79\pm0.71$& $42.78\pm0.012$ \\
{\it Swift}/XRT  &57564& $0.14\pm0.62$        & $1.66\pm0.03$& $4.51\pm0.64$& $43.24\pm0.008$& $2.45\pm0.04$& $4.07\pm0.72$& $42.95\pm0.011$ \\
{\it Swift}/XRT  &57991& $0.21\pm0.61$        & $1.35\pm0.03$& $2.21\pm0.63$& $43.09\pm0.016$& $2.43\pm0.03$& $8.06\pm0.78$& $43.00\pm0.007$ \\
{\it Swift}/XRT  &58359& $0.17\pm0.69$        & $1.81\pm0.03$& $10.1\pm0.67$& $43.53\pm0.006$& $3.23\pm0.03$& $4.23\pm0.71$& $42.89\pm0.013$ \\
{\it Swift}/XRT  &58676& $0.41\pm0.65$        & $1.73\pm0.03$& $7.05\pm0.61$& $43.41\pm0.007$& $3.17\pm0.04$& $7.01\pm0.72$& $43.11\pm0.009$ \\
{\it Swift}/XRT  &59087& $0.08\pm0.67$        & $1.34\pm0.02$& $10.8\pm0.64$& $43.42\pm0.003$& $2.78\pm0.04$& $0.46\pm0.70$& $42.77\pm0.021$ \\
\hline \multicolumn{9}{c}{\textbf{Ton S180}}\\ \hline
Instrument       & MJD & $N_{\rm H}$                & $\Gamma^{\rm PC}$& $Norm^{PC}$ & $\log(L^{\rm PC})$  & $\Gamma^{\rm SE}$& $Norm^{SE}$ & $\log(L^{\rm SE})$ \\
&     & $(10^{20}~cm^{-2})$  &              & $(10^{-3})$ &     (erg/s)     &              & $(10^{-3})$ & (erg/s)  \\
\hline
{\it XMM-Newton} &51892& $2.88\pm0.51$        & $2.10\pm0.03$& $1.96\pm0.58$& $43.91\pm0.002$& $4.05\pm0.03$& $2.64\pm0.53$& $43.95\pm0.001$ \\
{\it XMM-Newton} &52455& $1.13\pm0.58$        & $1.93\pm0.03$& $1.02\pm0.55$& $43.67\pm0.004$& $2.43\pm0.03$& $2.43\pm0.57$& $43.75\pm0.001$ \\
{\it XMM-Newton} &57206& $0.83\pm0.57$        & $2.06\pm0.02$& $1.52\pm0.51$& $43.81\pm0.002$& $3.98\pm0.02$& $1.68\pm0.51$& $43.74\pm0.002$ \\
{\it XMM-Newton} &57552& $0.70\pm0.55$        & $2.13\pm0.03$& $0.99\pm0.50$& $43.60\pm0.003$& $3.56\pm0.03$& $1.06\pm0.52$& $43.51\pm0.002$ \\
{\it Swift}/XRT  &57907& $2.07\pm0.52$        & $1.14\pm0.09$& $0.50\pm0.62$& $43.78\pm0.057$& $3.44\pm0.07$& $2.80\pm0.68$& $43.92\pm0.018$ \\
{\it Swift}/XRT  &59339& $2.95\pm0.54$        & $1.15\pm0.09$& $0.34\pm0.64$& $43.82\pm0.057$& $3.13\pm0.08$& $2.88\pm0.65$& $43.93\pm0.014$ \\
\hline \multicolumn{9}{c}{\textbf{UGC 6728}}\\ \hline
Instrument       & MJD & $N_{\rm H}$                & $\Gamma^{\rm PC}$& $Norm^{PC}$ & $\log(L^{\rm PC})$  & $\Gamma^{\rm SE}$& $Norm^{SE}$ & $\log(L^{\rm SE})$ \\
&     & $(10^{20}~cm^{-2})$  &              & $(10^{-3})$ &     (erg/s)     &              & $(10^{-3})$ & (erg/s)  \\
\hline
{\it XMM-Newton} &53789& $0.81\pm0.45$        & $1.39\pm0.02$& $1.01\pm0.48$& $41.72\pm0.005$& $3.01\pm0.02$& $1.47\pm0.47$& $41.52\pm0.004$ \\
{\it Swift}/XRT  &57579& $0.89\pm0.49$        & $1.14\pm0.08$& $0.53\pm0.59$& $41.42\pm0.021$& $2.82\pm0.06$& $1.03\pm0.56$& $38.74\pm0.021$ \\
{\it Swift}/XRT  &58039& $3.74\pm0.44$        & $1.66\pm0.07$& $3.50\pm0.57$& $42.04\pm0.008$& $4.57\pm0.01$& $0.20\pm0.57$& $41.78\pm0.067$ \\
{\it Swift}/XRT  &59097& $2.84\pm0.46$        & $1.25\pm0.08$& $0.61\pm0.64$& $41.80\pm0.040$& $3.73\pm0.07$& $2.17\pm0.68$& $41.76\pm0.037$ \\
{\it Swift}/XRT  &59414& $1.02\pm0.42$        & $1.23\pm0.09$& $0.86\pm0.47$& $41.72\pm0.033$& $3.06\pm0.05$& $2.61\pm0.49$& $41.71\pm0.022$ \\
\end{longtable}
	\end{center}

\bibliography{references}{}
\bibliographystyle{aasjournal}



\end{document}